\documentclass[twocolumn]{aastex631}

\usepackage{multirow}
\usepackage{amsmath}
\usepackage{xfrac}

\newcommand\lya{Lyman-$\alpha$}
\newcommand\hdb{HD~189733~b}

\shorttitle{Detection of hydrodynamic escape of carbon in HD~189733~\lowercase{b}}
\shortauthors{Dos Santos et al.}
\graphicspath{{./}{figures/}}

\begin{document}

\title{Hydrodynamic atmospheric escape in HD~189733~b: Signatures of carbon and hydrogen measured with the \emph{Hubble Space Telescope}}

\author[0000-0002-2248-3838]{Leonardo A. Dos Santos}
\affiliation{Space Telescope Science Institute, 3700 San Martin Drive, Baltimore, MD 21218, USA}
\email{ldsantos@stsci.edu}

\author[0000-0003-1756-4825]{Antonio Garc\'ia Mu\~noz}
\affiliation{Universit\'e Paris-Saclay, Universit\'e Paris Cit\'e, CEA, CNRS, AIM, 91191, Gif-sur-Yvette, France}

\author[0000-0001-6050-7645]{David K. Sing}
\affiliation{Department of Physics and Astronomy, Johns Hopkins University, Baltimore, MD 21218, USA}
\affiliation{Department of Earth \& Planetary Sciences, Johns Hopkins University, Baltimore, MD, USA}

\author[0000-0003-3204-8183]{Mercedes L\'opez-Morales}
\affiliation{Center for Astrophysics ${\rm \mid}$ Harvard {\rm \&} Smithsonian, 60 Garden St, Cambridge, MA 02138, USA}

\author[0000-0003-4157-832X]{Munazza K. Alam}
\affiliation{Carnegie Earth \& Planets Laboratory, 5241 Broad Branch Road NW, Washington, DC 20015, USA}

\author{Vincent Bourrier}
\affil{Observatoire astronomique de l'Universit\'e de Gen\`eve, Chemin Pegasi 51, 1290 Versoix, Switzerland}

\author{David Ehrenreich}
\affiliation{Observatoire astronomique de l'Universit\'e de Gen\`eve, Chemin Pegasi 51, 1290 Versoix, Switzerland}

\author[0000-0003-4155-8513]{Gregory W. Henry}
\affiliation{Center of Excellence in Information Systems, Tennessee State University, Nashville, TN  37209,  USA}

\author{Alain Lecavelier des Etangs}
\affiliation{Institut d'Astrophysique de Paris, CNRS, UMR 7095 \& Sorbonne Universit\'es UPMC Paris 6, 98bis bd Arago, 75014 Paris, France}

\author[0000-0001-5442-1300]{Thomas Mikal-Evans}
\affil{Max Planck Institute for Astronomy, K\"{o}nigstuhl 17, D-69117 Heidelberg, Germany}

\author[0000-0002-6500-3574]{Nikolay K. Nikolov}
\affiliation{Space Telescope Science Institute, 3700 San Martin Drive, Baltimore, MD 21218, USA}

\author[0000-0002-1600-7835]{Jorge Sanz-Forcada}
\affiliation{Departamento de Astrofi\'isica, Centro de Astrobiologi\'ia (CSIC-INTA), ESAC Campus, Camino bajo del Castillos s/n, 28692 Villanueva de la Ca\~nada, Madrid, Spain}

\author[0000-0003-4328-3867]{Hannah R. Wakeford}
\affiliation{University of Bristol, HH Wills Physics Laboratory, Tyndall Avenue, Bristol, UK}

\begin{abstract}

One of the most well-studied exoplanets to date, HD~189733~b, stands out as an archetypal hot Jupiter with many observations and theoretical models aimed at characterizing its atmosphere, interior, host star, and environment. We report here on the results of an extensive campaign to observe atmospheric escape signatures in HD~189733~b using the \emph{Hubble Space Telescope} and its unique ultraviolet capabilities. We have found a tentative, but repeatable in-transit absorption of singly-ionized carbon (C\,{\sc ii}, $5.2\% \pm 1.4\%$) in the epoch of June-July/2017, as well as a neutral hydrogen (H\,{\sc i}) absorption consistent with previous observations. We model the hydrodynamic outflow of HD~189733~b using an isothermal Parker wind formulation to interpret the observations of escaping C and O nuclei at the altitudes probed by our observations. Our forward models indicate that the outflow of HD~189733~b is mostly neutral within an altitude of $\sim 2$~R$_{\rm p}$ and singly ionized beyond that point. The measured in-transit absorption of C\,{\sc ii} at 1335.7~\AA\, is consistent with an escape rate of $\sim 1.1 \times 10^{11}$~g\,s$^{-1}$, assuming solar C abundance and outflow temperature of $12\,100$~K. Although we find a marginal neutral oxygen (O\,{\sc i}) in-transit absorption, our models predict an in-transit depth that is only comparable to the size of measurement uncertainties. A comparison between the observed \lya\ transit depths and hydrodynamics models suggests that the exosphere of this planet interacts with a stellar wind at least one order of magnitude stronger than solar. 

\end{abstract}


\section{Introduction} \label{sec:intro}

One of the most striking discoveries in the search for exoplanets is that they can orbit their host stars at extremely close-in distances, a fact that initially challenged our understanding of planetary formation outside the Solar System \citep[see, e.g., the recent review by][]{2021ARA&A..59..291Z}. In particular, hot-Jupiters were the first exoplanets to be found because they imprint strong transit and gravitational reflex signals in their host stars, despite their being intrinsically rare \citep{Yee2021}. Although as a community we ultimately aspire to find another planet similar to the Earth, hot-Jupiters stand out as an important stepping stone because they are excellent laboratories to test our hypotheses of how planetary systems form and evolve \citep[e.g.,][]{2021JGRE..12606629F}.

For small, short-period exoplanets, the impinging irradiation from their host stars and how it varies with time are some of the most important factors that drive the evolution of their atmospheres \citep[see, e.g.,][]{Owen2019}. That is because the incoming energetic photons (with wavelengths between X-rays and extreme-ultraviolet, or XUV) heat the upper atmosphere of the planet, which in turn expands and produces outflowing winds. If this outflow becomes supersonic, the atmospheric escape process is said to be hydrodynamic. Originally formulated by \citet{Watson1981} to describe the evolution of the early Earth and Venus, hydrodynamic escape has been observed in action in many hot exoplanets \citep[e.g.,][]{Vidal03, Vidal2004, Fossati2010, Sing2019}. Other factors such as composition, as in high mean molecular weight atmospheres, are also important in regulating the mass-loss rate of exoplanets \citep[e.g.,][]{GarciaMunoz21, Ito2021, Nakayama2022}. 

The hot-Jupiter HD~189733~b \citep{Bouchy2005} is a particularly well-studied exoplanet owing to: i) its proximity to the Solar System, ii) size and mass in relation to its host star, and iii) short orbital period --- see the stellar and planetary parameters in Table \ref{tab:params}; these were compiled following the most recent and most complete datasets available in the literature, aiming for precision and consistency. 

Previous optical observations of the atmosphere of HD~189733~b have shown that its transmission spectrum is consistent with the presence of high-altitude hazes \citep{Lecavelier2008a, 2011MNRAS.416.1443S, 2016Natur.529...59S}. In the near-infrared, the low amplitude of the water feature in transmission indicates a depletion of H$_2$O abundance from solar values, likely a result from its formation \citep{2014ApJ...791...55M, 2014ApJ...791L...9M}. Using both transit and eclipse data of this planet, \citet{2020ApJ...899...27Z} concluded that the C/O ratio of HD~189733~b is $\sim 0.66$, and that it has a super-solar atmospheric metallicity.

Using different observational and theoretical techniques, previous atmospheric escape studies of HD~189733~b found that the planet likely has high mass-loss rates in the order of $10^{10} - 10^{11}$~g\,s$^{-1}$, which is consistent with a hydrodynamic outflow \citep{Lecavelier2010, Lecavelier2012, 2013A&A...557A.124B, Salz16b, Lampon21}. In this regime, the outflow of H is so intense that it can drag heavier species, such as C and O, upwards to the exosphere of the planet, where these nuclei can quickly photoionize. In this context, \citet{2013A&A...553A..52B} reported on the detection of neutral oxygen (O\,{\sc i}) in the exosphere of HD~189733~b, which the authors attribute to atmospheric escape, but require super-solar abundances and super-thermal line broadening to be explained; they also report a non-detection of singly-ionized carbon (C\,{\sc ii}). More recently, \citet{2023A&A...671A.170C} ruled-out the presence of singly-ionized magnesium (Mg\,{\sc ii}) in the outflow of this planet and, although they reported a non-detection of Mg\,{\sc i}, they did not rule out the presence of this species.

In this manuscript, we report on a comprehensive analysis of all the far-ultraviolet (FUV) transit observations of HD~189733~b performed with the \emph{Hubble Space Telescope} and the Cosmic Origins Spectrograph (COS) instrument obtained to date. In Section \ref{sec:methods}, we describe the observational setup and data reduction steps; Section \ref{sec:results} contains the results of our data analysis in the form of spectroscopic light curves; in Section \ref{sect:models}, we discuss the models we used to interpret our results and how they compare with the literature; in Section \ref{sec:conclusions} we lay the main conclusions of this work.

\begin{deluxetable*}{llcl}
    \tablecaption{Stellar and planetary parameters, and transit ephemeris of the HD~189733~b system\label{tab:params}}
    \tablewidth{0pt}
    \tablehead{
    \colhead{Parameter} & \colhead{Unit} & \colhead{Value} & \colhead{Ref.}
    }
    \startdata
    Stellar radius                 & R$_\odot$         & $0.765^{+0.019}_{-0.018}$                 & \citet{2019PASP..131k5003A} \\
    Stellar mass                   & M$_\odot$         & $0.812^{+0.041}_{-0.038}$                 & \citet{2019PASP..131k5003A} \\
    Stellar effective temperature  & K                 & $5050 \pm 20$                             & \citet{2019PASP..131k5003A} \\
    Projected rotational velocity  & km s$^{-1}$       & $3.5 \pm 1.0$                             & \citet{Bonomo2017} \\
    Age                            & Gyr               & $\sim 1.2$                    & \citet{Sanz2010} \\
    Systemic radial velocity       & km s$^{-1}$       & $-2.204^{+0.010}_{-0.011}$                & \citet{2019PASP..131k5003A} \\
    Distance                       & pc                & $19.7638^{+0.0128}_{-0.0127}$             & \citet{Gaia2018} \\
    Spectral type                  &                   & K2V                                       & \citet{Gray2003} \\
    Planetary radius               & R$_\mathrm{Jup}$  & $1.119 \pm 0.038$                         & \citet{2019PASP..131k5003A} \\
    Planet-to-star ratio           & $R_{\rm p} / R_{\rm s}$ & $0.1504^{+0.0038}_{-0.0039}$ & \citet{2019PASP..131k5003A} \\
    Planetary mass                 & M$_\mathrm{Jup}$  & $1.166^{+0.052}_{-0.049}$                 & \citet{2019PASP..131k5003A} \\
    Planetary density              & g cm$^{-3}$       & $1.031^{+0.106}_{-0.090}$                 & \citet{2019PASP..131k5003A} \\
    Planetary eq. temperature      & K                 & $1209 \pm 11$                             & \citet{2019PASP..131k5003A} \\
    Orbital period                 & d                 & $2.218577^{+0.000009}_{-0.000010}$        & \citet{2019PASP..131k5003A} \\
    Semi-major axis                & au                & $0.03106^{+0.00051}_{-0.00049}$           & \citet{2019PASP..131k5003A} \\
    Orbital inclination            & $\deg$            & $85.690^{+0.095}_{-0.097}$                & \citet{2019PASP..131k5003A} \\
    Eccentricity                   &                   & $< 0.0039$                                & \citet{Bonomo2017} \\
    Transit center reference time  & BJD               & $2458334.990899^{+0.000726}_{-0.000781}$  & \citet{2019PASP..131k5003A} \\
    Transit dur. (1st-4th contact) & h                 & $1.84 \pm 0.04$                           & \citet{2019PASP..131k5003A} \\
    \enddata
    \tablecomments{Stellar and planetary parameters were obtained in the following DOI: \dataset[10.26133/NEA2]{https://doi.org/10.26133/NEA2}}
    \end{deluxetable*}

\section{Observations and data analysis} \label{sec:methods}

Several FUV transits of \hdb\ have been observed with \emph{HST} in the General Observer programs 11673 (PI: Lecavelier des Etangs), 14767 (PanCET program; PIs: Sing \& L\'opez-Morales), and 15710 (PI: Cauley). Another program (12984, PI: Pillitteri) also observed HD~189733 in the frame of star-planet interactions, but no in-transit exposures were obtained. In program 15710, which aimed at measuring transits simultaneously with \emph{HST} and ground-based facilities, two of three visits had guide star problems and are not usable; the third visit has only one exposure covering in-transit fluxes, and the remaining ones occur after the transit; this non-optimal transit coverage is likely the result of difficulties in coordinating \emph{HST} and ground-based observatories for simultaneous observations. We list the dataset identifiers and times of observation in Table \ref{tab:datasets_cos}\footnote{\footnotesize{These data are openly available in the following DOI: \dataset[10.17909/2dq3-g745]{https://doi.org/10.17909/2dq3-g745}.}}. Each identifier corresponds to one exposure of \emph{HST}, or one orbit.

The COS observations were set to spectroscopic element G130M centered at 1291 \AA\ and a circular aperture with diameter 2.5 arcsec, yielding wavelength ranges [1134, 1274] \AA\ and [1290, 1429] \AA. The data were reduced automatically by the instrument pipeline \citep[CALCOS version 3.3.11, which has corrected the bug with inflated uncertainties;][]{Johnson2021}. Several FUV transits of HD~189733~b have also been observed with the STIS spectrograph, but with a more limited wavelength range --- thorough analyses of the STIS datasets are discussed in \citet{Lecavelier2012}, \citet{Bourrier2013, Bourrier2020}, and \citet{Barth2021}. In this manuscript we focus only on the COS data, which cover more metallic emission lines than the STIS data. 

\begin{deluxetable*}{l c c c l}
\tablecaption{Observations log of HD~189733~b transits with \emph{HST}/COS. \label{tab:datasets_cos}}
\tablewidth{0pt}
\tablehead{
\colhead{Visit} & \colhead{Dataset} & \colhead{Start time (BJD)} & \colhead{Exp. time (s)} & \colhead{Phase}}
\startdata
\multirow{8}{*}{A} & \texttt{lb5k01ukq} & 2009-09-16 18:31:52.378 & 208.99 & Out of transit \\
& \texttt{lb5k01uoq} & 2009-09-16 19:50:49.344 & 889.18 & Out of transit \\
& \texttt{lb5ka1usq} & 2009-09-16 21:26:41.338 & 889.18 & Out of transit \\
& \texttt{lb5ka1uuq} & 2009-09-16 21:44:13.344 & 889.15 & Ingress \\
& \texttt{lb5ka1v2q} & 2009-09-16 23:02:33.331 & 889.15 & In transit \\
& \texttt{lb5ka1v4q} & 2009-09-16 23:20:05.338 & 889.18 & Egress \\
& \texttt{lb5ka1vmq} & 2009-09-17 00:38:25.325 & 889.18 & Post-transit \\
& \texttt{lb5ka1vpq} & 2009-09-17 00:55:57.331 & 889.18 & Post-transit \\
\hline
\multirow{5}{*}{B} & \texttt{ld9m50oxq} & 2017-06-24 08:03:55.843 & 2018.18 & Out of transit \\
& \texttt{ld9m50ozq} & 2017-06-24 09:24:16.877 & 2707.20 & Out of transit  \\
& \texttt{ld9m50p1q} & 2017-06-24 10:59:38.803 & 2707.17 & Ingress  \\
& \texttt{ld9m50p3q} & 2017-06-24 12:35:00.816 & 2707.17 & Egress  \\
& \texttt{ld9m50p5q} & 2017-06-24 14:10:23.866 & 2707.17 & Post-transit \\
\hline
\multirow{5}{*}{C} & \texttt{ld9m51clq} & 2017-07-03 05:00:55.786 & 2018.18 & Out of transit \\
& \texttt{ld9m51drq} & 2017-07-03 06:21:52.934 & 2707.20 & Out of transit \\
& \texttt{ld9m51duq} & 2017-07-03 07:57:12.960 & 2707.20 & Ingress \\
& \texttt{ld9m51dwq} & 2017-07-03 09:32:33.850 & 2707.20 & Egress \\
& \texttt{ld9m51dyq} & 2017-07-03 11:07:54.826 & 2707.20 & Post-transit \\
\hline
\multirow{5}{*}{D} & \texttt{ldzkh1ifq} & 2020-09-01 06:23:14.842 & 1068.16 & In transit \\
& \texttt{ldzkh1juq} & 2020-09-01 07:47:24.835 & 2060.19 & Post-transit \\
& \texttt{ldzkh1kbq} & 2020-09-01 09:22:43.824 & 2435.17 & Post-transit \\
& \texttt{ldzkh1l4q} & 2020-09-01 10:58:01.862 & 2600.19 & Post-transit \\
& \texttt{ldzkh1l7q} & 2020-09-01 12:33:20.851 & 2600.19 & Post-transit \\
\enddata
\end{deluxetable*}

\begin{figure*}[!ht]
\gridline{\fig{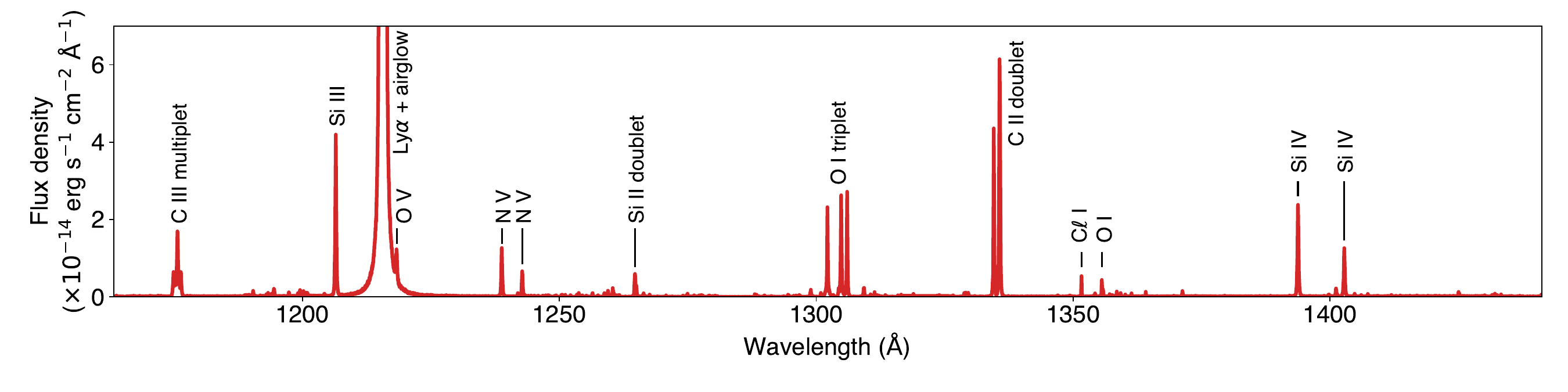}{1.0\textwidth}{(a)}}
\gridline{\fig{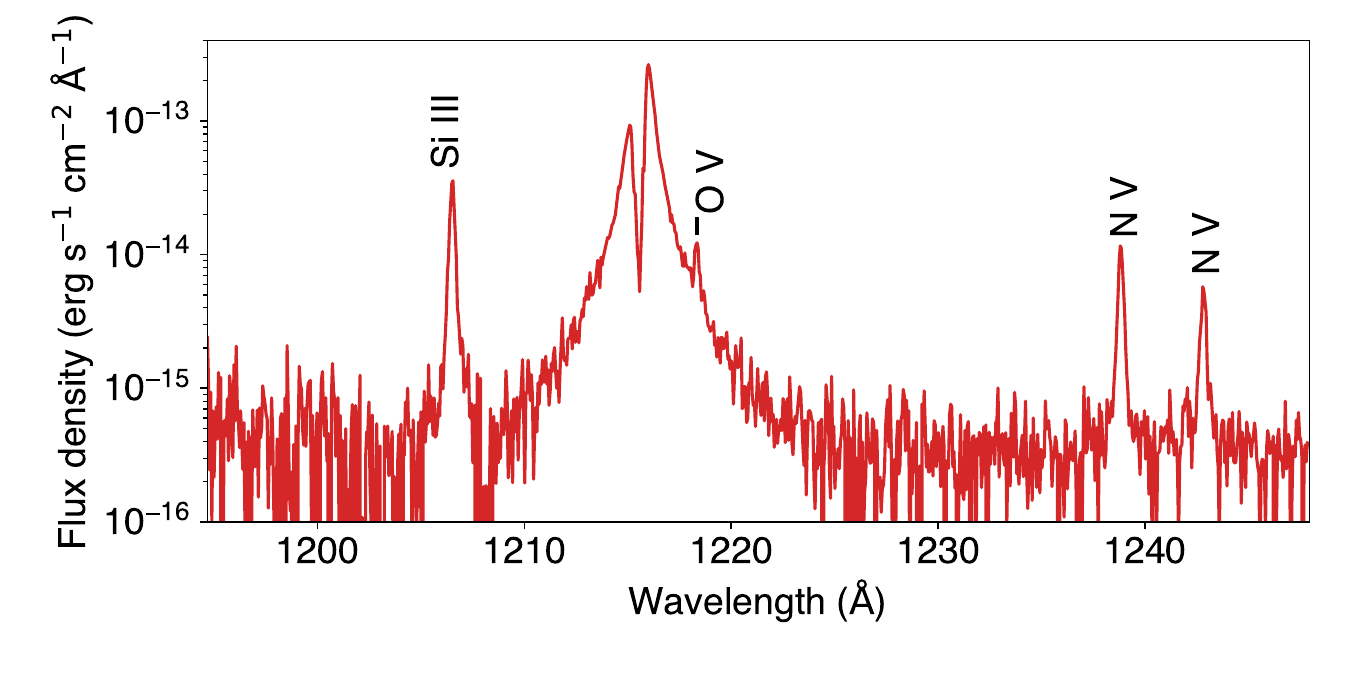}{0.49\textwidth}{(b)}
          \fig{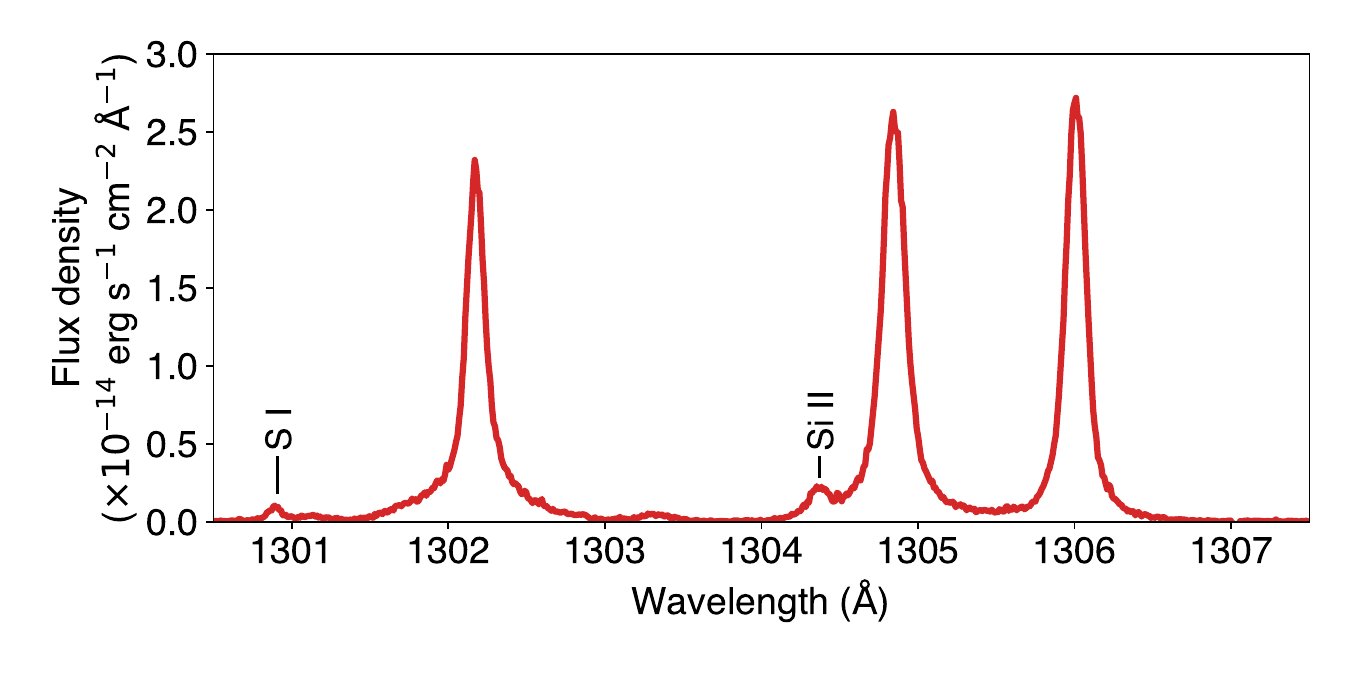}{0.49\textwidth}{(c)}}
\caption{(a) Far-ultraviolet spectrum of HD~189733 measured with \emph{HST}/COS in a combined exposure time of approximately 16 hours. (b) \emph{HST}/STIS spectrum near the Lyman-$\alpha$ line. (c) Zoom in the COS spectrum in wavelength range of the O\,{\sc i} triplet.  \label{fig:spectrum}}
\end{figure*}

We search for signals of atmospheric escape using the transmission spectroscopy technique. Due to the strong oscillator strengths of FUV spectral lines, we analyze stellar emission lines individually (see Figure \ref{fig:spectrum}). In this regime, one effective way of searching for excess in-transit absorption by an exospheric cloud around the planet is by measuring light curves of fluxes in the emission lines \citep[see, e.g.,][]{Vidal03, Ehrenreich15, DSantos2019}. Depending on the abundance of a certain species in the exosphere, an excess absorption of a few to several percent can be detected.

The C\,{\sc ii} lines are a doublet with central wavelengths at $1334.5$~{\rm \AA} and $1335.7$~{\rm \AA}\footnote{\footnotesize{More specifically, there is a third component blended with the second line at $1335.66$~{\rm \AA}, which would make this feature a triplet. However, this third component is one order of magnitude weaker than the second component.}}, both emitted by ions transitioning from the configuration 2s2p$^2$ to the ground and first excited states of the configuration 2s$^2$2p, respectively. The O\,{\sc i} lines are a triplet with central wavelengths at $1302.2$~{\rm \AA}, $1304.9$~{\rm \AA} and $1306.0$~{\rm \AA}, emitted by atoms transitioning from the configuration 2s$^2$2p$^3$($^4$S$^{\rm o}$)3s to the ground, first and second excited states of the configuration 2s$^2$2p$^4$, respectively. See the relative strengths of these spectral lines in Figure \ref{fig:spectrum}. As discussed in \citet{Bourrier2021}, insterstellar medium (ISM) absorption can in principle affect the observable flux of the C\,{\sc ii} lines. However, the effect is negligible for our analysis, which relies on a differential time-series analysis and not on the intrinsic stellar flux.

While analyzing the atomic oxygen (O\,{\sc i}) lines, care has to be taken because of geocoronal contamination. To get around this issue, we subdivided the \emph{HST} exposures into several subexposures, identifying which subexposures are contaminated, discarding them, and analyzing only the clean subexposures. Since the O\,{\sc i} contamination is correlated with the geocoronal emission levels in \lya, we identify the problematic subexposures using the \lya\ line, where the contamination is more obvious. When analyzing other emission lines that do not have geocoronal contamination, we do not discard any subexposures. In principle, the contamination in the O\,{\sc i} lines can also be subtracted using templates in a similar fashion as the \lya\ line \citep[see][]{Bourrier2018, Aguirre2023}. But the contrast between the airglow and the stellar emission of HD~189733 is too low to allow for a proper subtraction (see Appendix \ref{ag_app}), thus we opt to discard contaminated subexposures instead.

For relatively bright targets like HD~189733, the FUV continuum can be detected in COS spectroscopy despite the low signal-to-noise ratio (SNR). In our analysis, we measure the FUV continuum by integrating the COS spectra between the following wavelength ranges: [1165, 1173], [1178, 1188], [1338, 1346], [1372, 1392], [1413, 1424] and [1426, 1430]~\AA. These ranges strategically avoid strong emission lines and weaker ones that were identified by combining all available COS data (red spectrum in Figure \ref{fig:spectrum}).

\section{Results} \label{sec:results}

In the following discussion, we frequently mention fluxes during transit ingress and egress because those are the transit phases that are covered by Visits B and C (see Table \ref{tab:datasets_cos}). Visits A and D contain exposures near mid-transit, but the first visit has low SNR due to shorter exposure times and the latter does not have an adequate out-of-transit baseline. The light curves were normalized by the average flux in the exposures before the transit. As is customary in this methodology, we did not consider exposures after the transit as baseline for normalization because they may contain post-transit absorption caused by an extended tail. The transit depths quoted in this manuscript are measured in relation to the baseline out-of-transit flux. In this section, we deem signals as detections related to the planet HD~189733~b if they are repeatable during transits in the epoch of 2017 (Visits B and C).

\subsection{Exospheric oxygen and carbon}

We have found HD~189733~b to produce repeatable absorption levels of ionized carbon at blueshifted Doppler velocities. Furthermore, some of these signatures are asymmetric in relation to the transit center, indicative of departures from spherical symmetry. In Section \ref{sect:models}, we will see that no detectable atomic oxygen is expected in the exosphere of HD~189733~b.

For Visit A, we found that all exposures have low levels of geocoronal contamination except for the quarter of the following datasets: {\tt lb5k01umq}, {\tt lb5ka1uqq}, {\tt lb5ka1v0q} and {\tt lb5ka1vdq}; these subexposures with high contamination were discarded from the O\,{\sc i} analysis (see Appendix \ref{ag_app}). The exposures of Visits B, C and D were longer, so we divided them into five subexposures instead of four. For Visits B and C, we discard the last subexposure of every dataset; in the case of {\tt ld9m50oxq}, we also discard the fourth subexposure. In Visit D, we discard the first subexposure of all exposures, except {\tt ldzkh1ifq}.

\begin{figure*}[!ht]
    \gridline{\fig{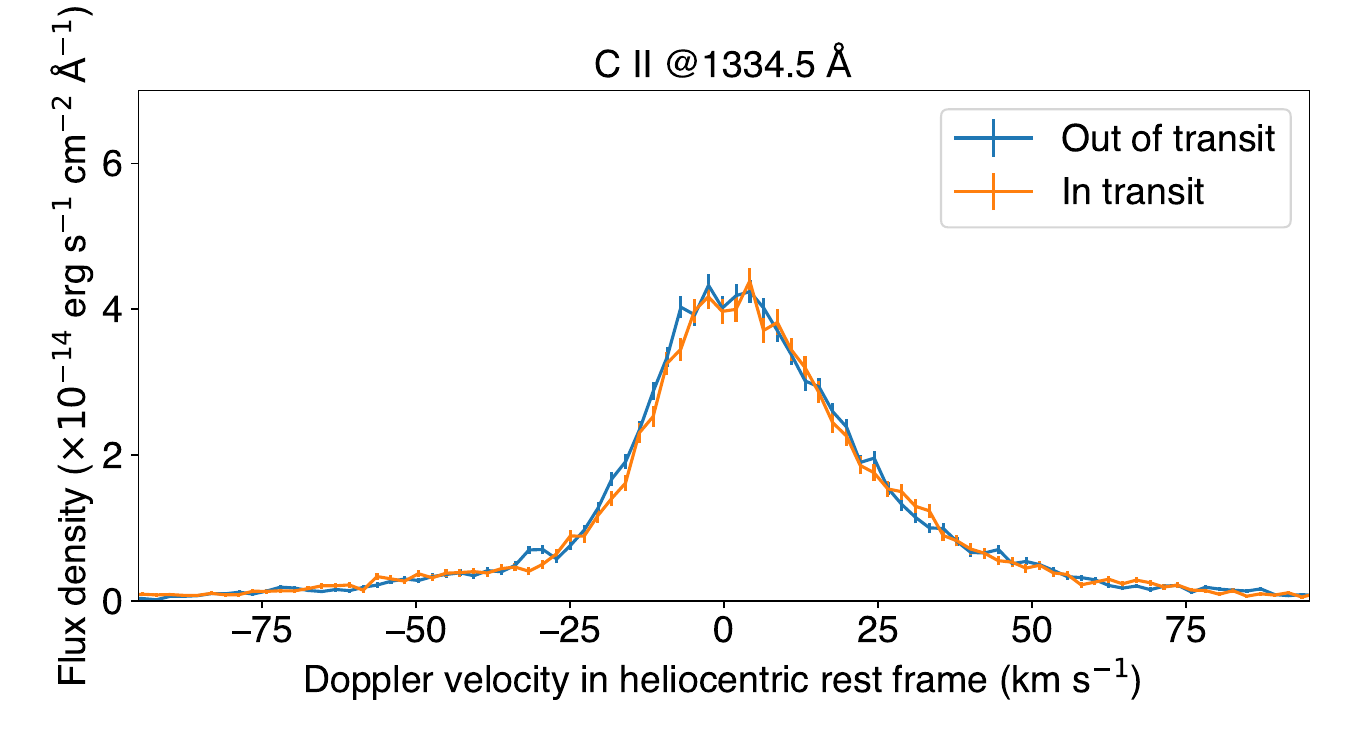}{0.48\textwidth}{(a)}
              \fig{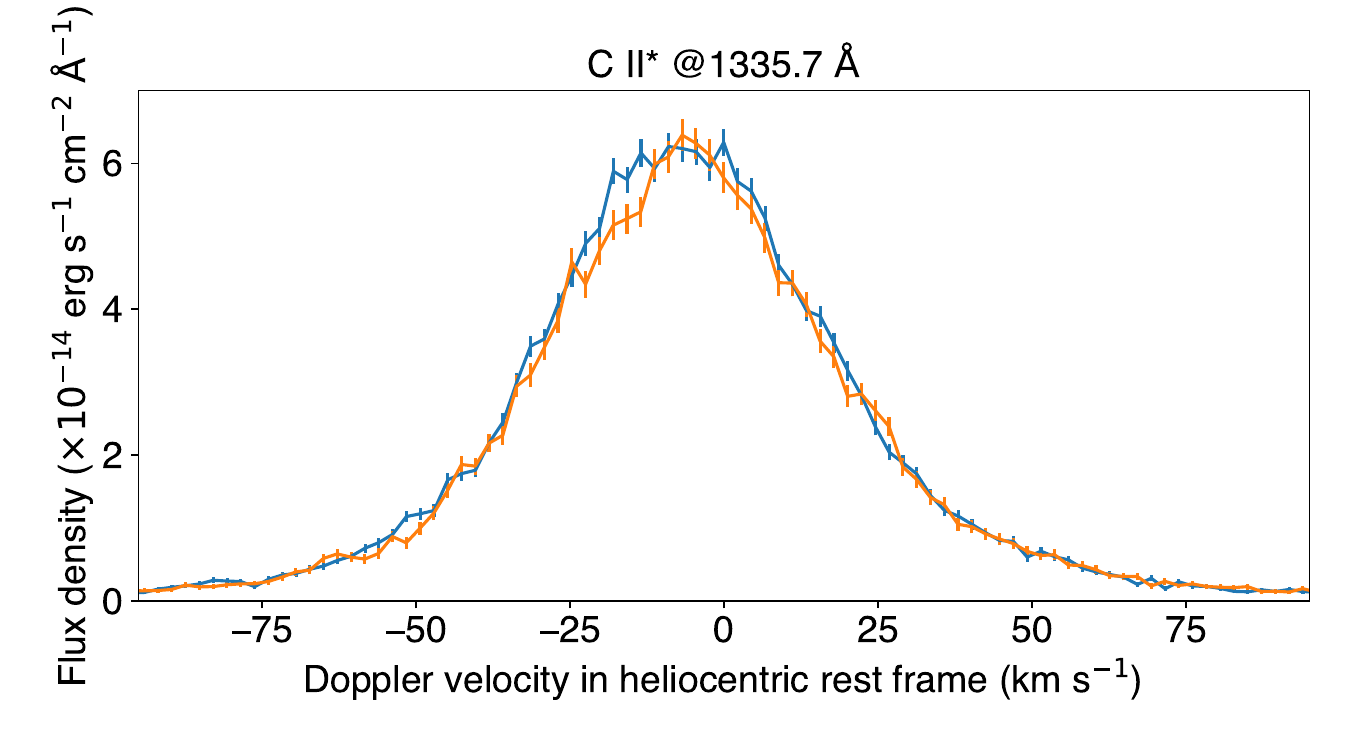}{0.48\textwidth}{(b)}}
    \gridline{\fig{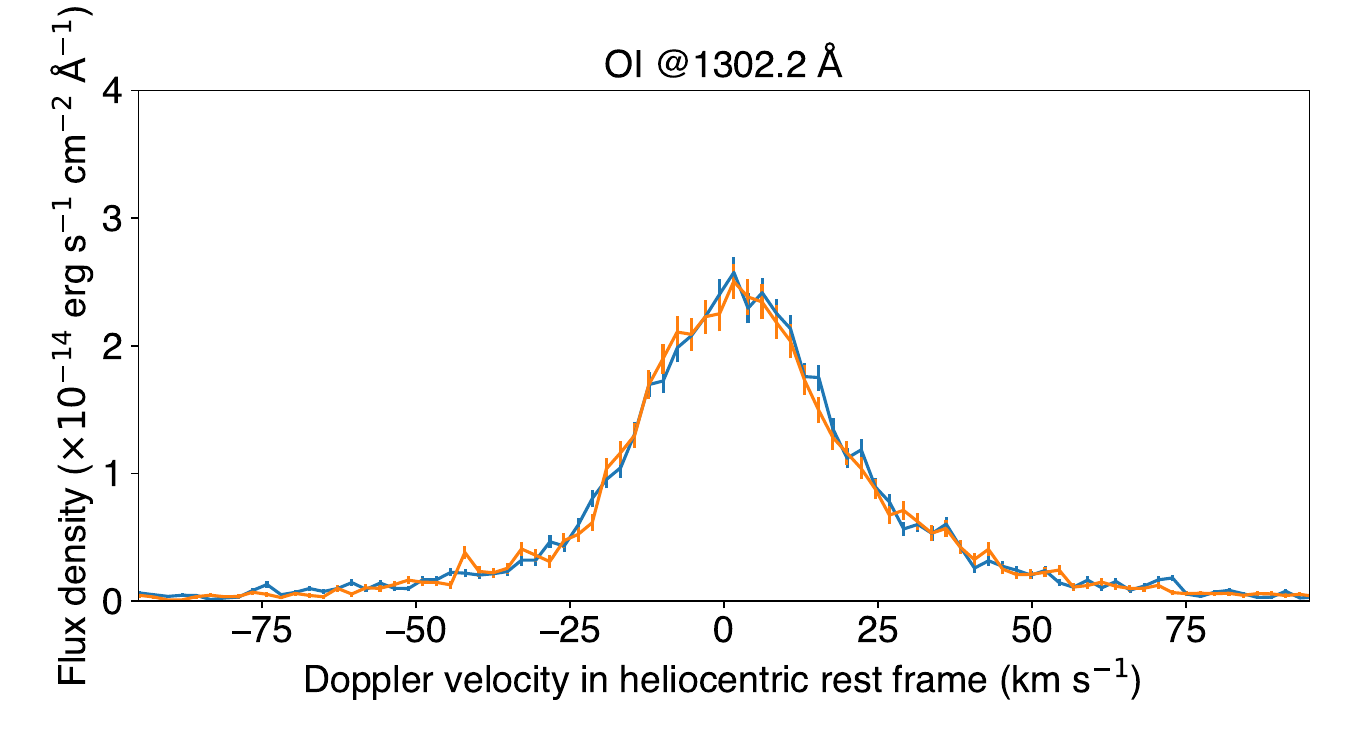}{0.48\textwidth}{(c)}
              \fig{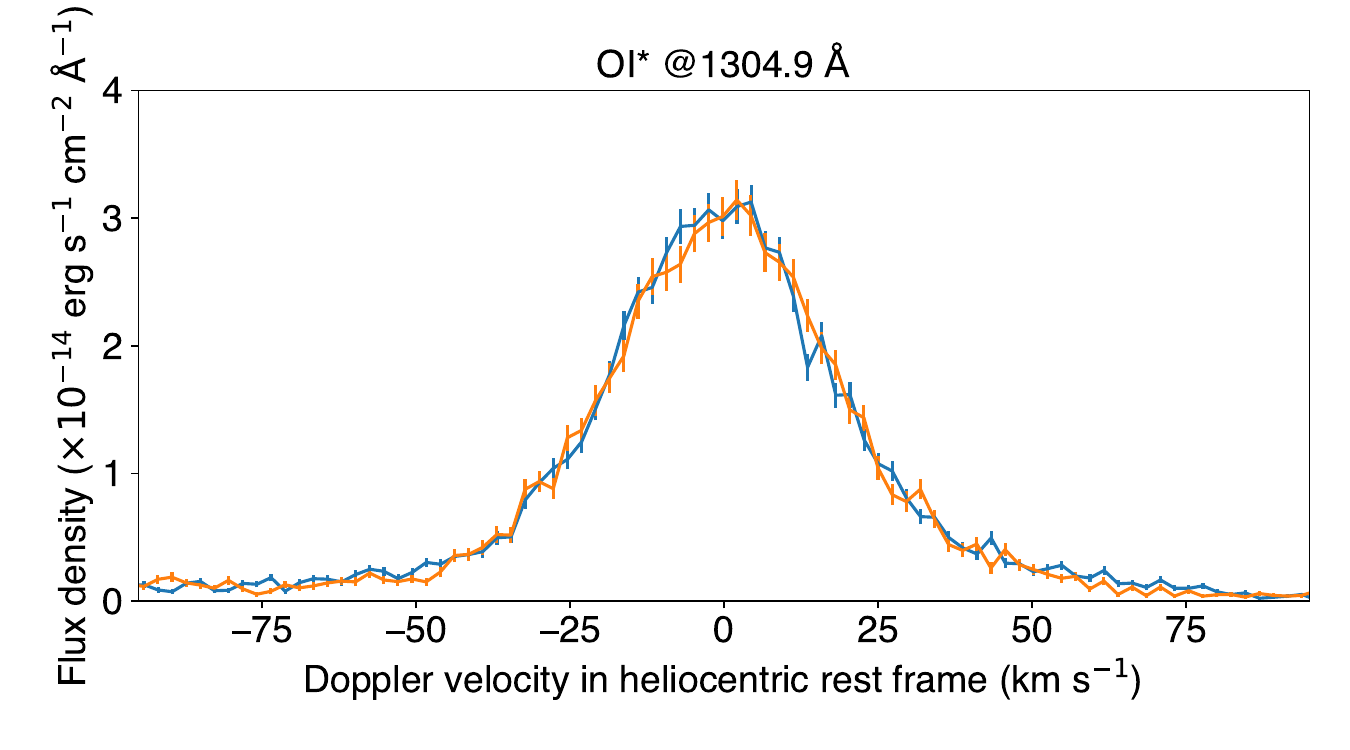}{0.48\textwidth}{(d)}}
    \gridline{\fig{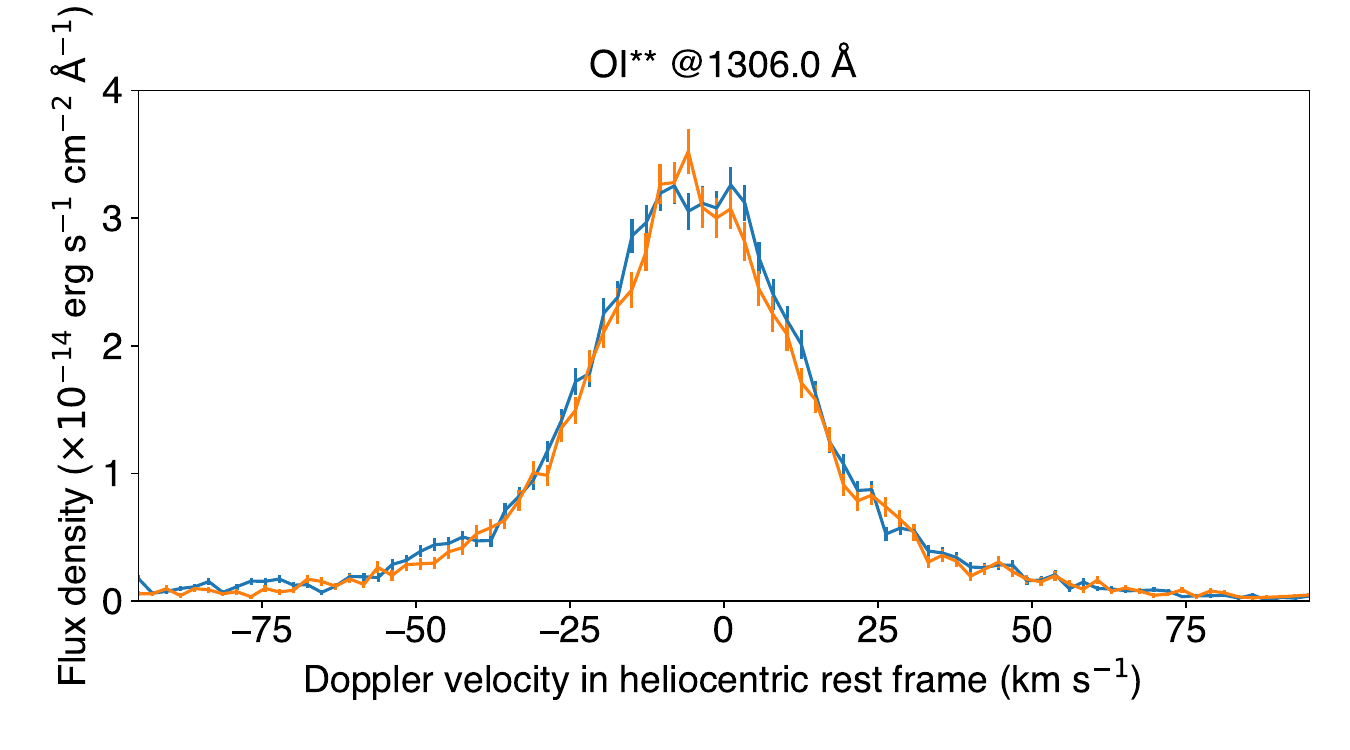}{0.48\textwidth}{(e)}}
\caption{In-transit and out-of-transit spectra of HD~189733 for the C\,{\sc ii} and O\,{\sc i} lines. In this plot we consider that the datasets {\tt ld9m50p1q}, {\tt ld9m50p3q}, {\tt ld9m51duq} and {\tt ld9m51dwq} are in transit, and the remaining datasets from Visits B and C are out of transit. We did not include Visits A and D in this combined spectrum. The in-transit absorption in the excited-state C\,{\sc ii}$^*$ line is seen in its blue wing.  \label{fig:cii_spec}}
\end{figure*}

Our analysis of the O\,{\sc i} light curves yields an in-transit absorption of $5.3\% \pm 1.9\%$ (2.8$\sigma$ significance; Doppler velocity range [-75, +75]~km\,s$^{-1}$) by combining Visits B and C when co-adding all O\,{\sc i} lines. In Visit A (epoch 2009), we do not detect a significant in-transit absorption, likely due to a combination of shorter exposure times and stellar variability (see Appendix \ref{app:add_lcs}). We deem the results of Visit D (epoch 2020) inconclusive due to a non-optimal out-of-transit baseline (see also Appendix \ref{app:add_lcs}). We show the in- and out-of-transit O\,{\sc i} spectra in the second and third rows of Figure \ref{fig:cii_spec}.

\begin{figure*}[!ht]
    \gridline{\fig{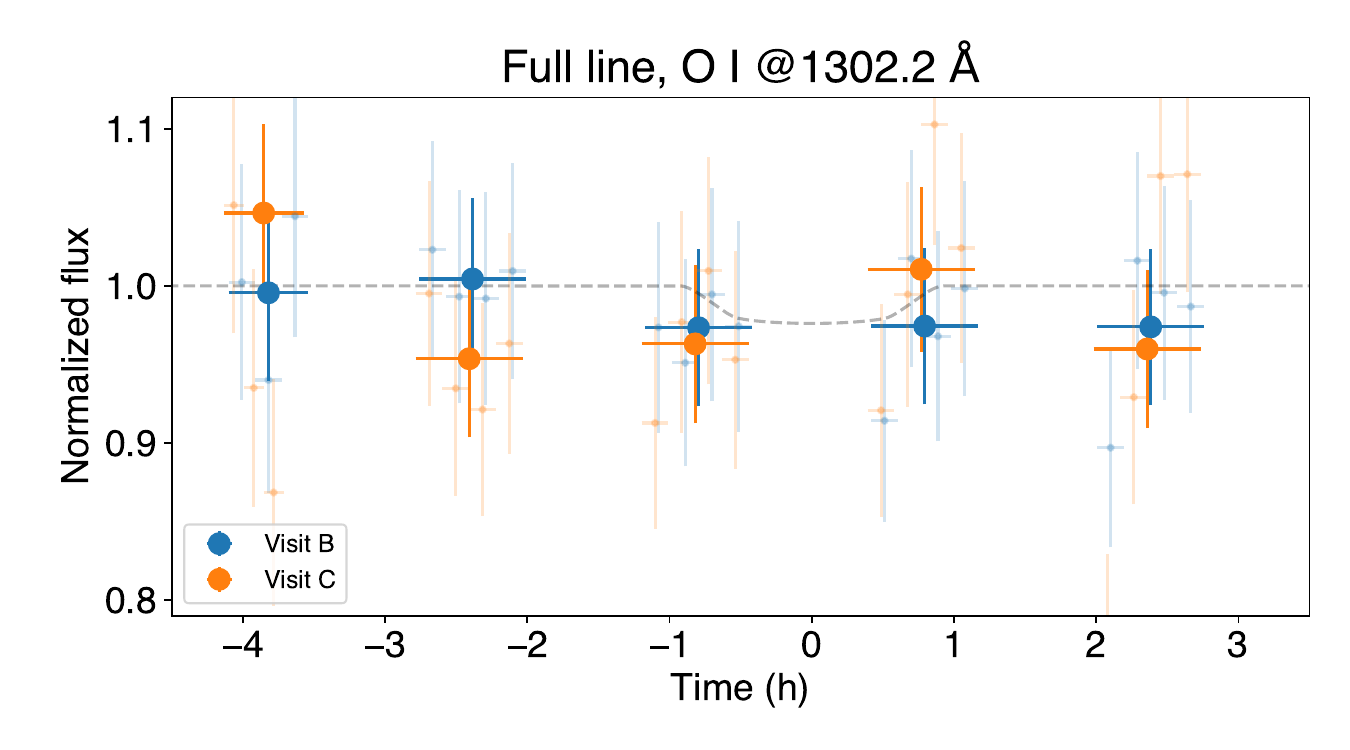}{0.48\textwidth}{(a)}
              \fig{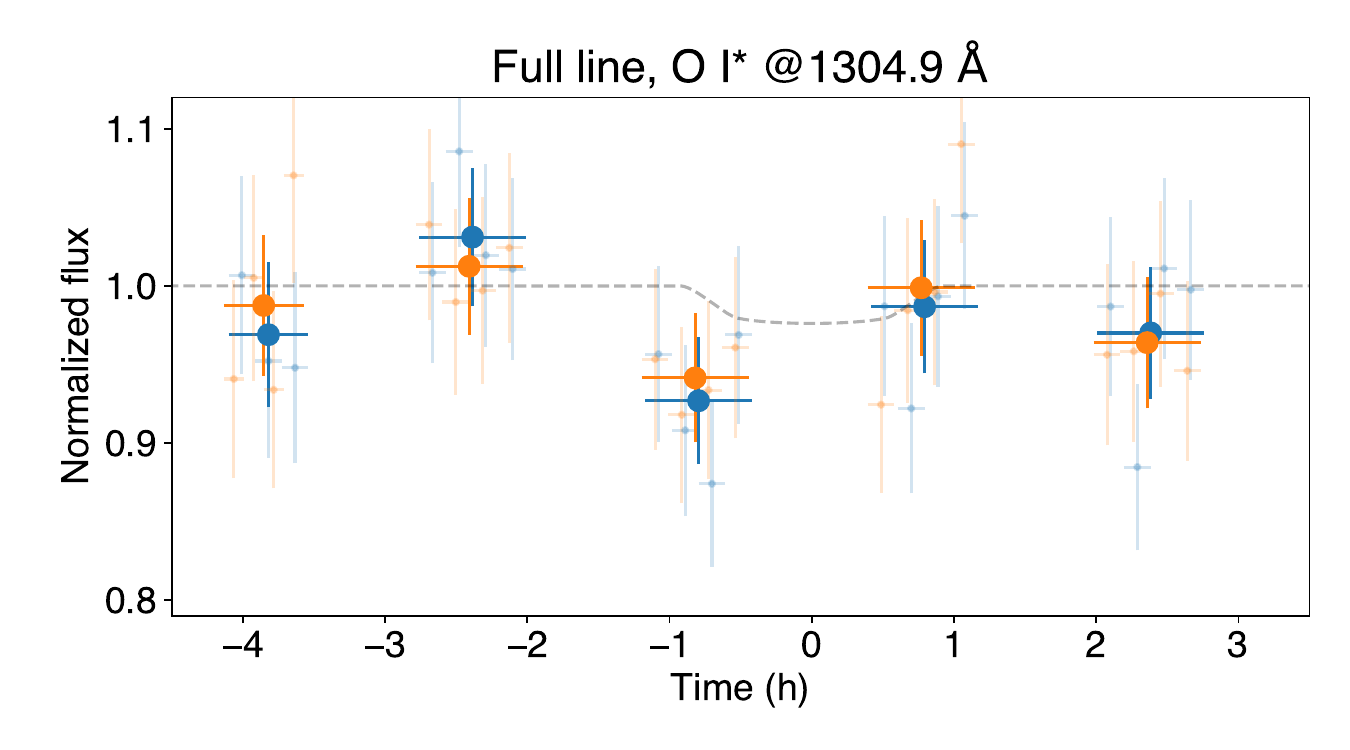}{0.48\textwidth}{(b)}}
    \gridline{\fig{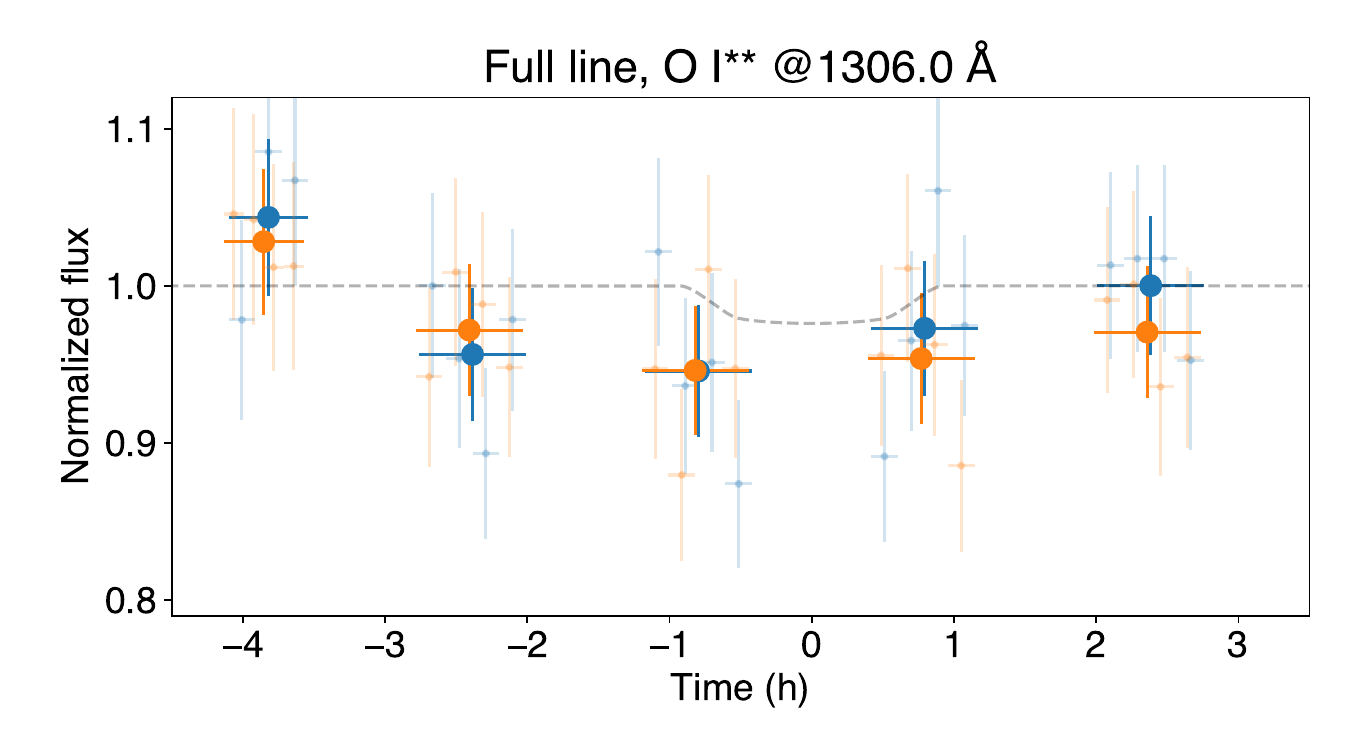}{0.48\textwidth}{(c)}
              \fig{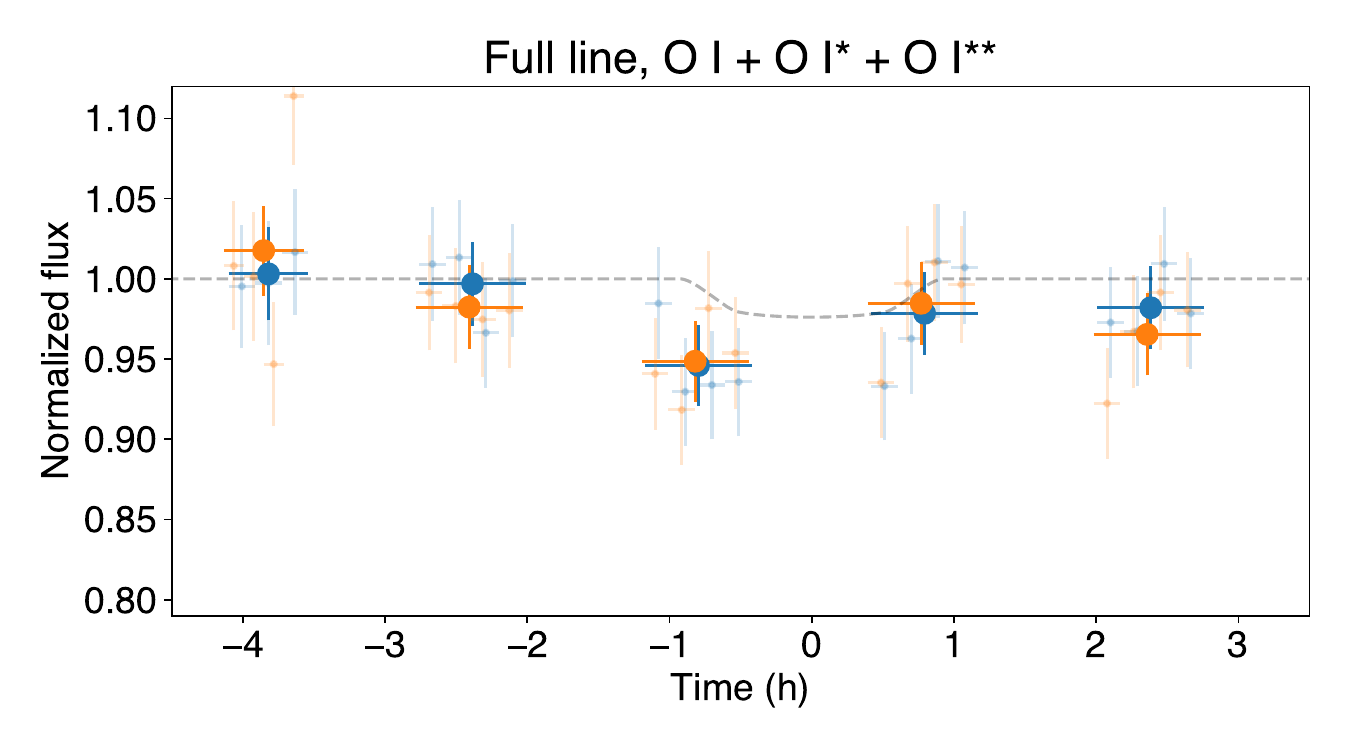}{0.48\textwidth}{(d)}}
\caption{Neutral oxygen (O\,{\sc i}) phase-folded transit light curves of HD~189733~b, split by the fine-structure lower level of the atom (here denoted by O\,{\sc i}, O\,{\sc i}$^*$ and O\,{\sc i}$^{**}$). The dashed gray curve represents the transit of the planet as seen in optical wavelengths. Visits B and C are denoted by blue and orange filled circles, respectively, which correspond to epoch 2017. The semi-transparent blue and orange measurements correspond to subexposures of Visits B and C. The light curves of Visits A and D, corresponding respectively to epochs 2009 and 2020, are shown in Appendix \ref{app:add_lcs}.  \label{fig:oxygen_lc_hd189}}
\end{figure*}

In addition to O\,{\sc i}, we also measure an in-transit absorption of singly-ionized carbon (C\,{\sc ii}, all lines co-added) in HD~189733~b, more specifically $7.4\% \pm 2.0\%$ and $3.0\% \pm 2.1\%$ for Visits B and C, respectively. By combining these two visits, we measure an absorption of $5.2\% \pm 1.4\%$ ($3.7\sigma$ detection; Doppler velocity range [-100, +100]~km\,s$^{-1}$). We report the ingress and egress absorption levels at different Doppler velocity ranges in Table \ref{tab:lc_results}. The signal is largely located in the blue wing, between velocities [-100, 0]~km\,s$^{-1}$, of the excited-state line at 1335.7~\AA\ (see the top row of Figure \ref{fig:cii_spec}); if the signal is indeed of planetary nature, this suggests that the material is being accelerated away from the host star \citep[as seen in][]{GarciaMunoz21}. The other emission lines in the COS spectrum (Si\,{\sc ii}, Si\,{\sc iv}, C\,{\sc iii} and N\,{\sc v}) do not show significant variability (see Appendix \ref{app:add_lcs}).

\begin{figure*}[!ht]
    \gridline{\fig{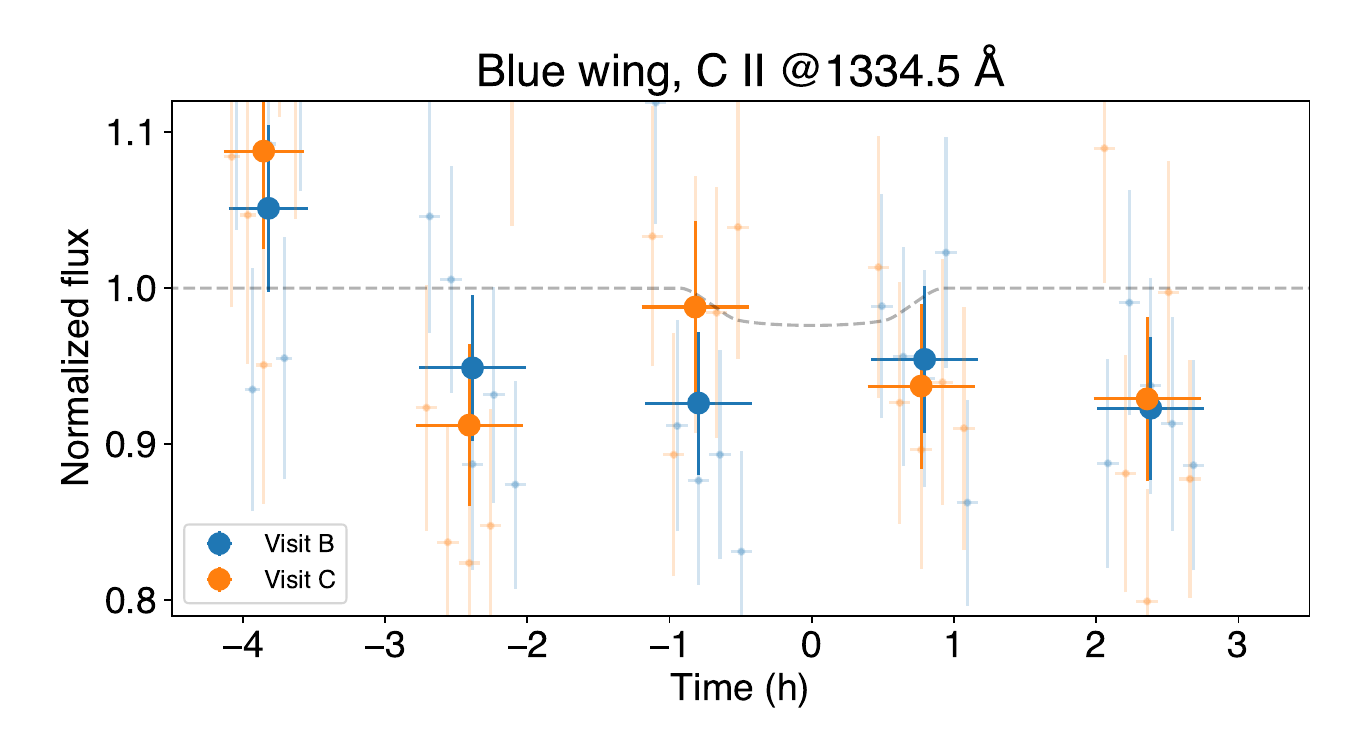}{0.48\textwidth}{(a)}
              \fig{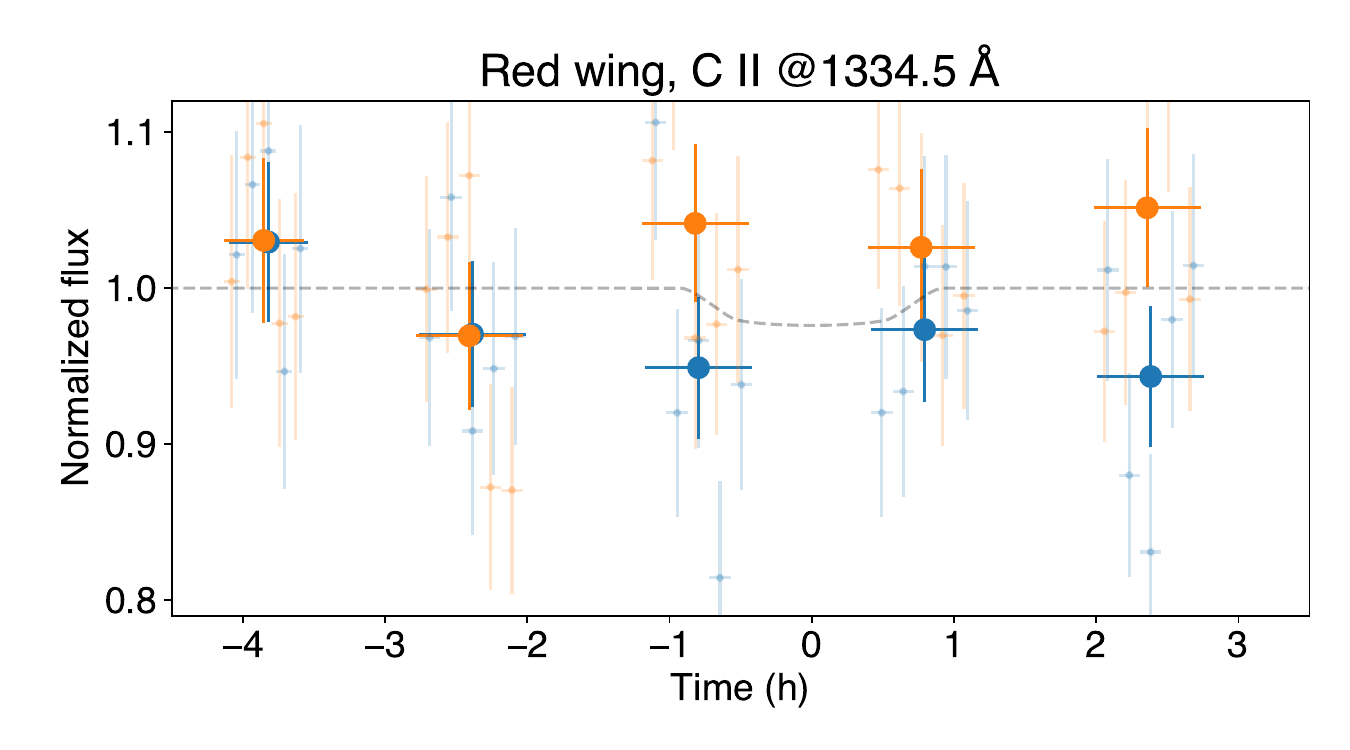}{0.48\textwidth}{(b)}}
    \gridline{\fig{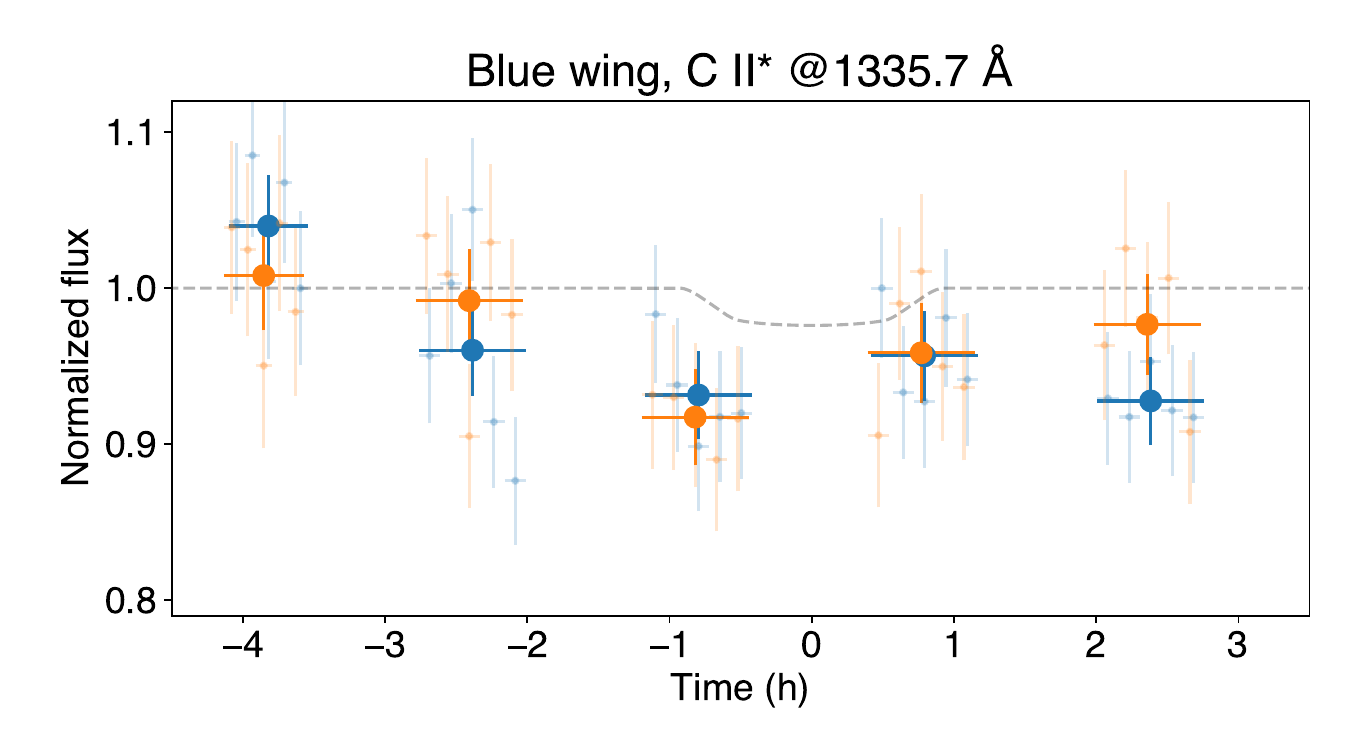}{0.48\textwidth}{(c)}
              \fig{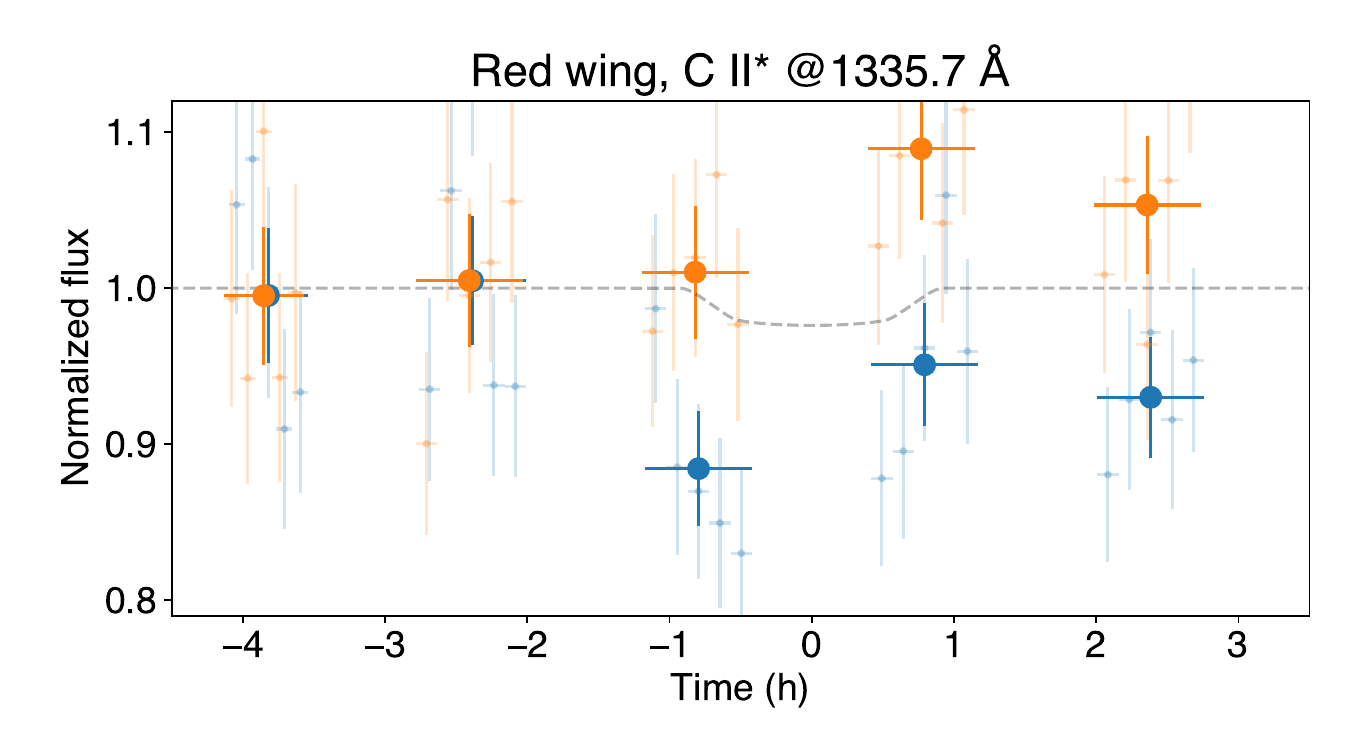}{0.48\textwidth}{(d)}}
    \gridline{\fig{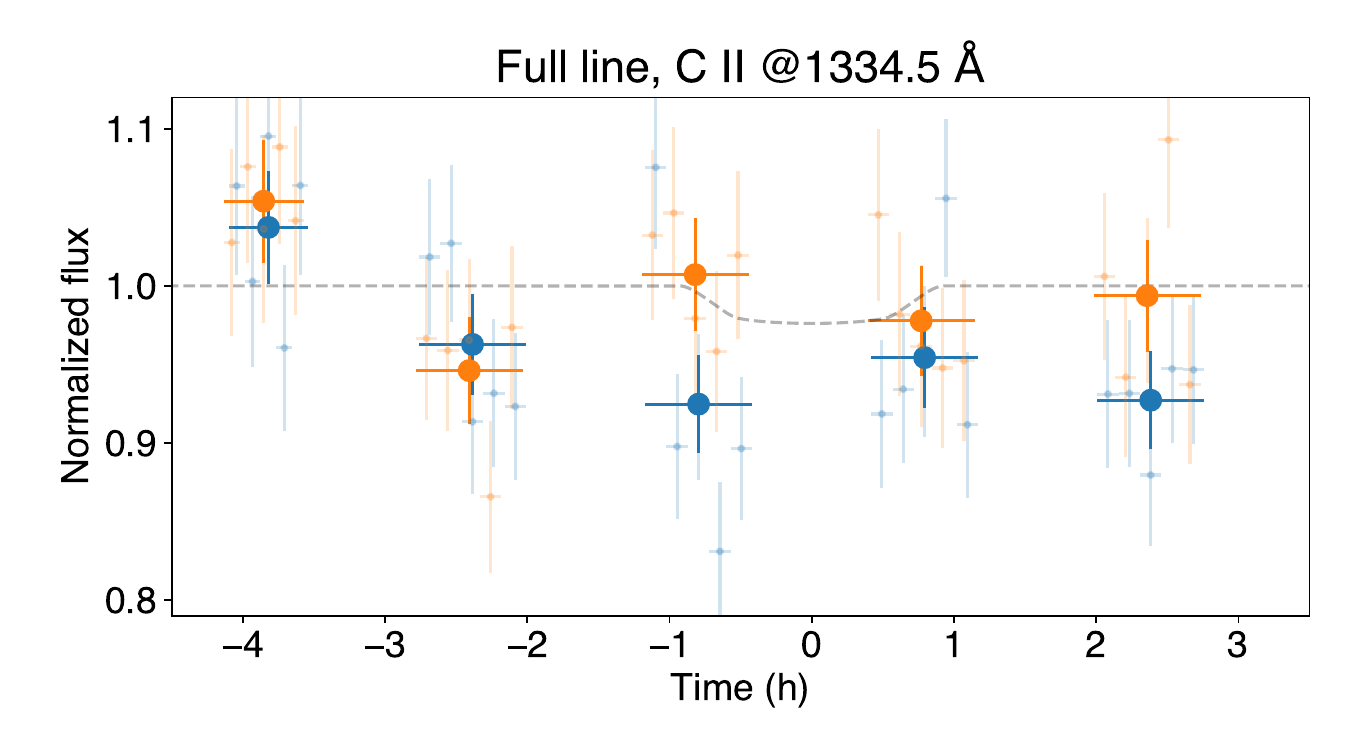}{0.48\textwidth}{(e)}
              \fig{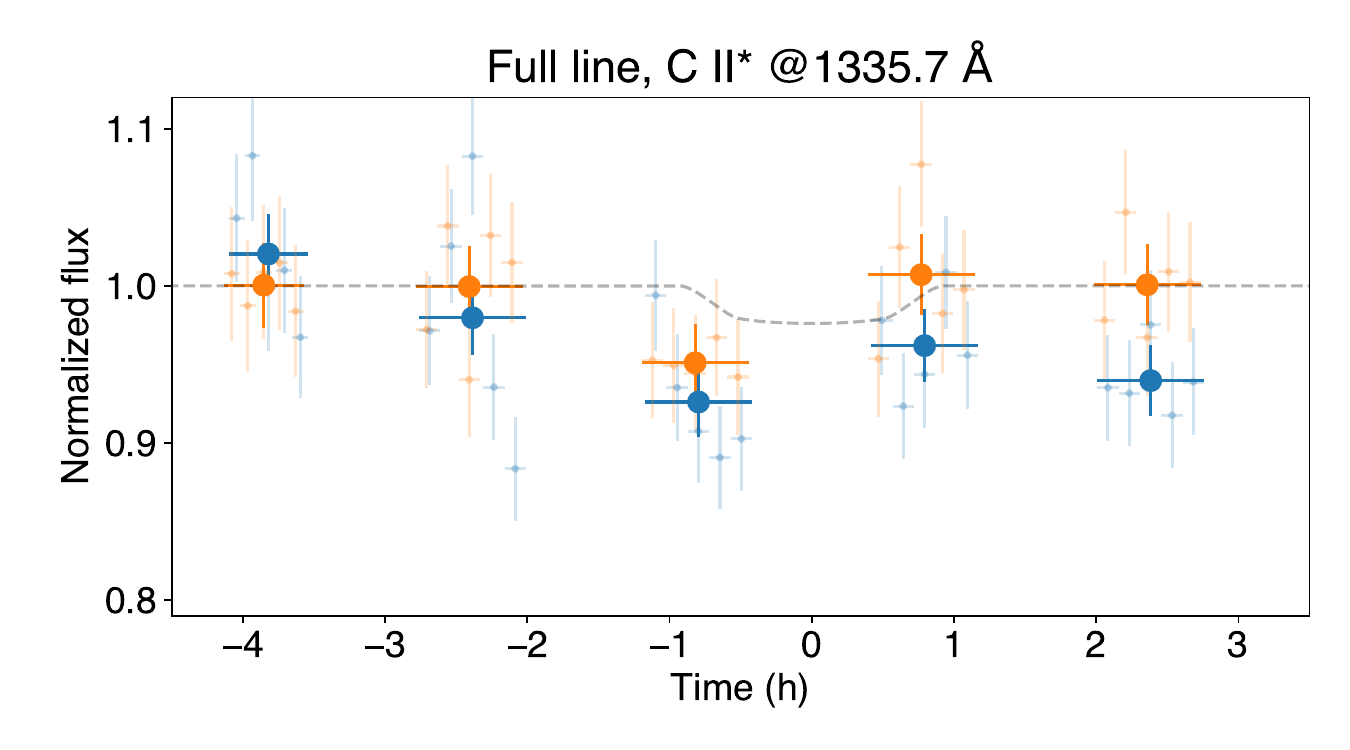}{0.48\textwidth}{(f)}}
\caption{Same as Figure \ref{fig:oxygen_lc_hd189}, but for singly-ionized carbon (C\,{\sc ii}). Here we separate the light curves in blue and red wings, which are calculated by integrating the flux between Doppler velocities [-100, 0]~km\,s$^{-1}$ and [0, +100]~km\,s$^{-1}$, respectively.  \label{fig:carbon_lc_hd189}}
\end{figure*}

\begin{deluxetable}{l c c c}
\tablecaption{Average in-transit absorption levels measured over Visits B and C. \label{tab:lc_results}}
\tablewidth{0pt}
\tablehead{
\colhead{Species} & \colhead{Blue wing} & \colhead{Red wing} & \colhead{Full line}
}
\startdata
\multicolumn{4}{l}{Ingress} \\ 
O\,{\sc i} & $2.3\% \pm 5.5\%$ & $5.1\% \pm 5.1\%$ & $3.2\% \pm 3.7\%$ \\
O\,{\sc i}$^*$ & $8.9\% \pm 4.2\%$ & $1.6\% \pm 4.7\%$ & $6.6\% \pm 3.1\%$ \\
O\,{\sc i}$^{**}$ & $3.4\% \pm 3.9\%$ & $9.9\% \pm 5.5\%$ & $5.4\% \pm 3.1\%$ \\
C\,{\sc ii} & $4.3\% \pm 3.7\%$ & $0.5\% \pm 3.4\%$  & $3.4\% \pm 2.4\%$  \\
C\,{\sc ii}$^*$ & $7.6\% \pm 2.3\%$ & $5.3\% \pm 3.0\%$  & $6.1\% \pm 1.8\%$  \\
\hline
\multicolumn{4}{l}{Egress} \\ 
O\,{\sc i} & $5.5\% \pm 5.5\%$ & $-0.1\% \pm 5.1\%$ & $0.7\% \pm 3.7\%$ \\
O\,{\sc i}$^*$ & $1.5\% \pm 4.2\%$ & $0.1\% \pm 4.7\%$ & $0.7\% \pm 3.1\%$ \\
O\,{\sc i}$^{**}$ & $0.0\% \pm 3.9\%$ & $8.6\% \pm 5.4\%$ & $3.7\% \pm 3.1\%$ \\
C\,{\sc ii} & $5.4\% \pm 3.7\%$ & $0.0\% \pm 3.4\%$  & $3.4\% \pm 2.4\%$  \\
C\,{\sc ii}$^*$ & $4.2\% \pm 2.2\%$ & $-2.0\% \pm 3.0\%$ & $1.5\% \pm 1.8\%$ \\
\enddata
\tablecomments{Visits A and D were excluded from this analysis because they were observed in different epochs from the PanCET visits (see details in Section \ref{sec:methods}).}
\end{deluxetable}

\subsection{Non-repeatable signals: stellar or planetary variability?}

The hot Jupiter HD~189733~b is known for orbiting a variable host star and for having variable signatures of atmospheric escape \citep[see, e.g.,][]{Bourrier2013, Salz18, Cauley18, Bourrier20, Zhang22, Pillitteri22}. Our analysis provides potential observational evidence for the variability of the upper atmosphere in this exoplanet, but it is difficult to disentangle it from variability in the host star emission-line flux. 

\begin{figure*}[!ht]
\gridline{\fig{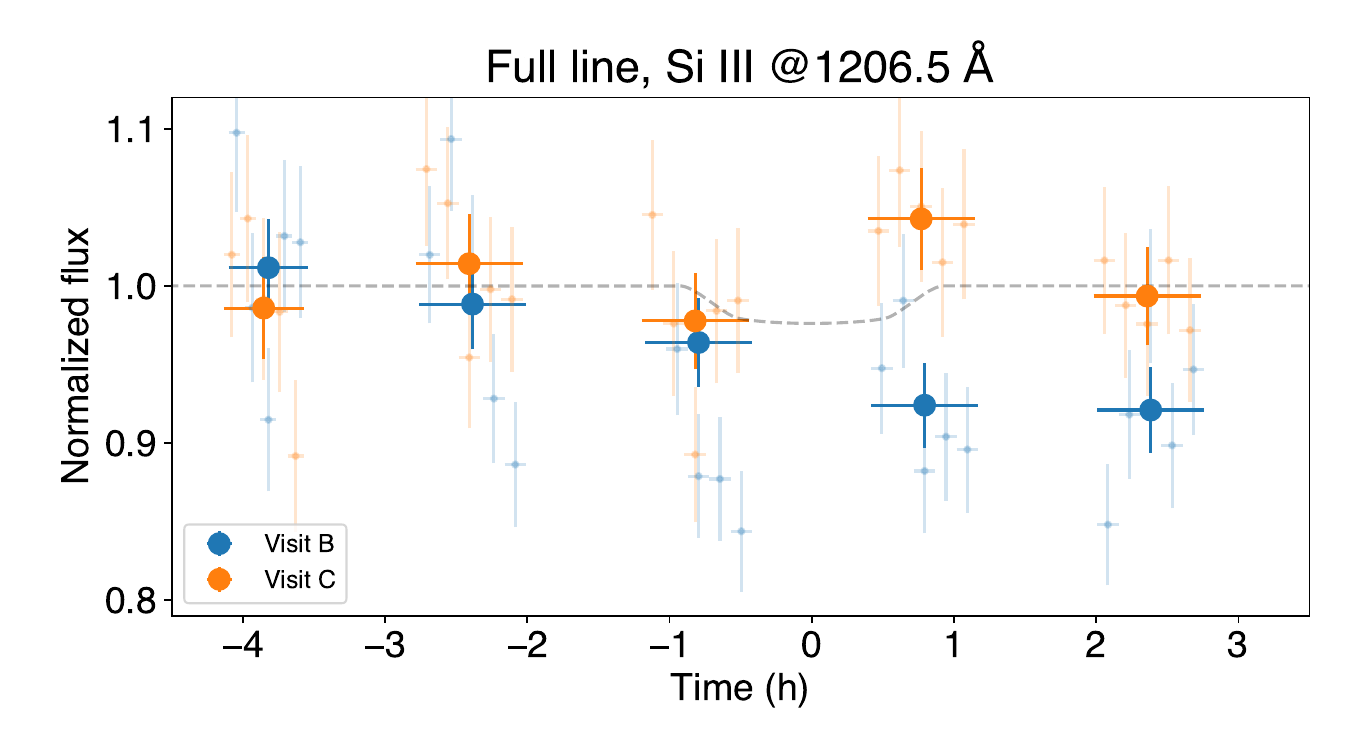}{0.48\textwidth}{(a)}
          \fig{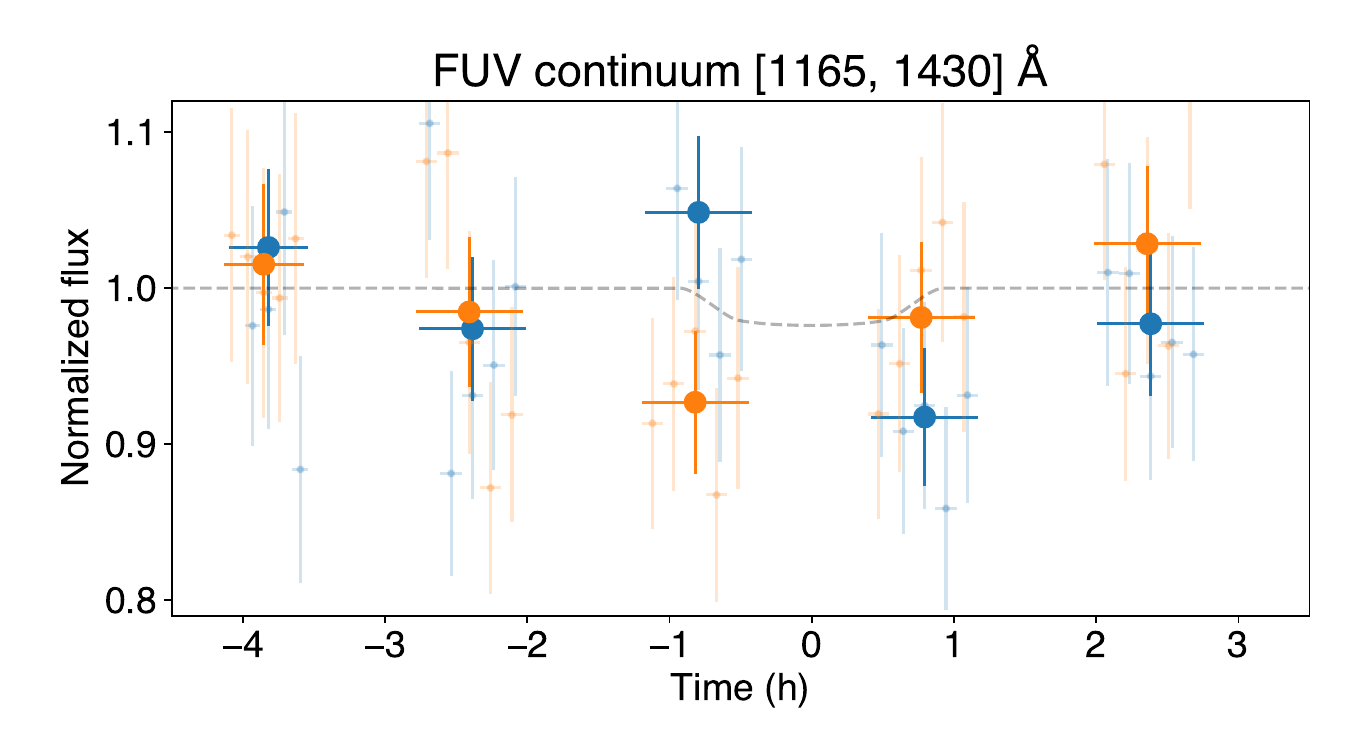}{0.48\textwidth}{(b)}}
\caption{{\it Left panel:} Transit light curve of doubly-ionized silicon, a line that is sensitive to stellar activity modulation. {\it Right panel:} Transit light curve of the FUV continuum (excluding emission lines, \lya\ wings and the detector gap). \label{fig:SiIII_lc}}
\end{figure*}

In Visit B, the Si\,{\sc iii} line at 1206.5~\AA\ shows a flux decrease of $7.6\% \pm 2.9\%$ near the egress of the transit (see left panel of Figure \ref{fig:SiIII_lc}). This line is well known for being a sensitive tracer of stellar activity \citep[e.g.,][]{2013A&A...553A..52B, Loyd2014, DSantos2019, Bourrier20}. However, it is difficult to determine whether this signal is due only to stellar activity or the presence of doubly-ionized Si in the upper atmosphere of HD~189733~b or a combination of both. Similar non-repeatable egress absorptions are seen in the C\,{\sc ii} line at 1335.7~\AA\ in Visit B. If it is due solely to stellar activity, it is possible that part of the in-transit absorption of C\,{\sc ii} observed in Visit B is also due to activity, since these lines are a moderate tracer of activity as well. With that said, it is not completely unexpected to see tails of ionized exospheric atoms after the egress of a highly-irradiated planet like HD~189733~b \citep{Owen2023}. If the egress Si\,{\sc iii} flux decrease seen in Visit B is indeed due to the presence of doubly-ionized Si in the exospheric tail of the planet, a non-detection in Visit C could be explained by: i) variability in the outflow velocities of HD~189733~b, ii) variability in its ionization level, or iii) variation in the stellar wind. Further modeling will be necessary to test these different hypotheses, and we leave it for future efforts.

Since the FUV continuum traces the lower chromosphere in solar-type stars \citep[e.g.,][]{Linsky2012a}, we also compute its light curve and search for signals of variability connected to stellar activity. For HD~189733, we measure an average out-of-transit FUV continuum flux density of $(1.18 \pm 0.05) \times 10^{-16}$ and $(1.11 \pm 0.05) \times 10^{-16}$~erg\,s\,cm$^{-2}$\,\AA$^{-1}$, respectively for visits B and C (see wavelength ranges in Section \ref{sec:methods}). The FUV continuum transit light curve is shown in the right panel of Figure \ref{fig:SiIII_lc}. We do not detect statistically significant variability of the normalized FUV continuum flux during Visits B and C; however, the uncertainties of the measured fluxes are slightly larger than the line fluxes measured for the C\,{\sc ii} light curves.

Following the methods described in \citet{DSantos2019}, we did not identify any flares in the datasets we analyzed, and found no evidence for rotational or magnetic activity modulation of FUV fluxes due to the relatively short baseline of observations available. The photometric monitoring of the host star (see Appendix \ref{app:phot}) suggests a rotational period of $12.25$~d. Visit B occurred during a time of maximum starspots, while Visit C occurred between times of maximum and minimum spottedness. In the \emph{HST} data, we found that Visit B has higher absolute fluxes of metallic lines than Visit C by a factor of $\sim$10\% (except for Si\,{\sc ii}; see Appendix \ref{app:add_lcs}). On the other hand, for the ground-based photometry in $b$ and $y$ bands, we observe a $\Delta {\rm mag}$ of $\sim$0.02, which corresponds to a difference in flux of approximately 1.9\% in the optical.

\subsection{A repeatable hydrogen signature}

\begin{figure*}[!ht]
\plottwo{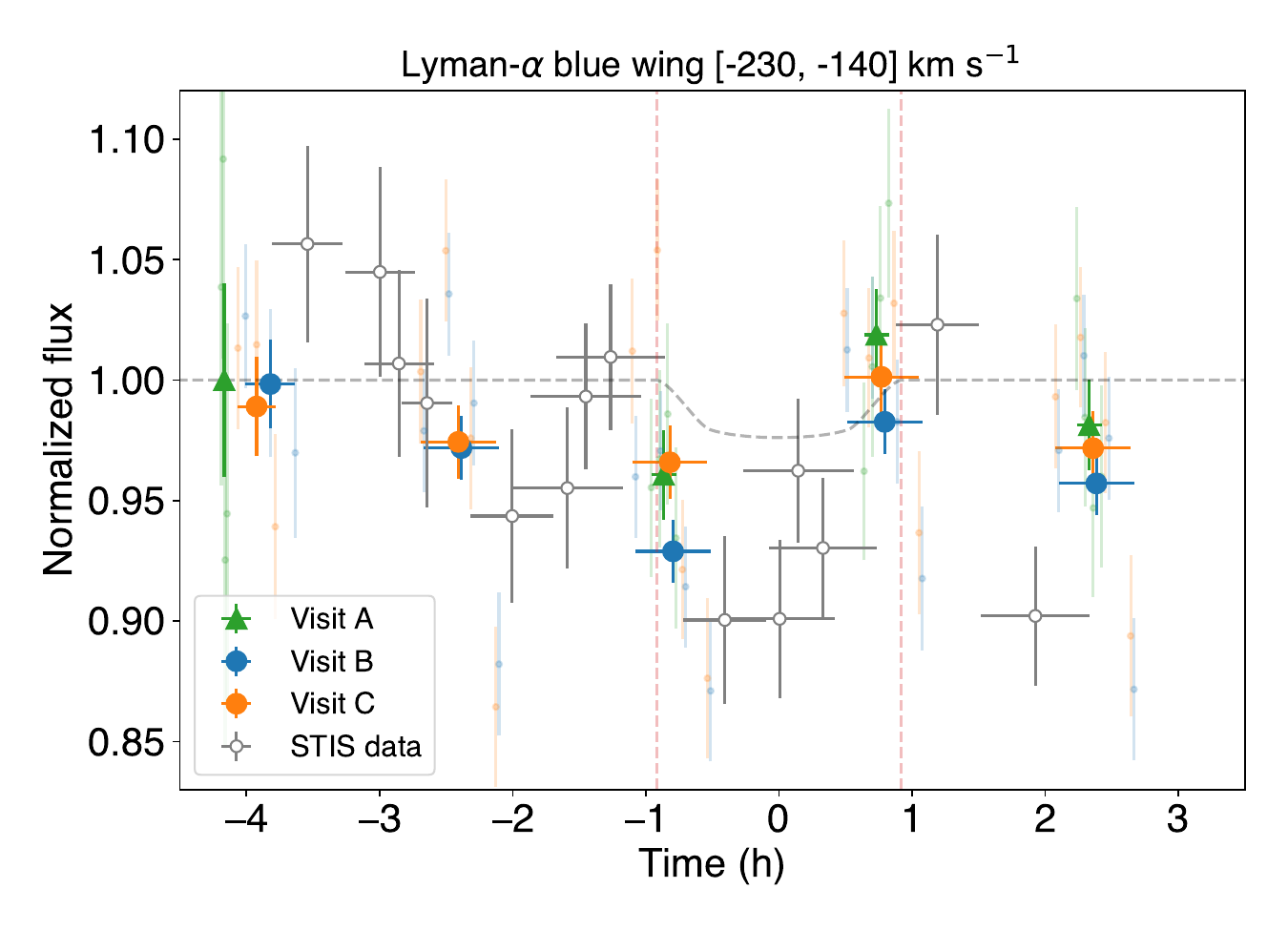}{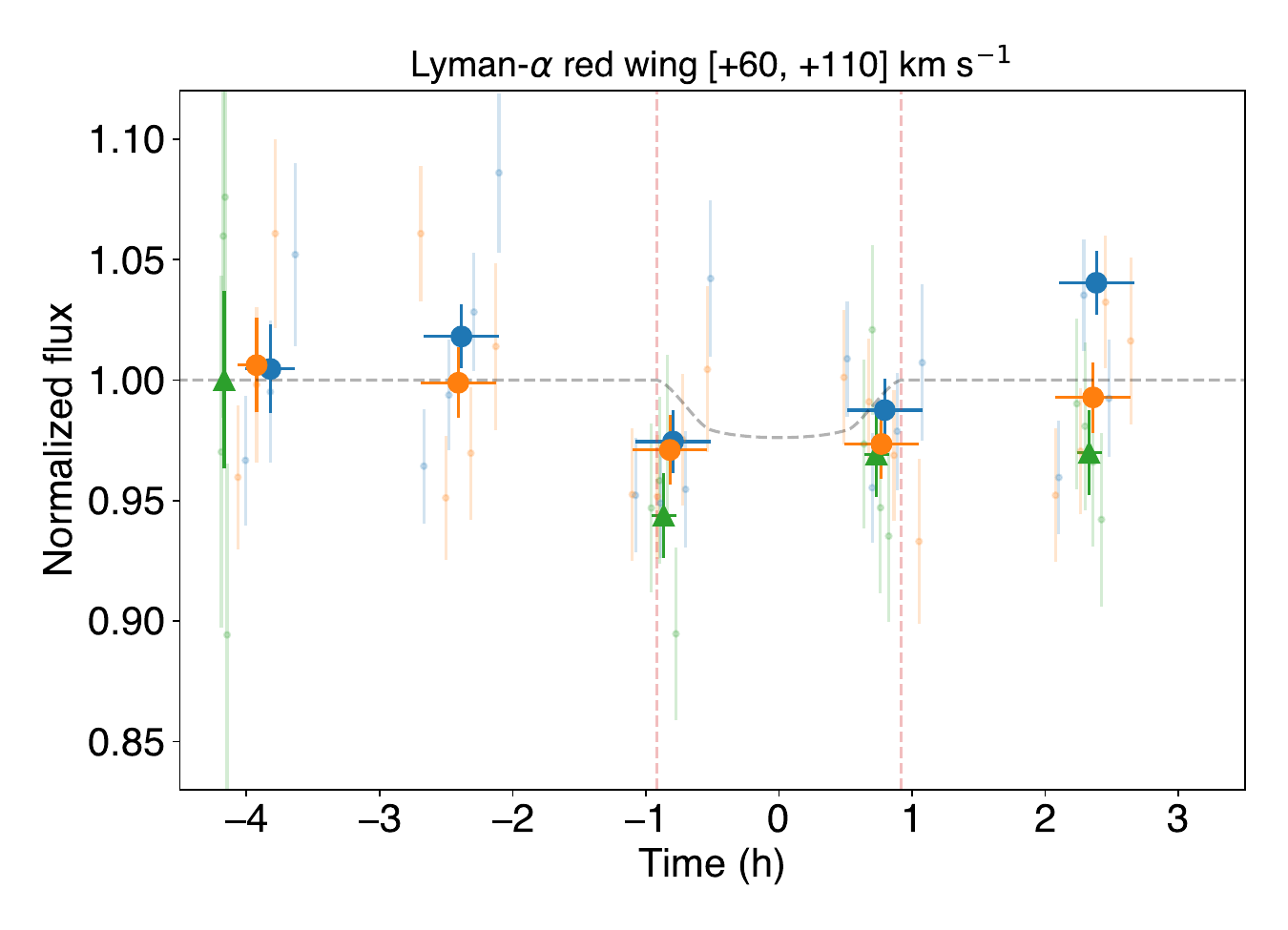}
\caption{Lyman-$\alpha$ transit light curves of HD~189733~b measured with COS (full symbols) and STIS (open circles). The green symbols correspond to Visit A, and the blue and orange symbols correspond to Visits B and C, similarly to the other light curves shown in this manuscript. We detect a repeatable absorption in the Lyman-$\alpha$ blue wing (left panel) attributed to the presence of escaping H in Visits A and B, which are consistent with the one previsouly measured with STIS \citep[see][]{Bourrier20}. No in-transit absorption is detected in Visit C, reinforcing that the exospheric H in the planet is variable. We also observe a variable in-transit absorption in the red wing with COS (right panel).  \label{fig:lya_lc_hd189}}
\end{figure*}

The COS spectra we analyze also contain information about the stellar Lyman-$\alpha$ line, despite the strong geocoronal contamination. We used the same technique described in \citet{DSantos2019} to remove the geocoronal contamination (see Appendix \ref{ag_app}) and analyzed the time series of the cleaned \lya\ line for both its blue and red wings (respectively [-230, -140]~km\,s$^{-1}$ and [+60, +110]~km\,s$^{-1}$, based on the results of \citealt{Lecavelier2012}). Some of the exposures in Visit A are not suitable for decontamination due to low SNR and had to be discarded. The resulting light curves are shown in Figure \ref{fig:lya_lc_hd189}.

We found that the blue wing of the \lya\ line shows a repeatable absorption during the ingress of HD~189733~b, with transit depths consistent between all visits. The light curves measured with COS are also consistent with that observed with STIS \citep[open circles in Figure \ref{fig:lya_lc_hd189}; see also][]{Bourrier20}. This signal at high velocities in the blue wing suggests that the exospheric H of HD~189733~b is accelerated away from the host star, an effect that has been extensively studied in, e.g., \citet{2013A&A...557A.124B}, \citet{2020A&A...638A..49O} and \citet{2022ApJ...927..238R}. The time series of the \lya\ red wing also shows an absorption during the transit of HD~189733~b. These results are consistent with the observations reported in \citet{Lecavelier2012}.

In addition, we also found a hint of a post-transit absorption in the blue wing two hours after the transit mid-time, which could indicate the presence of a long neutral H tail as predicted by \citet{Owen2023}. At this moment, it is unclear how consistent this would be with the doubly-ionized Si tail hinted in the left panel of Figure \ref{fig:SiIII_lc}, so we would benefit from future work in more detailed hydrodynamic modeling of HD~189733~b involving H and metallic species.

As we shall see in Section \ref{sect:models}, our simplified one-dimensional modeling suggests that the exosphere of HD~189733~b is mostly ionized for all the species we simulated (H, He, C and O). The detection of neutral H, however, is not inconsistent with this scenario, since even a small fraction of neutral of H can yield a detectable signal due to the large absolute abundances of H atoms in the outflow.

The global 3D hydrodynamics simulations presented by \citet{2022ApJ...927..238R} predict different levels of absorption in the blue and red wings of the \lya\ line depending on the strength of the stellar wind (SW). According to their models, weaker winds ($\dot{M}_{\rm sw} < 10^{13}$~g\,s$^{-1}$)\footnote{\footnotesize{For comparison, the solar wind $\dot{M}$ is $\sim 10^{12}$~g\,s$^{-1}$ \citep{Hundhausen97}.}} tend to produce deeper red wing transits and shallower blue wing absorption in the Doppler velocity ranges we analyzed. On the other hand, stronger winds ($\dot{M}_{\rm sw} > 10^{13}$~g\,s$^{-1}$) produce transit depths that are similar to those that we measured in both blue and red wings. These results highlight the importance of \lya\ transits to study not only exoplanet outflows, but also their interaction with the stellar wind.

\section{Modeling the escape of carbon and oxygen}\label{sect:models}

To interpret our observations, we produce forward models of the escaping atmosphere of HD~189733~b using the code \texttt{p-winds} \citep[version 1.4.4\footnote{\footnotesize{DOI: 10.5281/zenodo.7814782}};][]{DSantos22}. The code calculates the structure of the planet's upper atmosphere by assuming that the outflow can be simulated as an isothermal, one-dimensional Parker wind \citep{Parker58}. We further assume that the planet's atmosphere has a H/He fraction of 90/10, and that C and O are trace elements. In this version of \texttt{p-winds}, the code computes the ionization-advection balance using the photoionization, recombination and charge transfer reactions listed in Table 1 of \citet{Koskinen13}. We further include the charge transfer reaction between C$^{2+}$ and He$^0$ from \citet{Brown72}. A more detailed description of this version of the code is present in Appendix \ref{p-winds_app}. We caution that our simulation is simplified in that we assume that the outflow is one-dimensional; a more accurate model for an asymmetric transit would require three-dimensional modeling. Since the focus of this manuscript is on reporting the results of the observation, we provide here only this simplified modeling approach and leave more detailed modeling for future work.

The distribution of ions in the upper atmosphere of an exoplanet is highly dependent on the incident high-energy spectral energy distribution (SED) from the host star. For the purposes of this experiment, we will use the high-energy SED of HD~189733 that was estimated in \citet{Bourrier2020}, based on X-rays and \emph{HST} observations.

We calculate the expected distributions of C\,{\sc ii} and O\,{\sc i} as a function of altitude by assuming that the planet has solar C and O abundances, the escape rate\footnote{\footnotesize{When referring to mass-loss rates in this manuscript, we are referring to the sub-stellar escape rate. The term ``sub-stellar" is used when assuming that the planet is irradiated over 4$\pi$~sr. This assumption is usually employed in one-dimensional models like {\tt p-winds}. In reality, planets are irradiated over $\pi$~sr only, and the total mass-loss rate is obtained by dividing the sub-stellar rate by four.}} is $1.7 \times 10^{10}$~g\,s$^{-1}$ and the outflow temperature is $11\,800$~K based on the comprehensive 1D hydrodynamic simulations of \citet{Salz16b}; we refer to this setup as Model 1 or M1. We also produce a forward model assuming solar C and O abundances, an escape rate of $1.1 \times 10^{11}$~g\,s$^{-1}$ and outflow temperature of $12\,400$~K, which were estimated by fitting the metastable He transmission spectrum of HD~189733~b \citep{Lampon21}; we refer to this setup as Model 2 or M2. Finally, we also calculate a forward model based on the photochemical formalism presented in \citet{GMunoz2007}, which yields an escape rate of $1.8 \times 10^{10}$~g\,s$^{-1}$ and a maximum thermospheric temperature of $12\,000$~K; we refer to this setup as Model 3 or M3.

\begin{figure*}[ht]
    \gridline{\fig{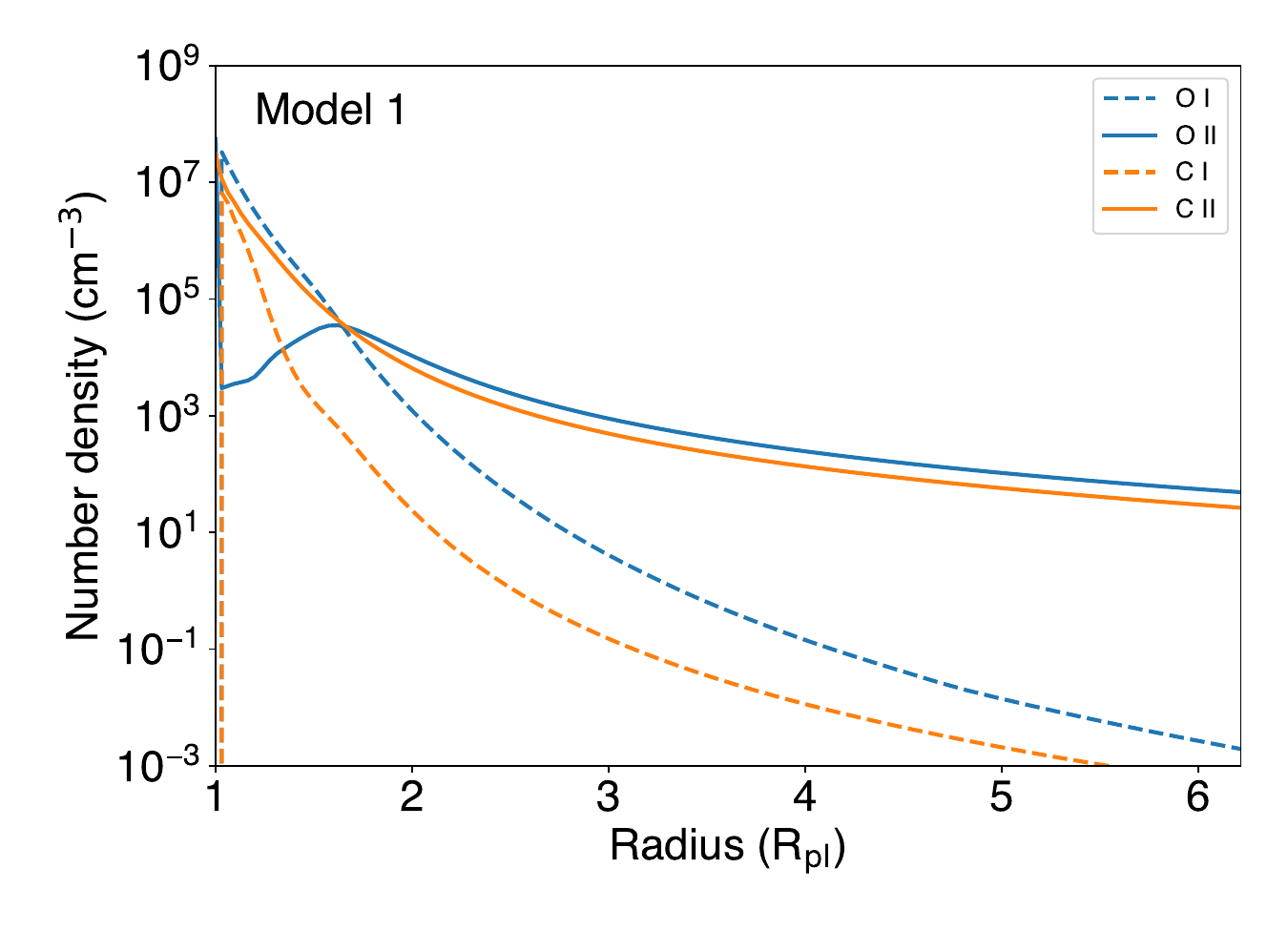}{0.49\textwidth}{(a)}
              \fig{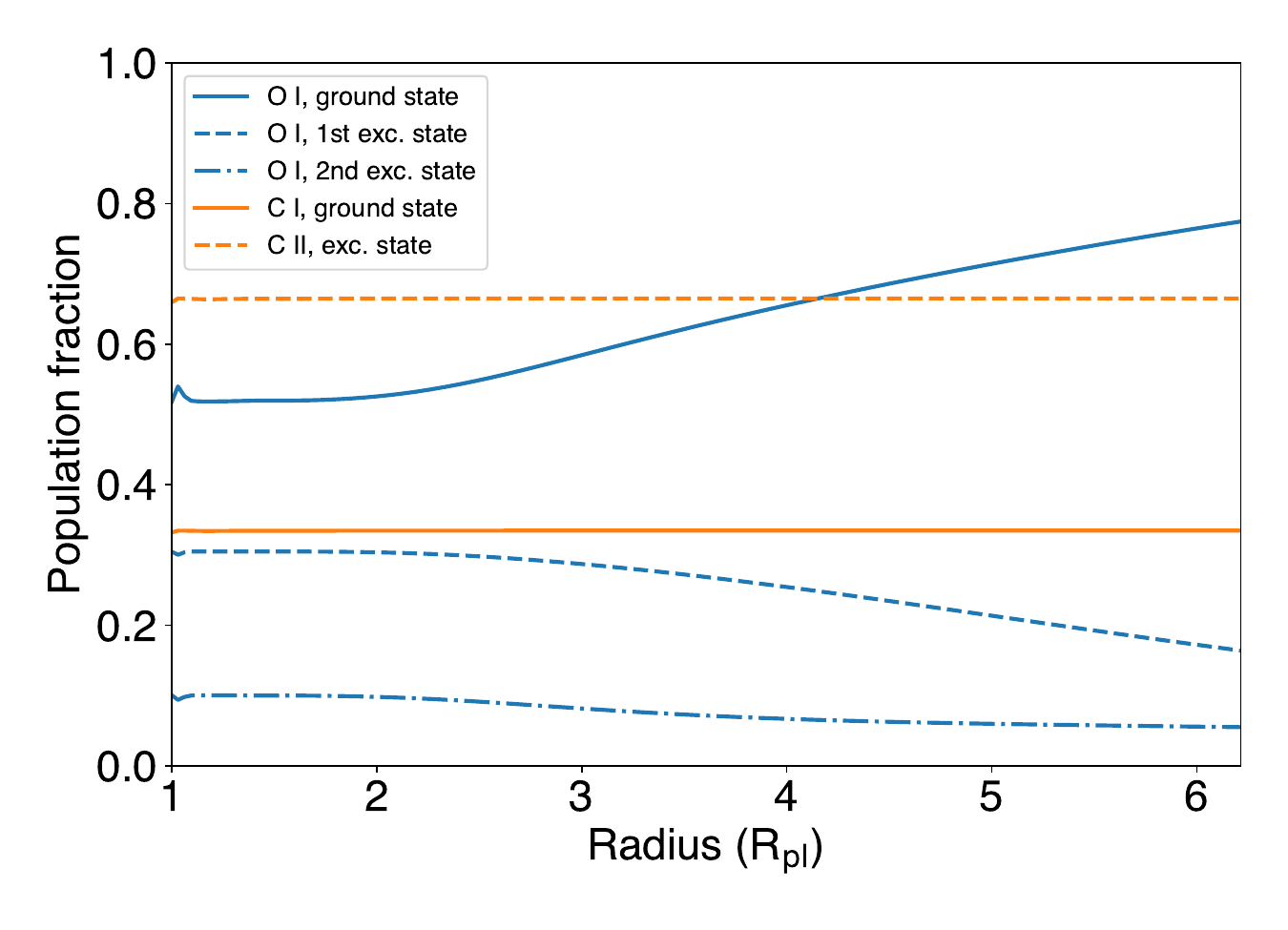}{0.49\textwidth}{(b)}}
    \gridline{\fig{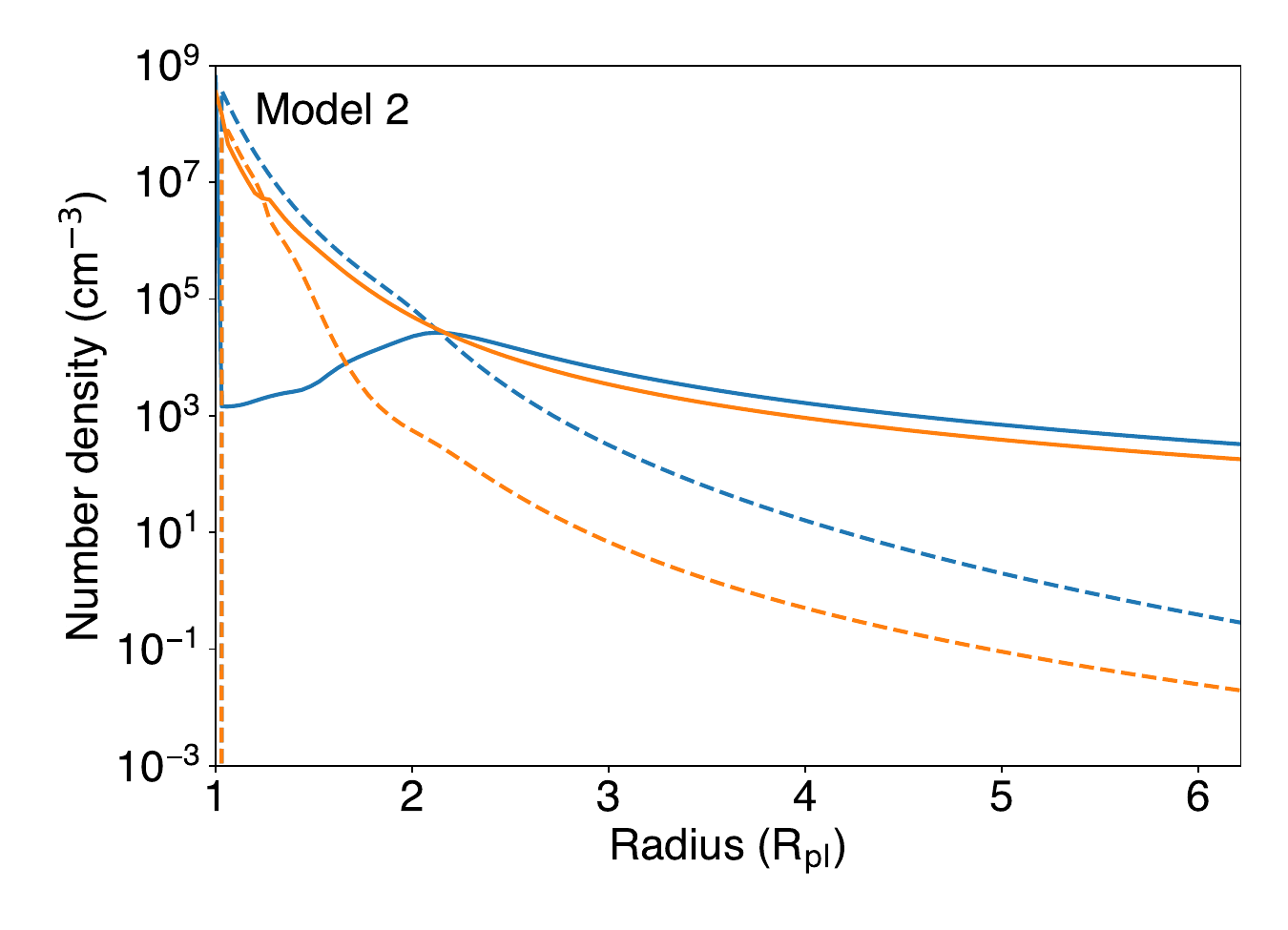}{0.49\textwidth}{(c)}
              \fig{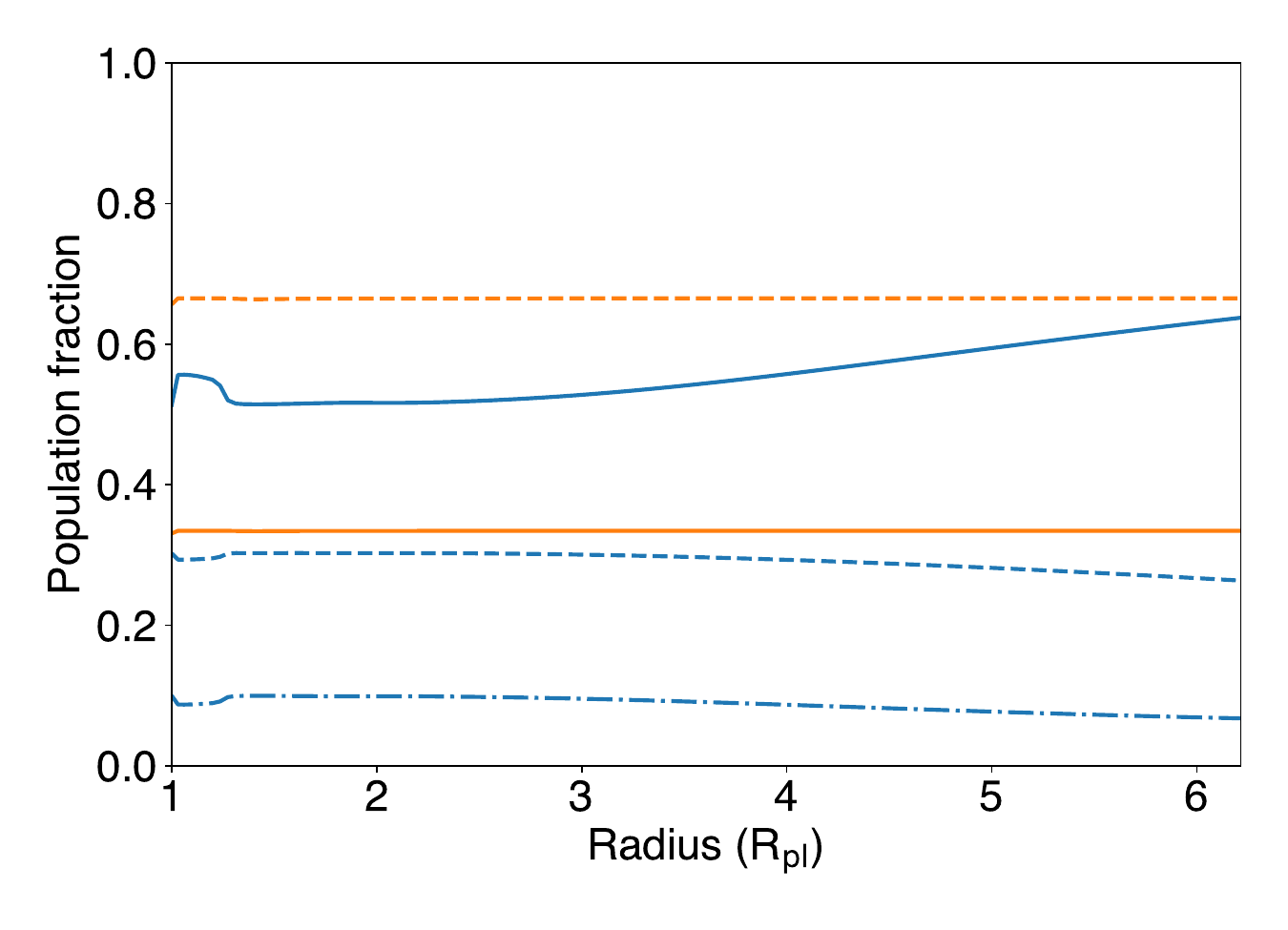}{0.49\textwidth}{(d)}}
    \gridline{\fig{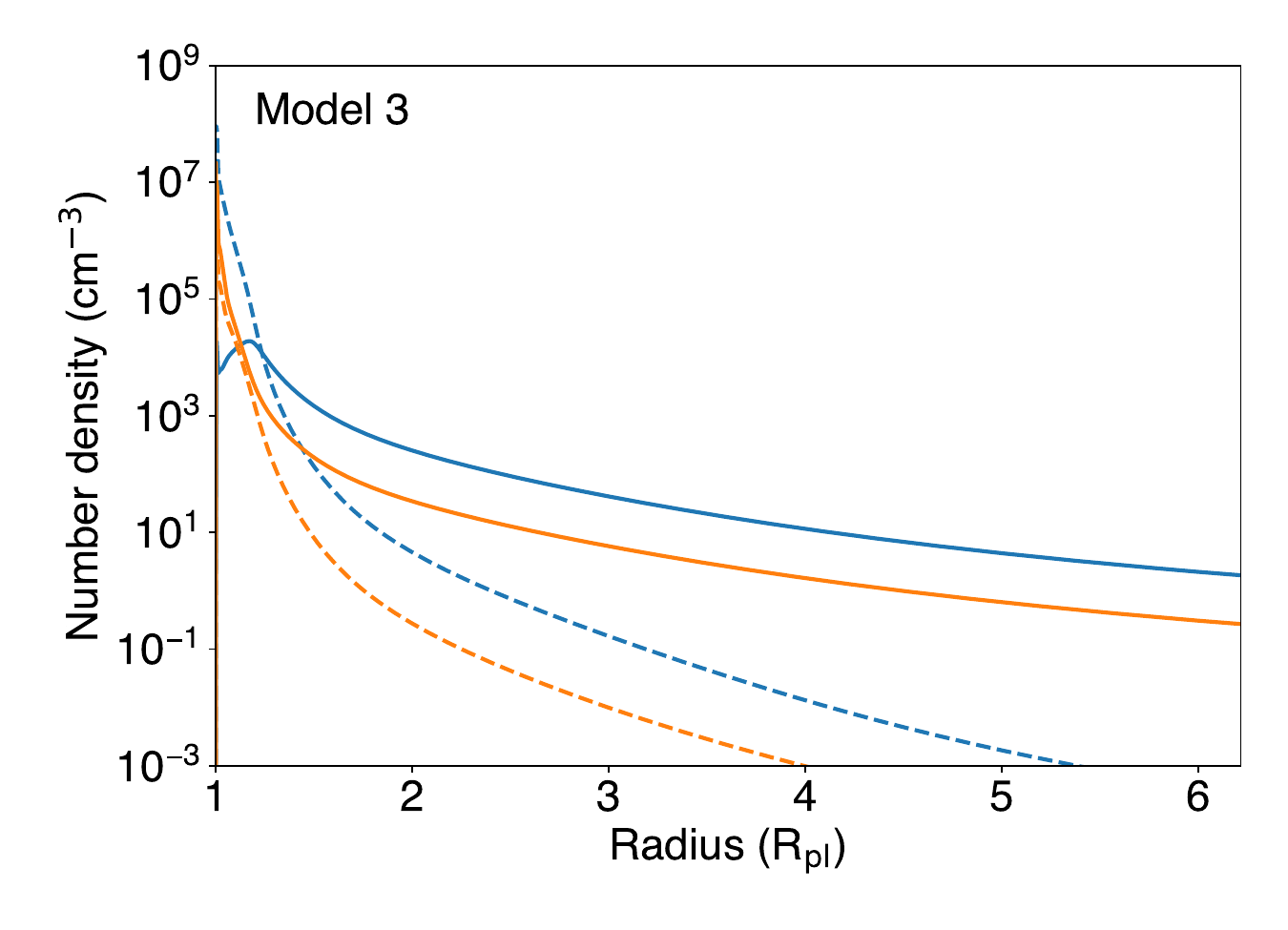}{0.49\textwidth}{(e)}
              \fig{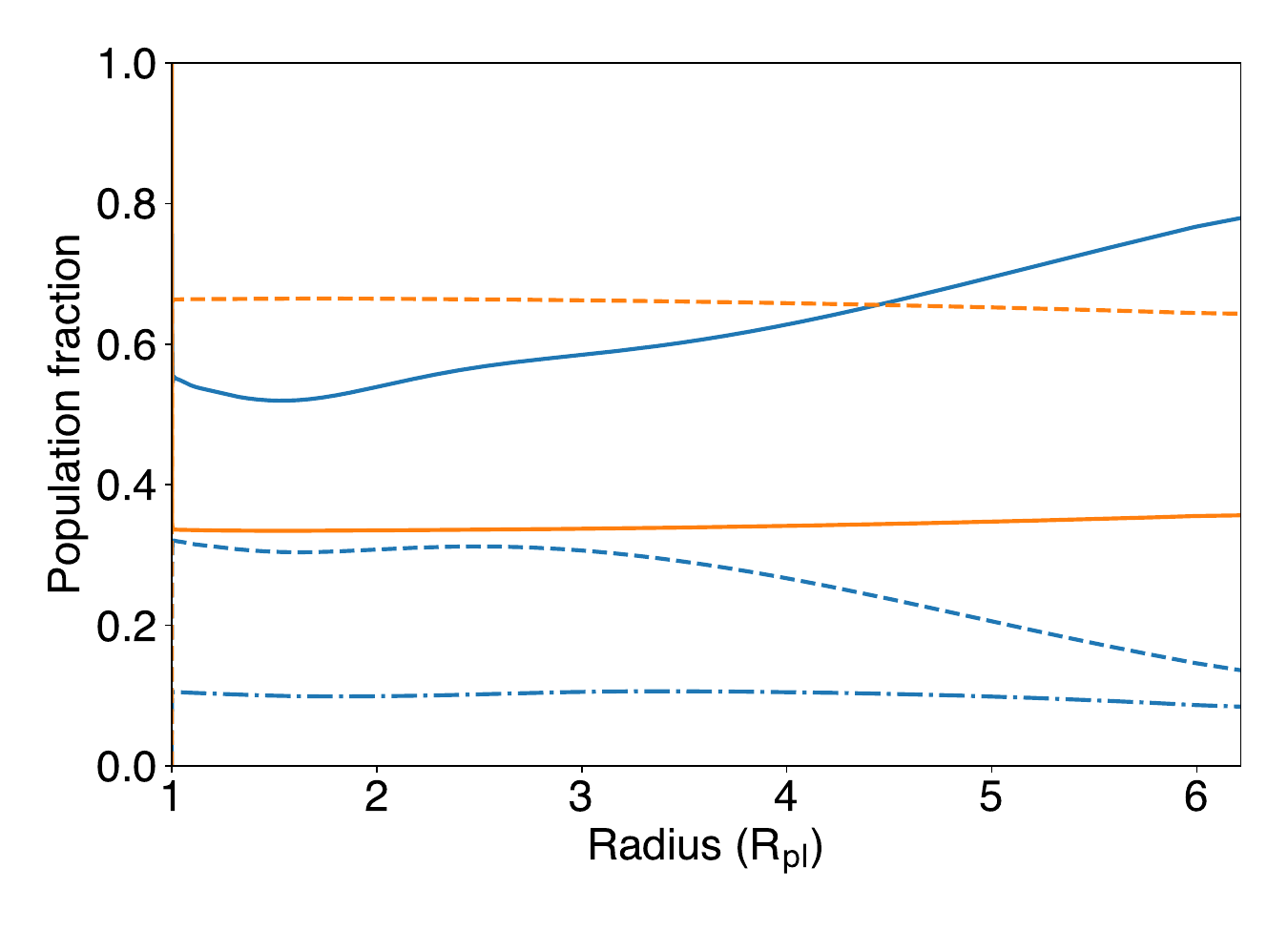}{0.49\textwidth}{(f)}}
\caption{(a) Distribution of neutral (dashed lines) and singly-ionized (full lines) carbon (orange) and oxygen (blue) in the upper atmosphere of HD~189733~b. The outflow of HD~189733~b is mostly singly-ionized, and no doubly-ionized C is produced. (b) Distribution of the different excitation levels of C and O, with the full, dashed, and dot-dashed lines indicating, respectively, the ground, first excited and second excited states, respectively. Panels (a) and (b) assume the escape rate and outflow temperature estimated by \citet{Salz16b}. Panels (c) and (d) are the same as (a) and (b), but assuming the escape rate and outflow temperature estimated by \citet{Lampon21}. Panels (e) and (f) correspond to the photochemical model described in \citet{GarciaMunoz21}.  \label{fig:M1_M2_pop}}
\end{figure*}

Our results for M1 and M2 show that the upper atmosphere of HD~189733~b is mostly neutral within altitudes of $\sim 2$~R$_{\rm p}$ for all species: H, He, C and O; beyond this point, the atmosphere is predominantly singly ionized (see left panels of Figure \ref{fig:M1_M2_pop}). For M3, we found that the outflow is mostly singly ionized and is neutral only near the base of the wind. The population of ground and excited states C and O is sensitive to the assumed mass-loss rate and outflow temperature because they control how many electrons are available to collide and excite the nuclei, as well as the energy of the electrons (see right panels of Figure \ref{fig:M1_M2_pop}). 

\begin{figure}[ht]
\plotone{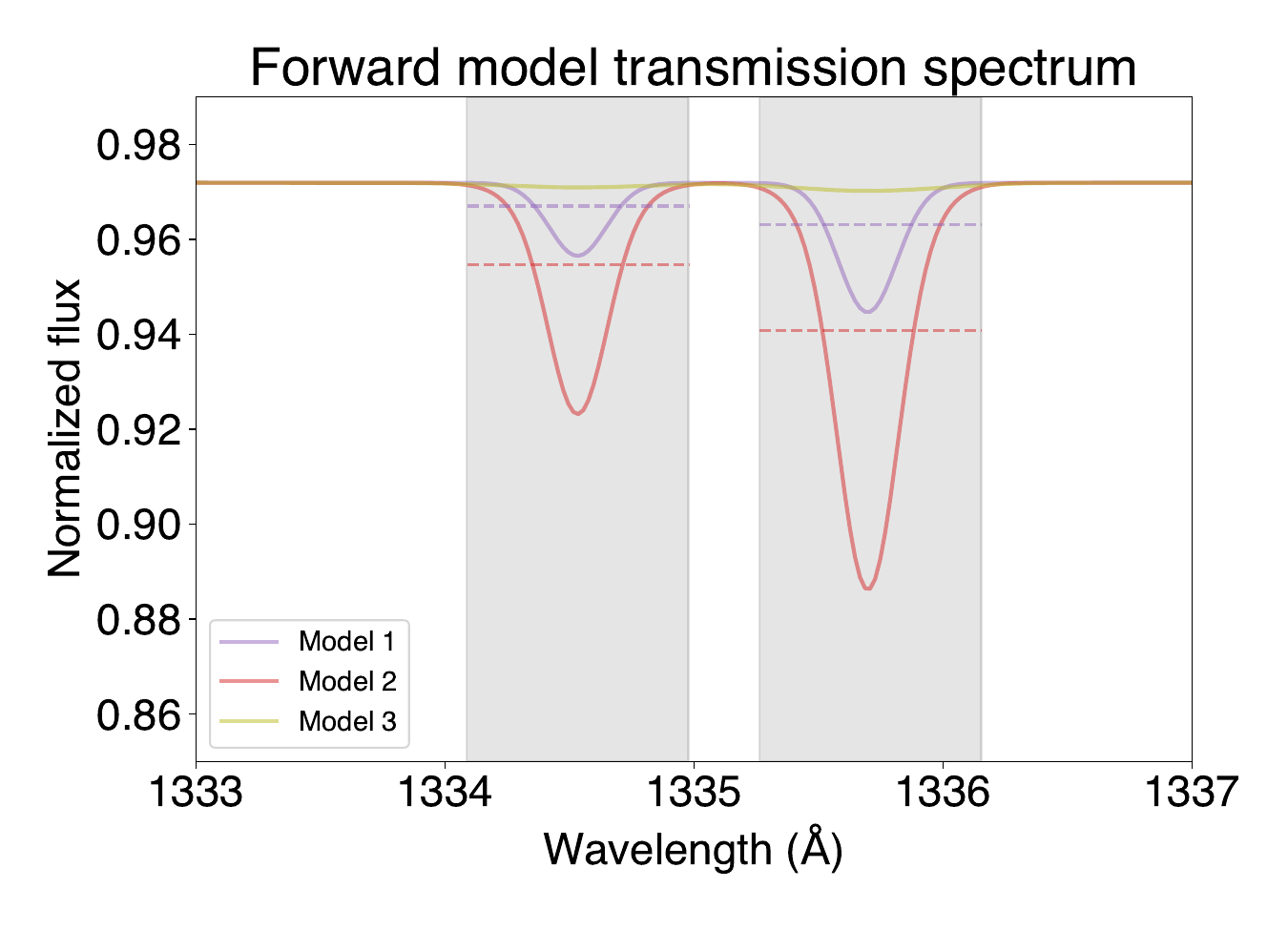}
\caption{Simulated ingress transmission spectra of HD~189733~b near the C\,{\sc ii} triplet in the UV. The shaded region delimits the wavelength ranges in which the flux density is integrated to calculate the light curves seen in Figure \ref{fig:carbon_lc_hd189}. The dashed horizontal lines indicate the average ingress absorption in the wavelength range delimited by the shaded region and corresponds to the absorption levels in Table \ref{tab:abs_models}. See a description of the different models in the main text.   \label{fig:t_spec_m1_m2}}
\end{figure}

We used the density profiles of O\,{\sc i} and C\,{\sc ii}, as well as the ground/excited state fractions to calculate the expected transmission spectra of HD~189733~b near the ingress of the planet, where we found the strongest signals of a possible in-transit planetary absorption (see Figure \ref{fig:carbon_lc_hd189}). To this end, we used the instrumental line spread function of COS with the G130M grating centered at $\lambda_0 = 1291$~\AA\ obtained from STScI\footnote{\footnotesize{\url{https://www.stsci.edu/hst/instrumentation/cos/performance/spectral-resolution}}}. The resulting theoretical transmission spectra are shown in Figure \ref{fig:t_spec_m1_m2}. 

In order to compare these predictions with our observations, we take the average in-transit absorption within limited integration ranges where there is a detectable stellar flux in each wavelength bin. These ranges are the Doppler velocities [-100, +100]~km\,s$^{-1}$ in the stellar rest frame for C\,{\sc ii} and [-75, +75]~km\,s$^{-1}$ for O\,{\sc i}. The results are shown in Table \ref{tab:abs_models}.

\begin{deluxetable}{l c c c}
\tablecaption{Wavelength-averaged ingress absorption calculated for our models. \label{tab:abs_models}}
\tablewidth{0pt}
\tablehead{
\colhead{Species} & \colhead{Model 1} & \colhead{Model 2} & \colhead{Model 3} 
}
\startdata
O\,{\sc i} (all lines) & $3.0\%$ & $3.3\%$ & $3.0\%$ \\
C\,{\sc ii} & $3.3\%$ & $4.5\%$ & $2.9\%$ \\
C\,{\sc ii}$^*$ & $3.6\%$ & $5.9\%$ & $2.9\%$ \\
\enddata
\end{deluxetable}

As expected for the lower mass-loss rate inferred by \citet[][$1.7 \times 10^{10}$~g\,s$^{-1}$]{Salz16b} and by M3 ($1.8 \times 10^{10}$~g\,s$^{-1}$) as compared to the estimate of \citet{Lampon21}, the excess absorptions in M1 and M3 are shallower than that of M2 and are inconsistent with the C\,{\sc ii} transit depths that we observe with COS. We find that an isothermal Parker wind model with an escape rate of $1.1 \times 10^{11}$~g\,s$^{-1}$, estimated from metastable He spectroscopy \citep{Lampon21}, yields a C\,{\sc ii} transit depth that is consistent with the COS observations. This mass-loss rate is also consistent with the simple estimates from \citet{Sanz2011}, based on the energy-limited formulation \citep{Salz16a}. As seen in Table \ref{tab:abs_models}, all models predict an O\,{\sc i} transit depth of approximately $3\%$, which is comparable to the sizes of the uncertainties of the COS transit depths.

Considering that M2 has a solar H fraction of 90\% and it yields an average neutral fraction of 26\%, we estimate that the sub-stellar escape rate of neutral H is $2.6 \times 10^{10}$~g\,s$^{-1}$. Since the planet is irradiated only over $\pi$~sr, we divide the sub-stellar rate by four, yielding $6.4 \times 10^9$~g\,s$^{-1}$. This escape rate of neutral H is comparable to that estimated by \citet{Lecavelier2012} and \citet{2013A&A...557A.124B} from Ly$\alpha$ transit observations. Our simulations are also consistent with the hydrodynamic models calculated by \citet{2013A&A...553A..52B}, which predict an O\,{\sc i} transit depth of about $3.5\%$. However, \citet{2013A&A...553A..52B} had claimed that super-solar O abundances or super-thermal broadening or the absorption lines are required to fit the $\sim 6.4\%$ transit depth they had measured for O\,{\sc i}. Since we did not find such a deep O\,{\sc i} transit in our analysis, no changes in the assumptions of our models were necessary.

\section{Conclusions} \label{sec:conclusions}

We reported on the analysis of several \emph{HST} transit spectroscopy observations of HD~189733~b in the FUV. We found a tentative, but repeatable absorption of $6.1\% \pm 1.8\%$ in the singly-ionized carbon line in the first excited state during the June-July/2017 epoch. This signal is in tension with the 2009 observations of this planet, which found no significant in-transit absorption of C\,{\sc ii}. In addition, we found a less significant ingress absorption in the neutral oxygen lines of $5.3\% \pm 1.9\%$.

Our analysis yielded hints of a C\,{\sc ii} and Si\,{\sc iii} post-transit tail, but they are not repeatable across the visits in question. We could not draw a definitive conclusion whether these non-repeatable signals are due to planetary or stellar variability. Although we were able to measure the FUV continuum flux of HD~189733 using COS, its light curves show no signal of significant in-transit absorption or variability. A comparison between absolute FUV fluxes and nearly simultaneous ground-based photometry in the $b$ and $y$ bands suggest that FUV emission lines tend to increase in flux by a factor of 10\% when the star is 1.9\% fainter in the optical due to starspots.

Using a geocoronal decontamination technique, we analyzed the \lya\ time series and found a repeatable in-transit absorption in both the blue and red wings of the stellar emission. This result is consistent with previous studies using the STIS spectrograph. A comparison with hydrodynamics models in the literature shows that the \lya\ absorption levels we found are consistent with an outflow that interacts with a stellar wind at least 10 times stronger than solar.

We interpreted the tentative C\,{\sc ii} and O\,{\sc i} signals using the one-dimensional, isothermal Parker-wind approximation of the Python code {\tt p-winds}, which was originally created for metastable He observations. We adapted this code to include the photochemistry of C and O nuclei (see Appendix \ref{p-winds_app}). This adaptation is publicly available as {\tt p-winds} version 1.4.3. Based on our modeling, we conclude that the mass-loss rate of HD~189733~b is consistent with those inferred by the previous observational estimates of \citet[][neutral H escape rate of $\sim 10^9$~g\,s$^{-1}$]{2013A&A...557A.124B} and \citet[][total escape rate of $1.1 \times 10^{11}$~g\,s$^{-1}$]{Lampon21}, assuming solar abundances for the planet. Interestingly, for exoplanets that we detect both C and O escaping, we may be able to measure the C/O ratio of the outflow and compare them with estimates obtained with near-infrared transmission spectra measured with \emph{JWST}.

We will benefit from future modeling efforts to address the following open questions: i) What levels of stellar variability in its wind and high-energy input are necessary to produce variability in the planetary outflow? And how can we observationally disentangle them? ii) Does HD~189733~b possess a post-transit tail with neutral H and ionized C and Si?

\begin{acknowledgments}
LADS thanks the postdoctoral researchers of STScI, Aline Vidotto and Ofer Cohen for their helpful input during the writing of this manuscript and acknowledges the often-overlooked labor of the custodial, facilities and security staff at STScI -- this research would not be possible without them. LADS further thanks Shreyas Vissapragada and Michael Gully-Santiago for contributing to the {\tt p-winds} code, and Dion Linssen, Lars Klijn and Yassin Jaziri for helping find bugs in the code. The authors thank the anonymous referee for the kind and valuable feedback given to this manuscript. ALDE acknowledges support from the CNES (Centre national d'\'etudes spatiales, France). This work is part of the \emph{HST} Panchromatic Comparative Exoplanetary Treasury (PanCET) Program GO-14767. Astronomy at Tennessee State University is supported by the State of Tennessee through its Centers of Excellence program. This research was enabled by the financial support from the European Research Council (ERC) under the European Union's Horizon 2020 research and innovation programme (projects: {\sc Four Aces} grant agreement No 724427 and {\sc Spice Dune} grant agreement No 947634), and it has been carried out in the frame of the NCCR PlanetS supported by the Swiss National Science Foundation under grants 51NF40$\_$182901 and 51NF40$\_$205606. This research is based on observations made with the NASA/ESA Hubble Space Telescope. The data are openly available in the Mikulski Archive for Space Telescopes (MAST), which is maintained by the Space Telescope Science Institute (STScI). STScI is operated by the Association of Universities for Research in Astronomy, Inc. under NASA contract NAS 5-26555. STScI stands on the traditional and unceded territory of the Piscataway-Conoy and Susquehannock peoples. This research made use of the NASA Exoplanet Archive, which is operated by the California Institute of Technology, under contract with the National Aeronautics and Space Administration under the Exoplanet Exploration Program.
\end{acknowledgments}

\vspace{5mm}
\facilities{HST(STIS), HST(COS).}

\software{NumPy \citep{harris2020array}, SciPy \citep{2020SciPy-NMeth}, Astropy \citep{astropy:2018}, Jupyter \citep{jupyter}, Matplotlib \citep{Hunter:2007}, {\tt p-winds} \citep{DSantos22}, ChiantiPy \citep{Landi12}, {\tt batman} \citep{Kreidberg2015}.
          }

\appendix

\section{Airglow removal from COS spectroscopy}\label{ag_app}

\begin{figure*}[ht]
    \gridline{\fig{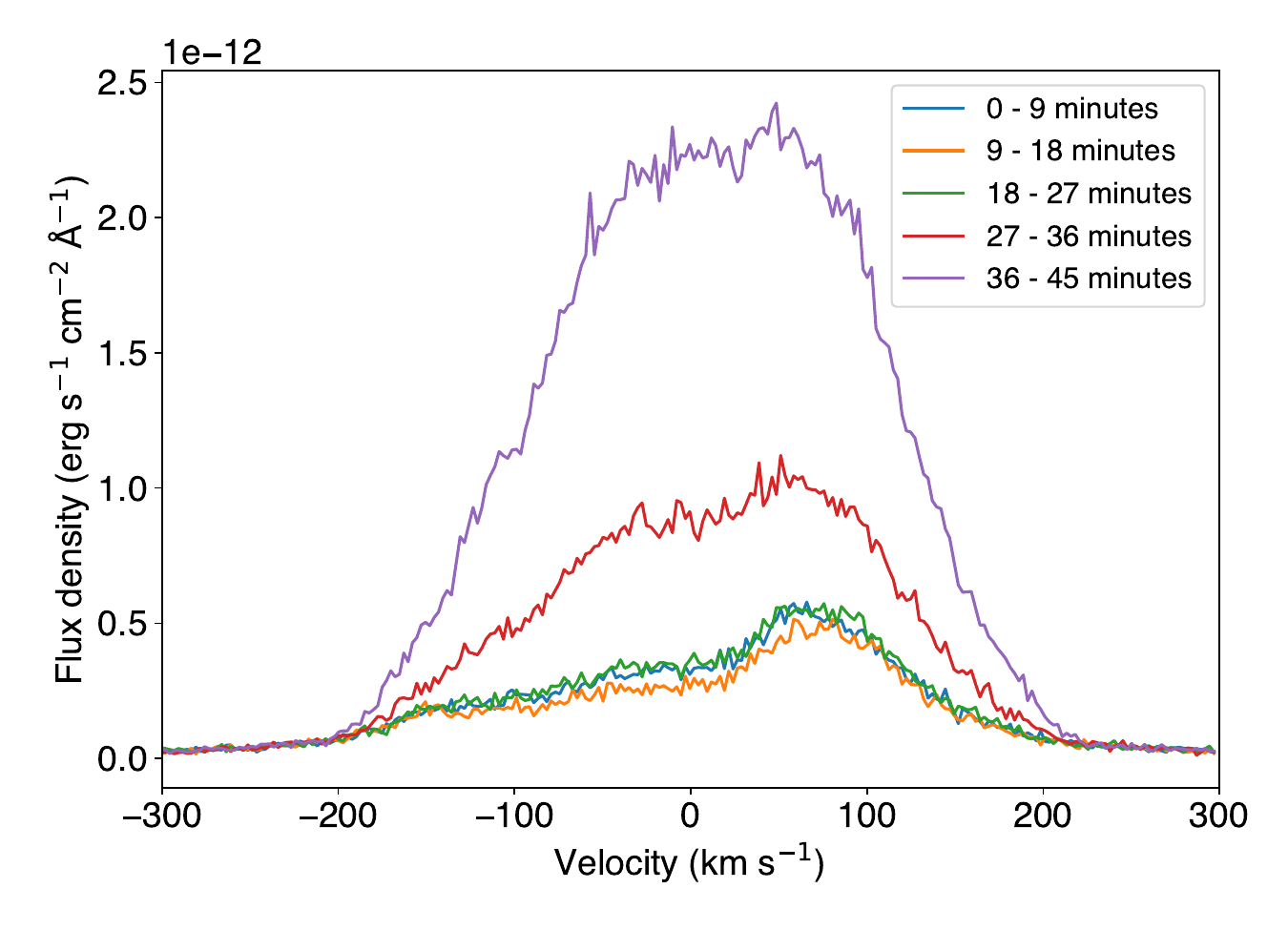}{0.49\textwidth}{(a)}
              \fig{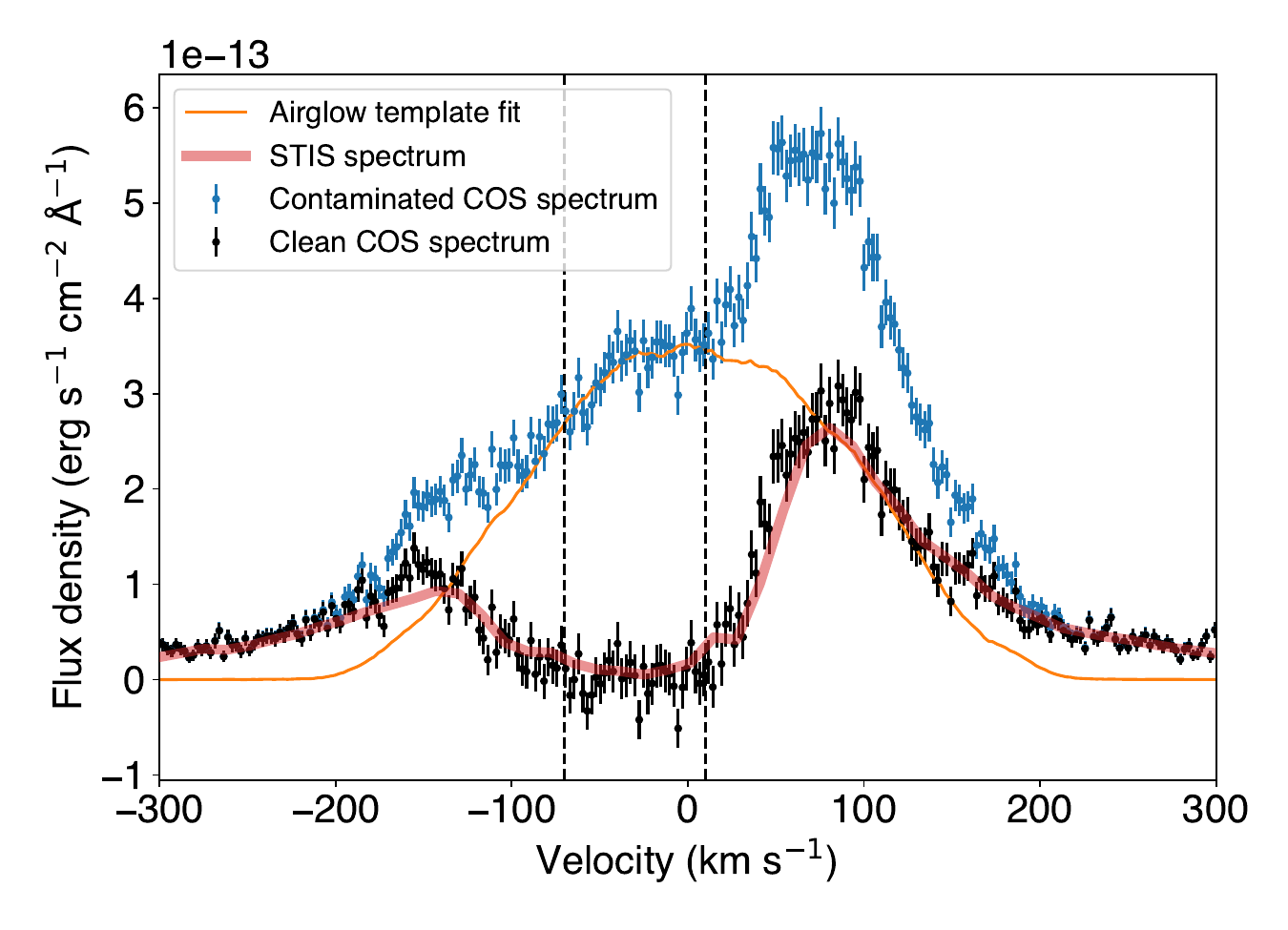}{0.49\textwidth}{(b)}}
\caption{(a) Example of the varying levels of geocoronal contamination depending on the time of the subexposure in a given orbit; we show here the sub-exposures from dataset \texttt{ld9m50ozq}. (b) Example of one \lya\ geocoronal contamination removal using airglow templates. The sub-exposure shown here corresponds to the first quarter of dataset \texttt{ld9m50ozq}. Velocities are in the heliocentric rest frame.  \label{fig:ag_fit}}
\end{figure*}

The geocoronal contamination in COS spectra extends over a wide wavelength range near the \lya\ and O\,{\sc i} lines and, as opposed to STIS, it is not easily subtracted by the instrument's pipeline. The reason for that is because of a combination of a large angular size of its circular aperture and the fact that it does not acquire angularly-resolved spectra away from the science target that would serve as a sky background \citep{Green2012}. However, it is possible to subtract this contamination with some careful analysis after the data reduction. A detailed description of this process is discussed in \citet{DSantos2019} \citep[see also][]{Aguirre2023}. Briefly, for \lya, it consists of identifying a wavelength range where we do not expect stellar emission, such as the core of the stellar \lya\ line (which is absorbed by the interstellar medium and thus contain only geocoronal emission). Then, we fit an airglow template\footnote{\footnotesize{COS airglow templates are publicly available in \url{https://www.stsci.edu/hst/instrumentation/cos/calibration/airglow}.}} with varying amplitudes and wavelength shifts to this region without stellar emission. In the case of our observation, we found that the best range to fit the airglow templates was [-70, +10]~km\,s$^{-1}$ in the heliocentric rest frame. We illustrate this process in Figure \ref{fig:ag_fit}.

\begin{figure}[ht]
\plotone{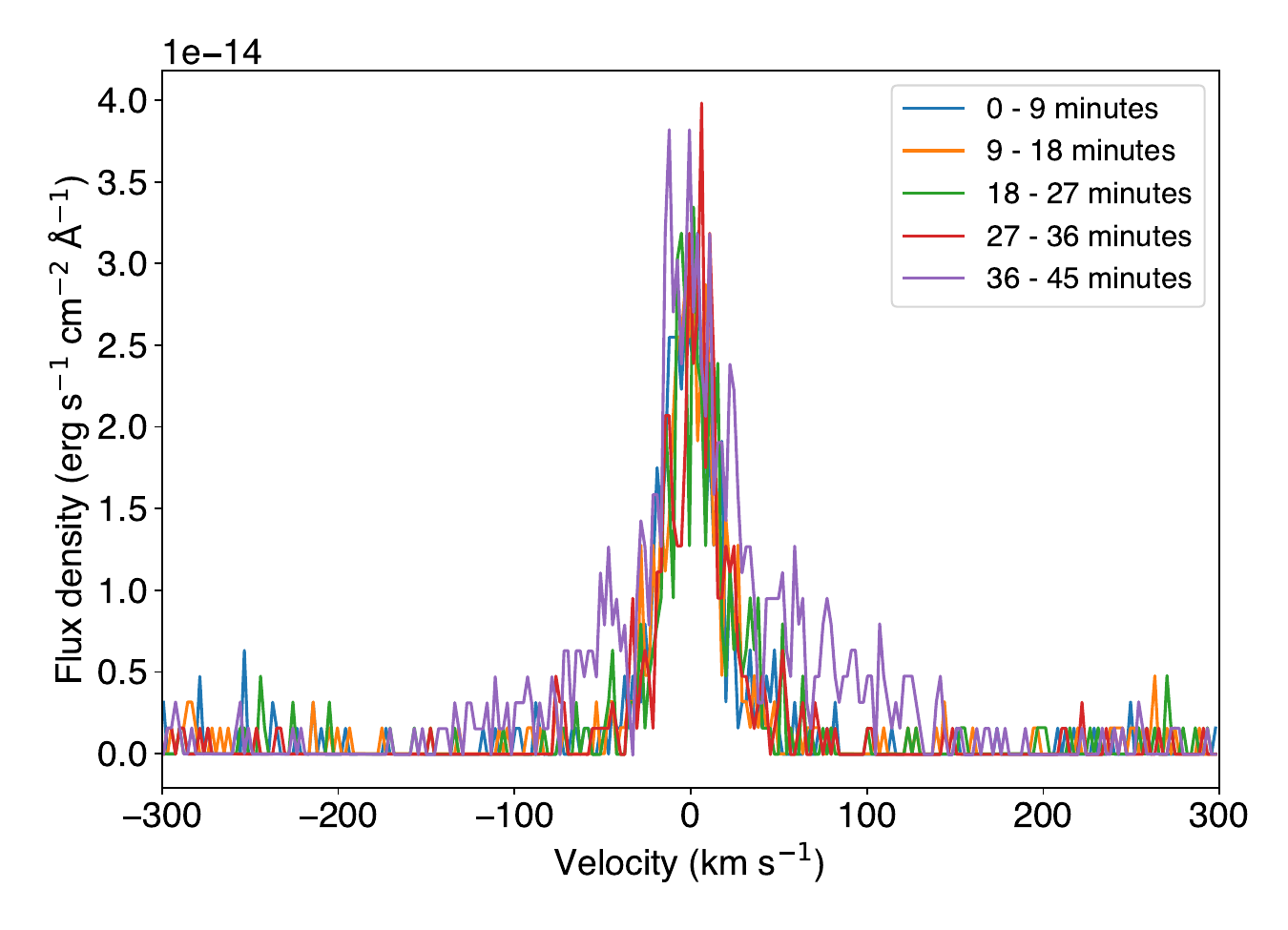}
\caption{Example of O\,{\sc i} geocoronal contamination in five subexposures of dataset \texttt{ld9m50ozq}, observed as a wider contribution to narrow stellar emission line. By comparison with the left panel of Figure \ref{fig:ag_fit}, we conclude that the level of O\,{\sc i} contamination correlates with the \lya\ geocoronal contamination.  \label{fig:oi_v3}}
\end{figure}

We verified that, for Visits A, B and C, each COS full exposure is comprised of one or more sub-exposures that have relatively low levels of contamination (see the blue, orange and red spectra in the left panel of Figure \ref{fig:ag_fit}). We use these sub-exposures to build clean O\,{\sc i} profiles, since the geocoronal contamination in these lines are negligible in the regime of low airglow (see an example in Figure \ref{fig:oi_v3}).

\section{Additional light curves}\label{app:add_lcs}

\begin{figure*}[ht]
    \gridline{\fig{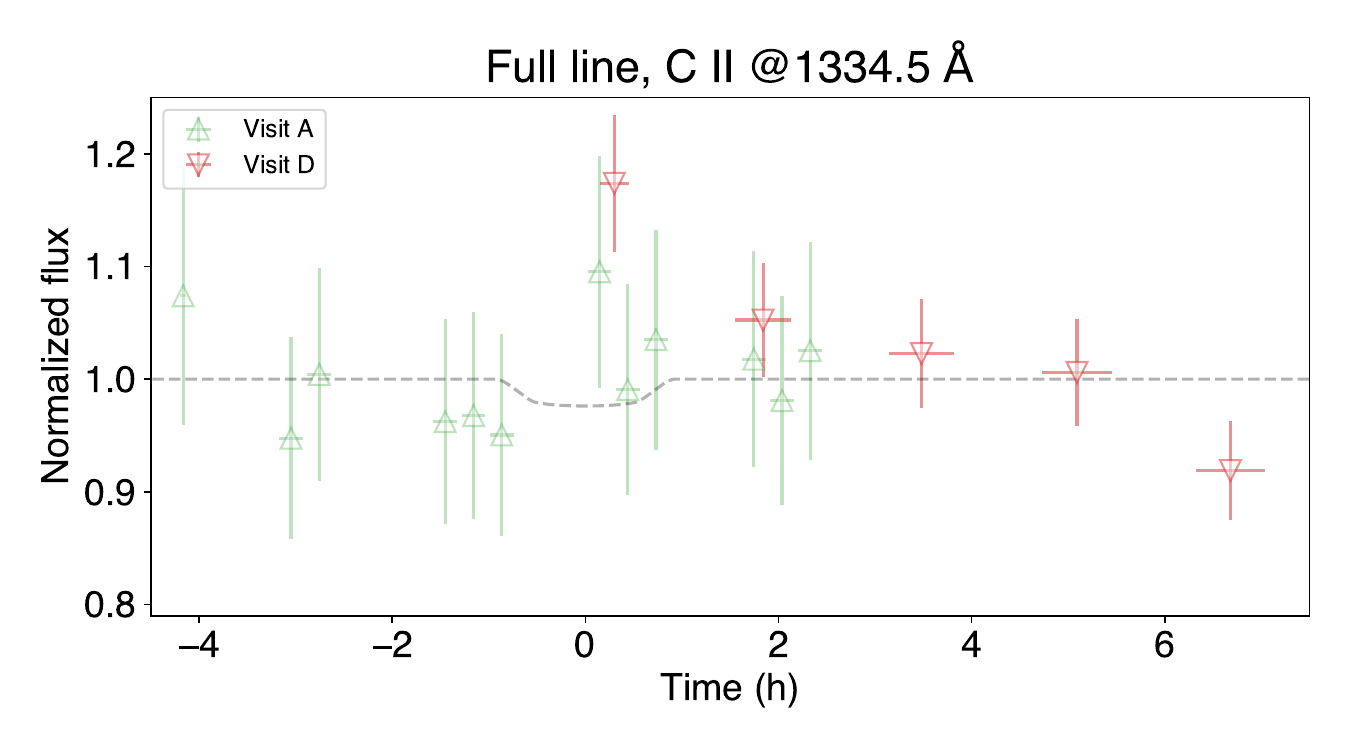}{0.49\textwidth}{(a)}
              \fig{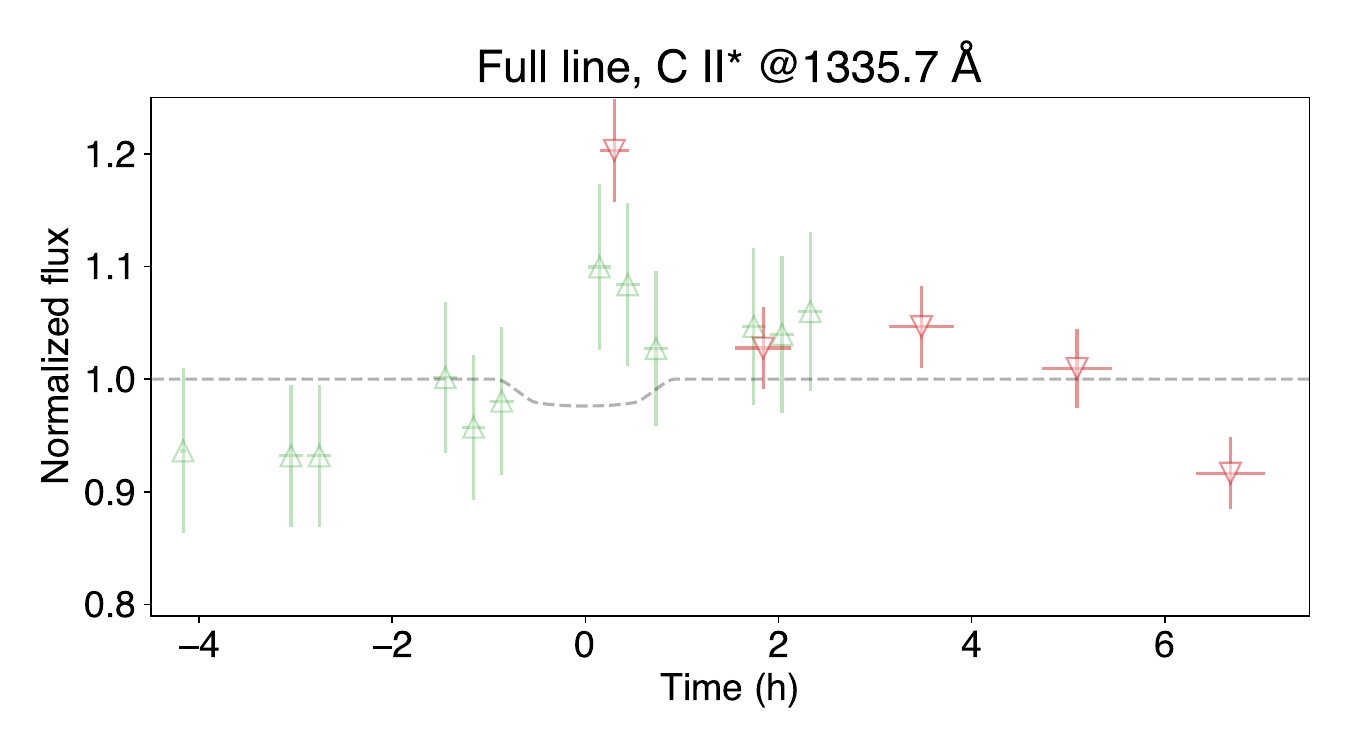}{0.49\textwidth}{(b)}}
    \gridline{\fig{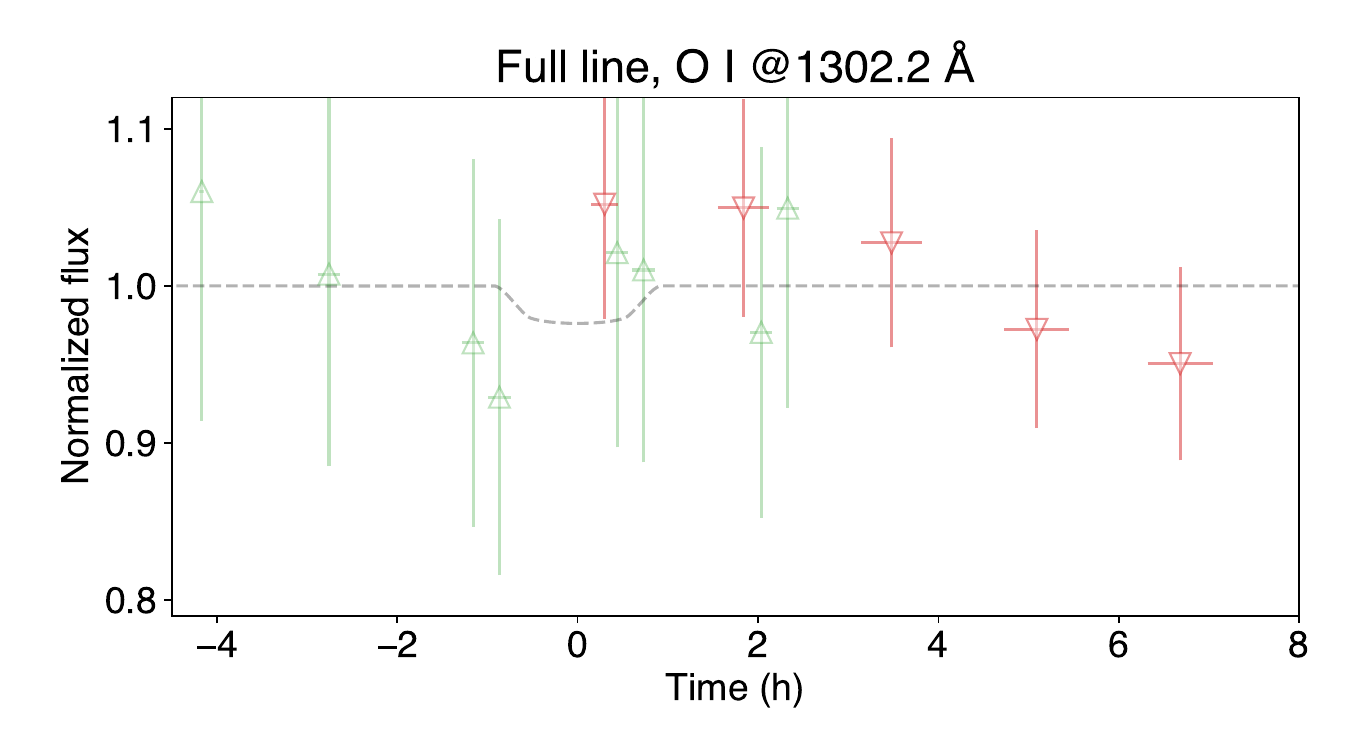}{0.49\textwidth}{(c)}
              \fig{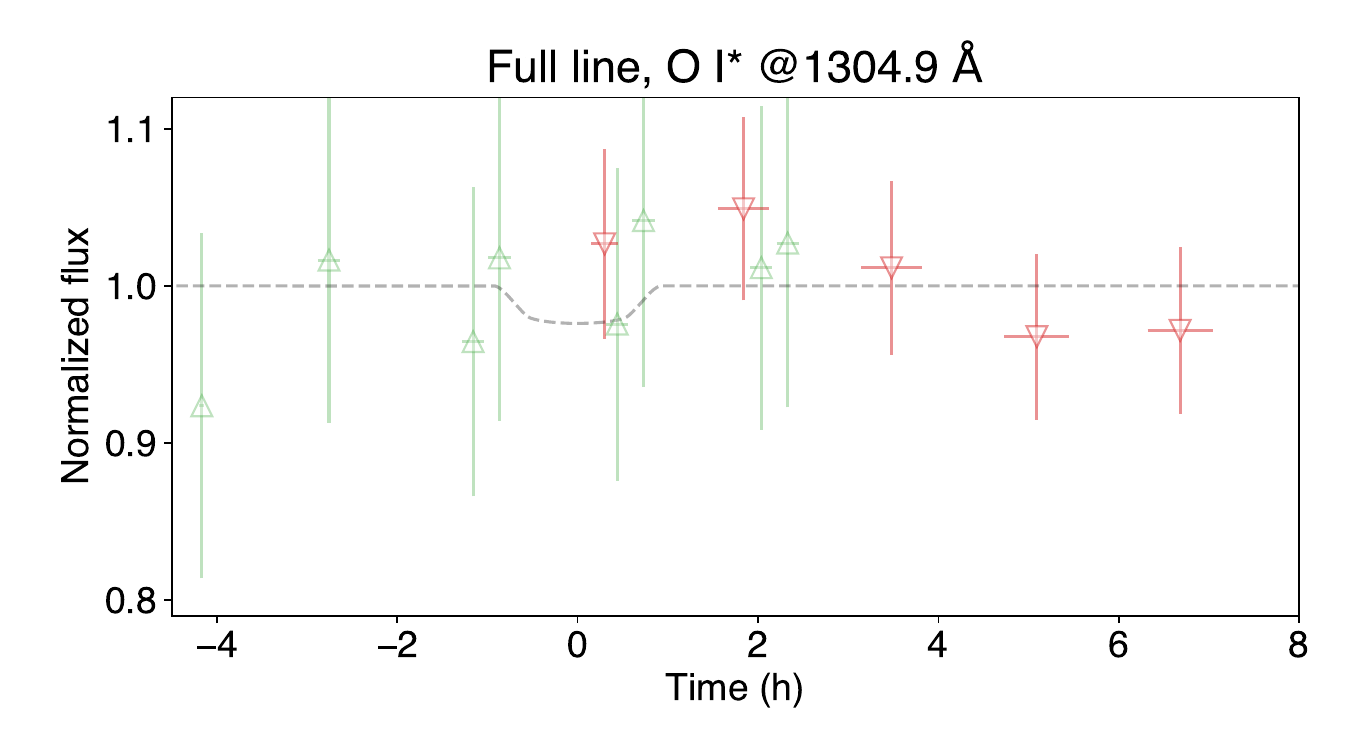}{0.49\textwidth}{(d)}}
    \gridline{\fig{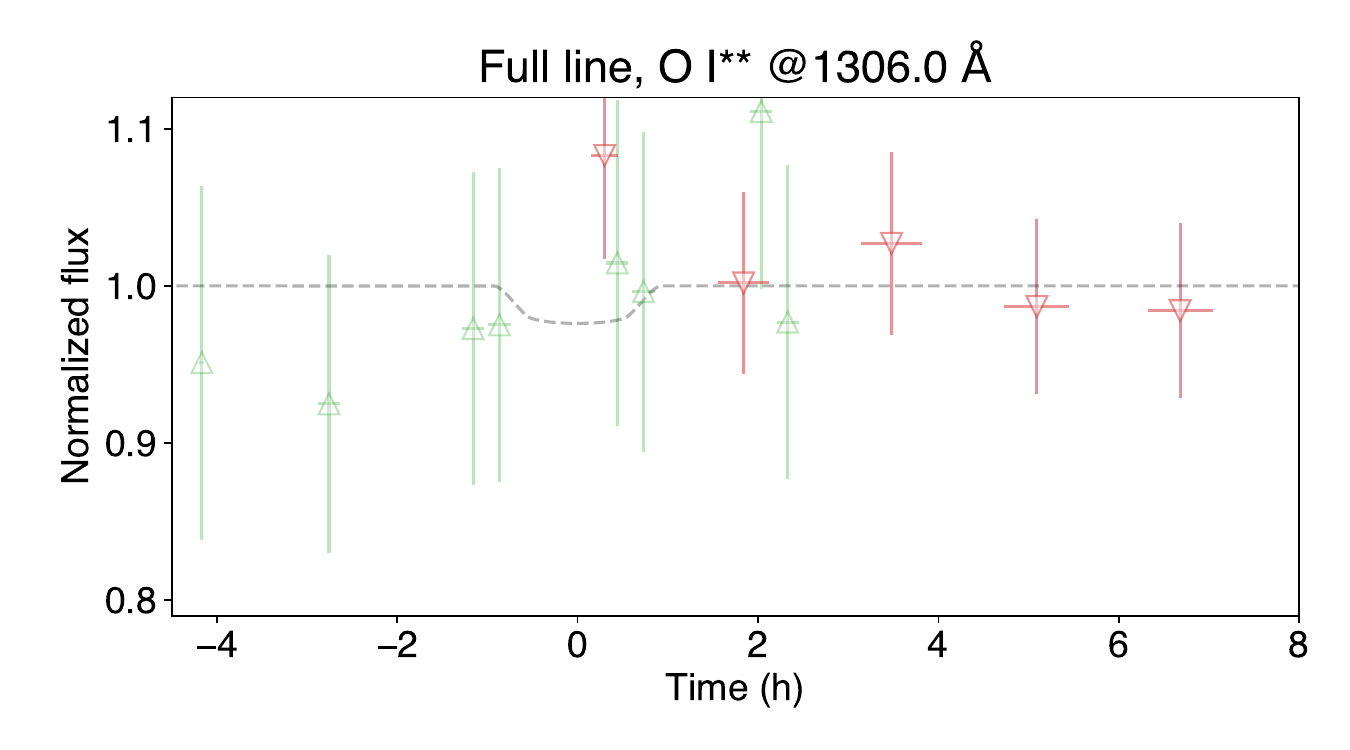}{0.49\textwidth}{(e)}}
\caption{Transit light curves of HD~189733~b in Visits A and D for the O\,{\sc i} and C\,{\sc ii} lines present in COS spectra.  \label{fig:add_lcs_1}}
\end{figure*}

\begin{figure*}[ht]
    \gridline{\fig{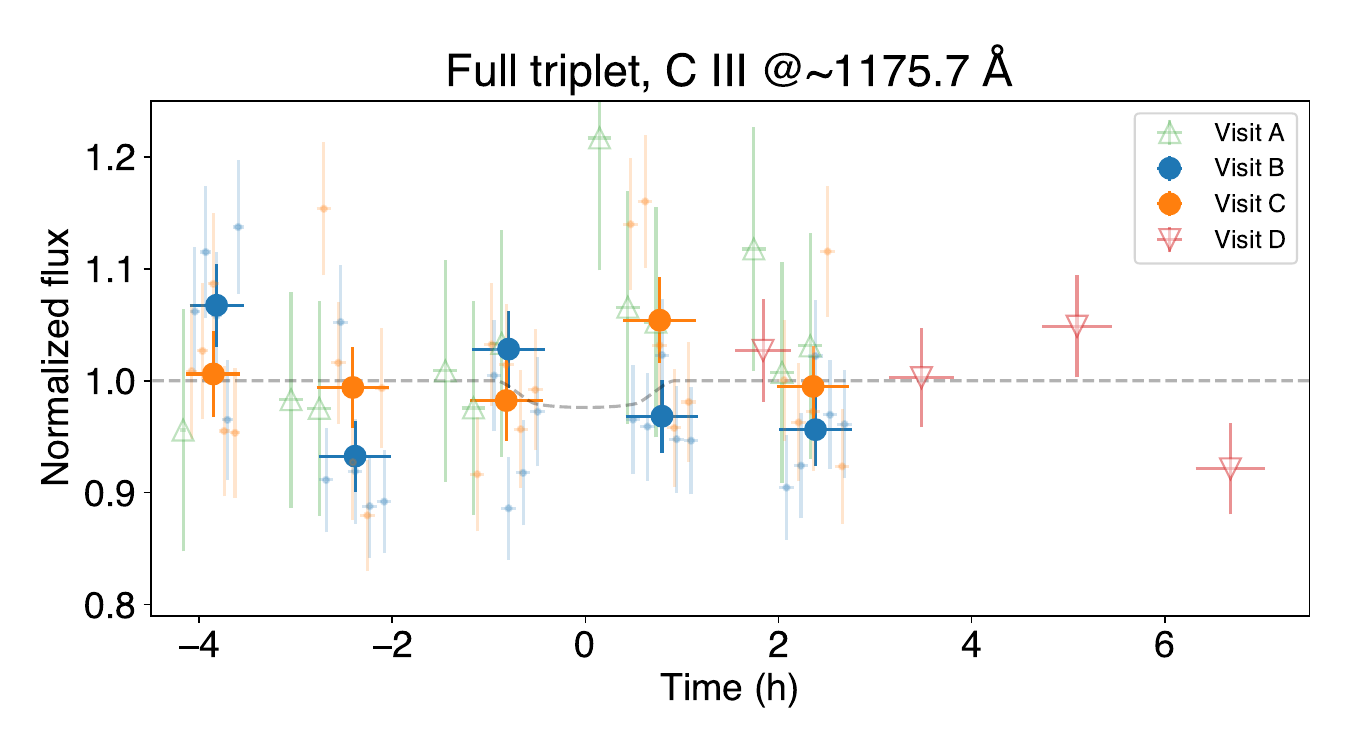}{0.49\textwidth}{(a)}
              \fig{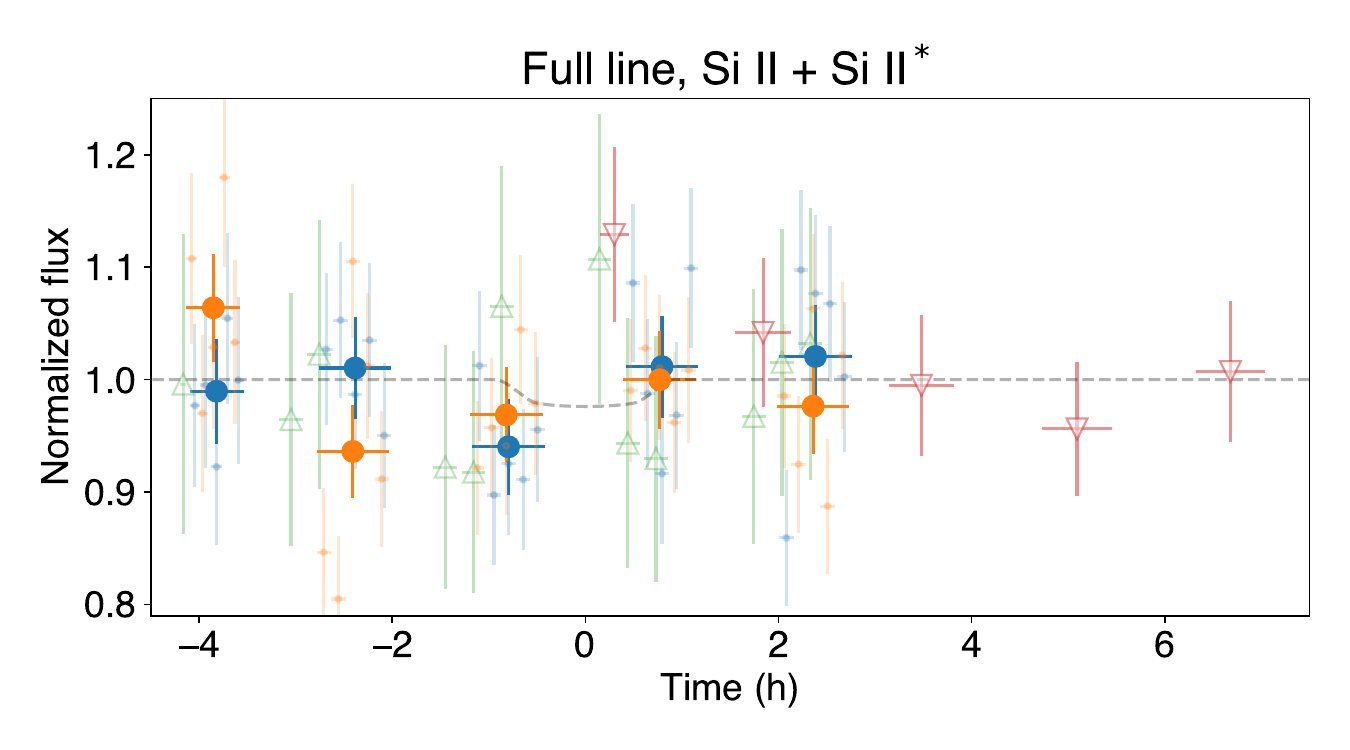}{0.49\textwidth}{(b)}}
    \gridline{\fig{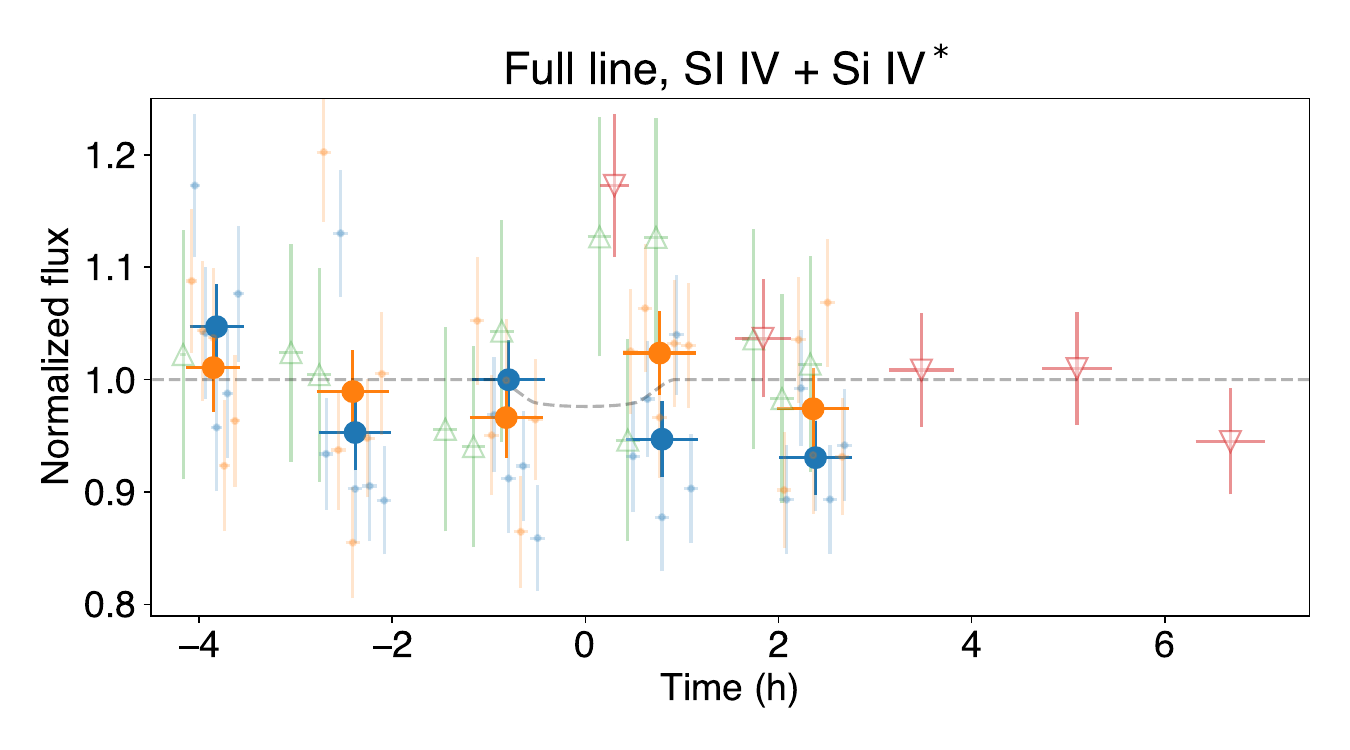}{0.49\textwidth}{(c)}
              \fig{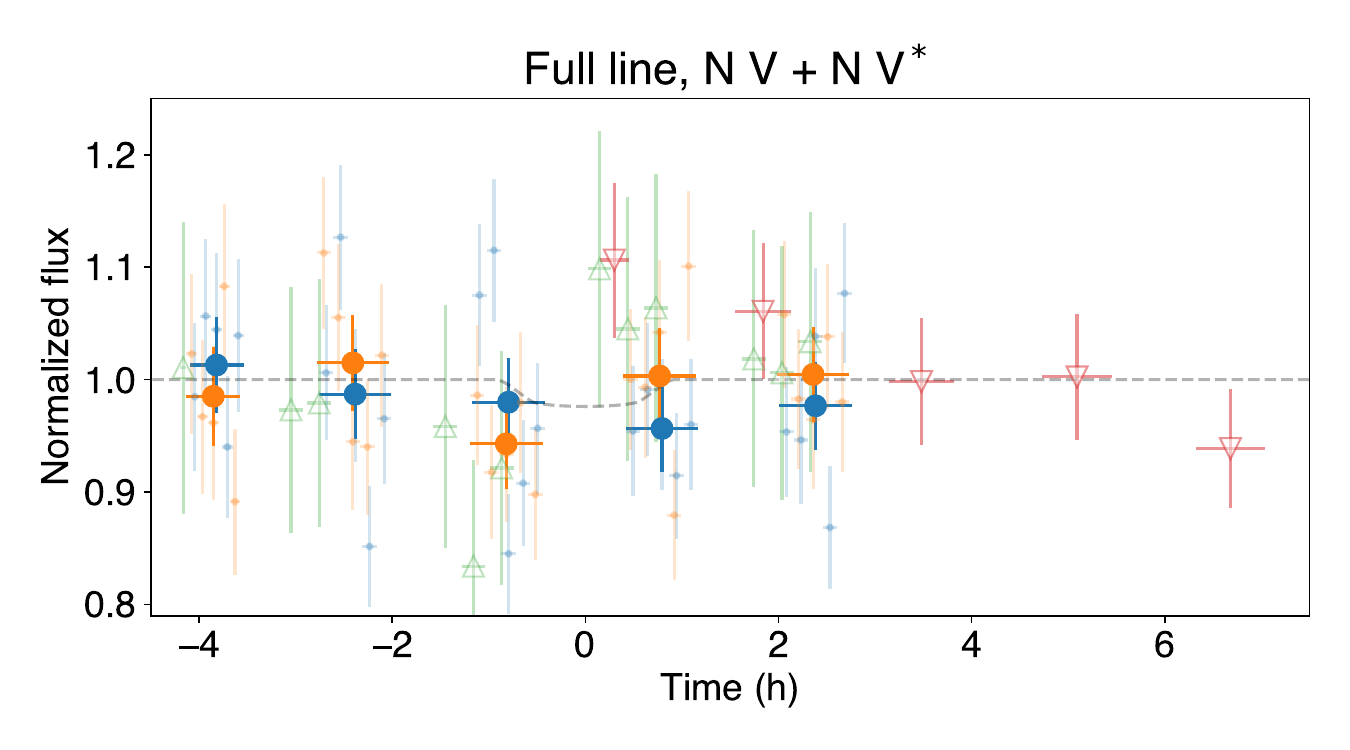}{0.49\textwidth}{(d)}}
\caption{Additional transit light curves of HD~189733~b for other lines of metallic species present in COS spectra.  \label{fig:add_lcs_2}}
\end{figure*}

\begin{figure*}[ht]
    \gridline{\fig{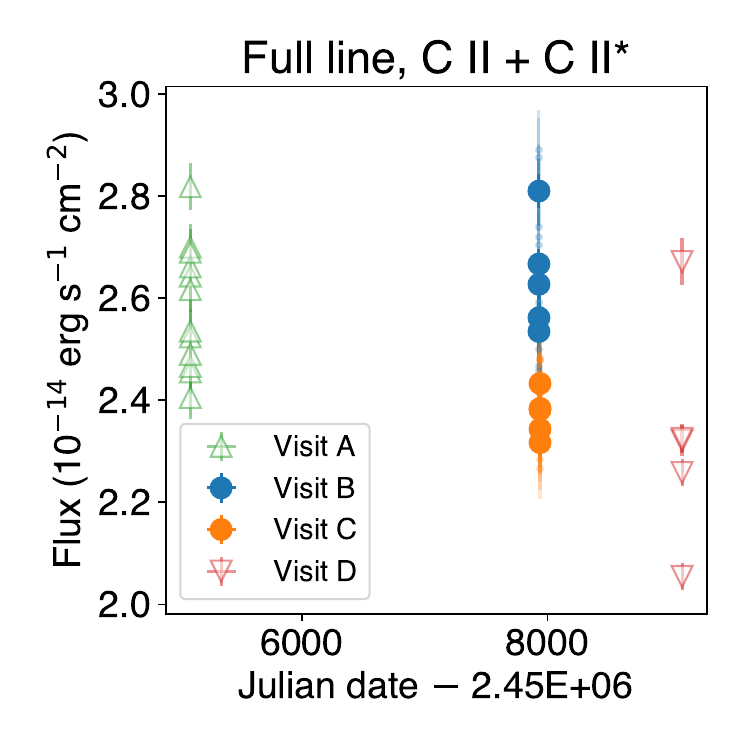}{0.32\textwidth}{(a)}
              \fig{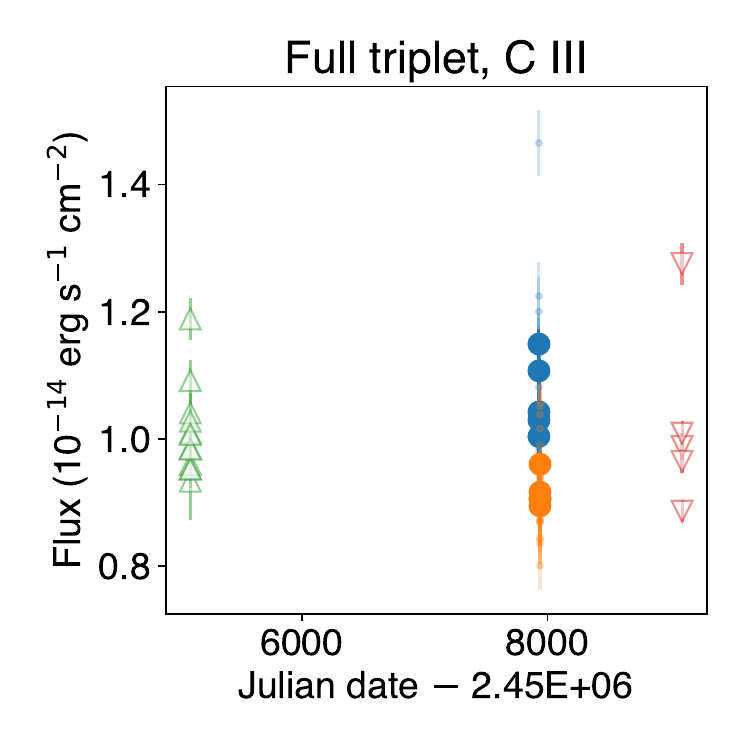}{0.32\textwidth}{(b)}
              \fig{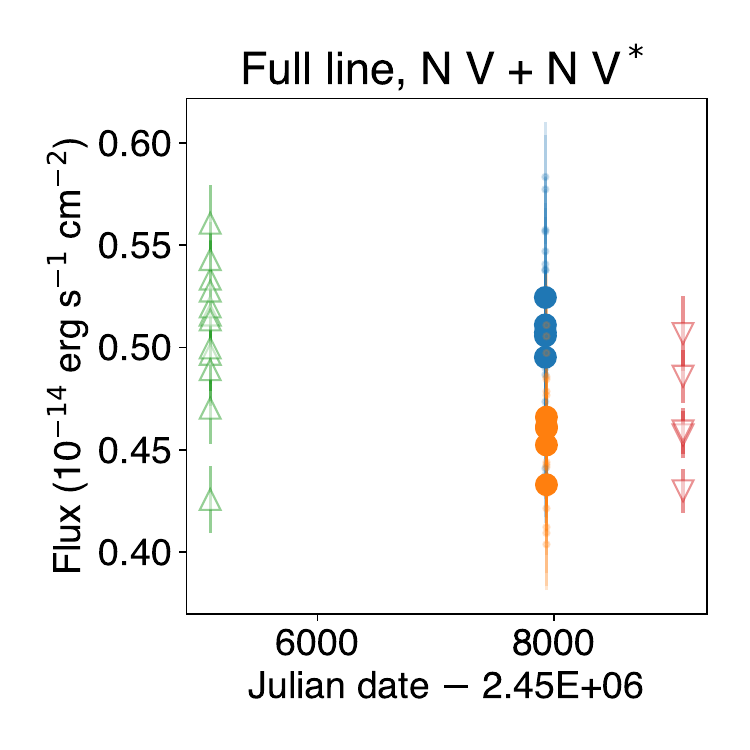}{0.32\textwidth}{(c)}}
    \gridline{\fig{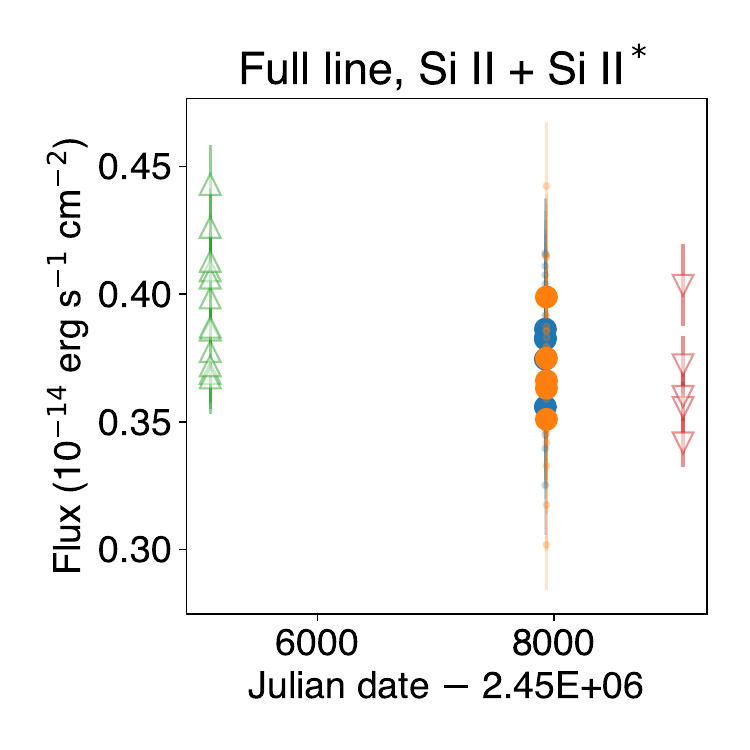}{0.32\textwidth}{(d)}
              \fig{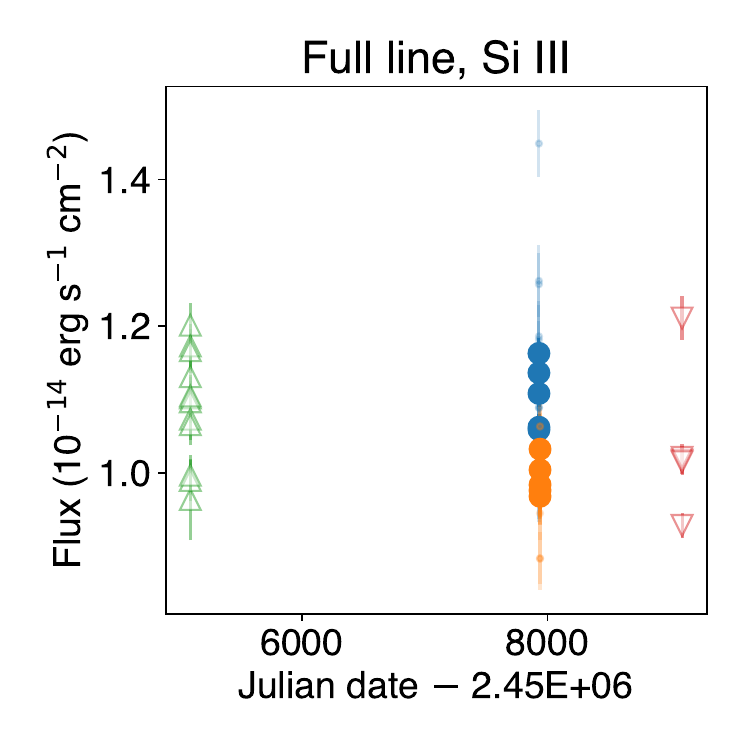}{0.32\textwidth}{(e)}
              \fig{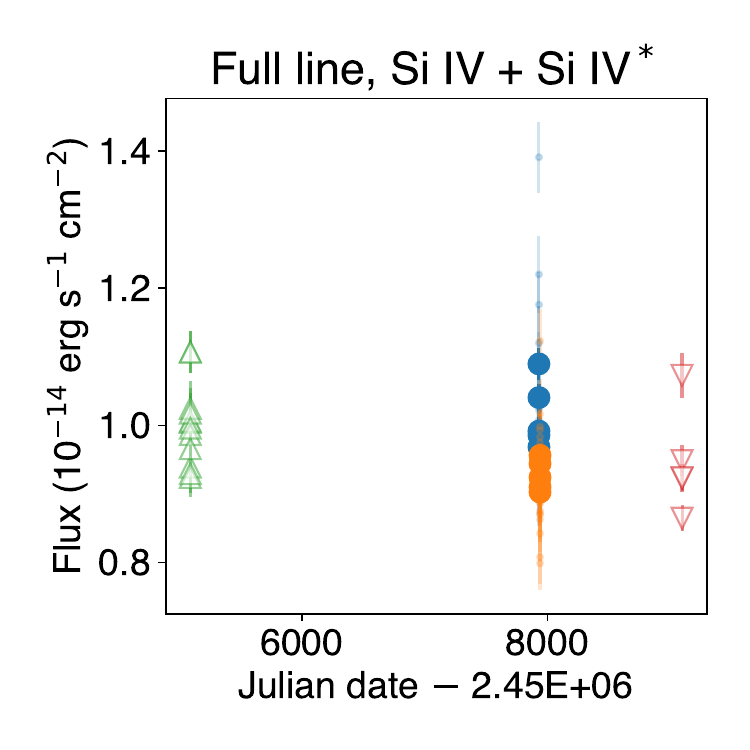}{0.32\textwidth}{(f)}}
\caption{Absolute flux light curves of HD~189733 for lines of metallic species present in COS spectra.  \label{fig:add_lcs_3}}
\end{figure*}

We include here light curves of O\,{\sc i} and C\,{\sc ii} for Visits A and D (see Figure \ref{fig:add_lcs_1}), as well as light curves of C\,{\sc iii}, Si\,{\sc ii}, Si\,{\sc iv} and N\,{\sc v} for all visits (see Figure \ref{fig:add_lcs_2}). The absolute-flux light curves of the host star are shown in Figure \ref{fig:add_lcs_3}.

\section{Ground-based photometric monitoring of HD~189733}\label{app:phot}

We acquired photometric observations of~HD 189733 during 2017A with the T10 0.80 m automatic photoelectric telescope (APT) at Fairborn Observatory in Arizona. The T10 APT is equipped with a two-channel photometer that uses two EMI 9124QB bi-alkali photomultiplier tubes to measure stellar brightness simultaneously in the Str\"omgren {\it b} and {\it y} passbands.

The photometry of HD~189733 was measured differentially with respect to the nearby comparison star HD~191260. To improve the photometric precision of the individual nightly observations, we combine the differential {\it b} and {\it y} magnitudes into a single pseudo-bandpass $(b + y)/2$. The typical precision of a single observation, as measured from pairs of constant comparison stars, typically ranges between 0.0010 mag and 0.0015 mag on good nights. The T8 APT, which is identical to T10, is described in \citet{Henry1999}.

\begin{figure}[ht]
\plotone{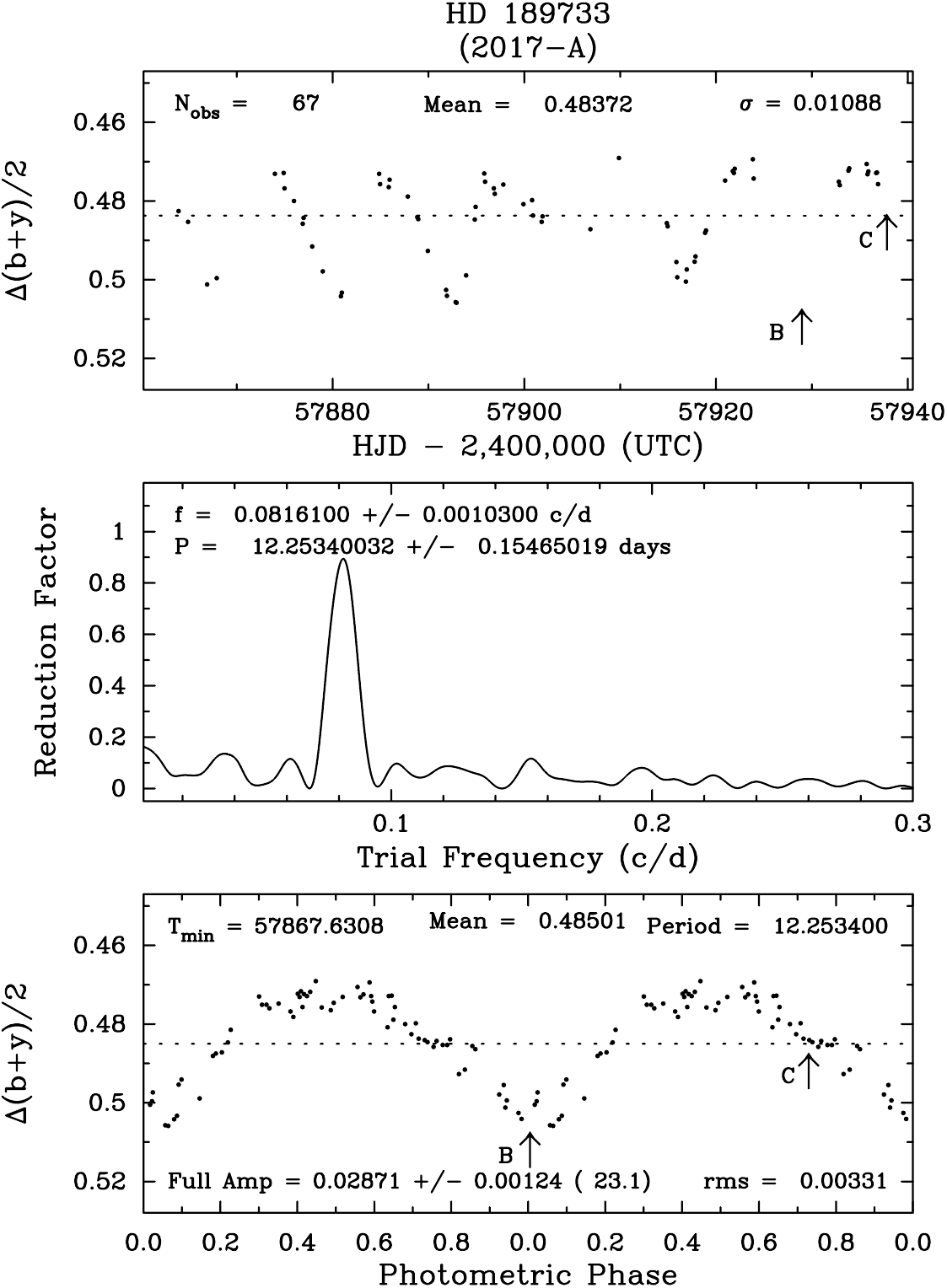}
\caption{{\it Top panel:} Ground-based differential photometry of HD~189733 in the epoch of 2017A. {\it Middle panel:} Frequency spectrum of the rotational modulation of HD~189733. {\it Bottom panel:} Same as top panel, but phase-folded to the rotational period. The times corresponding to Visits B and C are marked with arrows.  \label{fig:phot}}
\end{figure}

We show the differential photometry of HD~189733 from season 2017A in Figure \ref{fig:phot} (top and bottom panels). Visits B and C, which are the ones discussed in this manuscript, are marked with arrows. Visit B occurred when HD~189733 was in a light curve minimum and therefore the most spotted phase. Visit C occurred after the following maximum when the star was approximately 0.025 mag brighter than during Visit B, and therefore less spotted. We identify a rotational period of $12.25 \pm 0.15$~d in the middle panel of Figure \ref{fig:phot}.

\section{Extension of {\tt \lowercase{p-winds}} for exospheric C and O}\label{p-winds_app}

The isothermal, one-dimensional Parker wind code {\tt p-winds} was originally developed to simulate transmission spectra of metastable He in the upper atmosphere of evaporating exoplanets \citep{DSantos22, Vissa22a, Kirk22}. In the current development version (1.4.3), we implemented the modules {\tt carbon} and {\tt oxygen} that can calculate the distribution of neutral, singly- and doubly-ionized C nuclei, as well as neutral and singly-ionized O nuclei. Future versions of the code will include other species relevant for observations of atmospheric escape, such as Si, Fe and Mg. The development version of {\tt p-winds} also implements Roche lobe effects, as described in \citet{Vissa22b} and \citet{Erkaev07}.

The list of new reactions implemented on {\tt p-winds} to allow the modeling of C and O are shown in Table \ref{tab:reactions}; these are in addition to the reactions described in \citet{DSantos22}. To calculate the photoionization rates $\Phi$ as a function of radius $r$, we use the following equation: 

\begin{equation}
    \Phi(r) = \int^{\lambda_0}_{0} \frac{\lambda}{hc}\,f_\lambda\,\sigma_\lambda\,e^{-\tau_\lambda(r)}\,d\lambda \mathrm{,}
\end{equation}where $\lambda_0$ is the wavelength corresponding to the ionization energy of a given species and $f_\lambda$ is the incident flux density. $\sigma_\lambda$ is the photoionization cross section taken from the references listed in Table \ref{tab:reactions}. $\tau_\lambda$ is the optical depth for a given species and is calculated as follows:

\begin{equation}\label{tau_eq}
    \tau_{\lambda}(r) = \int_{r}^{\infty} \sigma_\lambda\,n(r)\,dr \mathrm{,}
\end{equation}where $n$ is the number density of a given species.

Following the formulation of \citet{Oklopcic18}, we calculate the fractions of ionized C and O using the steady-state advection and ionization balance combined with mass conservation to obtain the following equations:

\begin{equation}\label{eq:CII}
    \begin{split}
    v(r) \frac{df_{\rm C II}}{dr} ={} & f_{\rm C I} \left(\Phi_{\rm C I} e^{-\tau_{\rm C I}} + n_e\,q_{\rm E1} + n_{\rm H II}\,q_{\rm T2} + n_{{\rm He II}}\,q_{\rm T4}\right) \\
     & - f_{\rm C II} \left(n_e\,q_{\rm R2} + n_{\rm H II}\,q_{\rm T1} \right) + f_{\rm C III} \left(n_e\,q_{\rm R3} + n_{\rm H I}\,q_{\rm T3} + n_{\rm He I}\,q_{\rm T5} \right) \mathrm{,}
    \end{split}
\end{equation}

\begin{equation}\label{eq:CIII}
    v(r) \frac{df_{\rm C III}}{dr} = f_{\rm C II} \left(\Phi_{\rm C II} e^{-\tau_{\rm C II}} + n_e\,q_{\rm E2}\right) - f_{\rm C III} \left(n_e\,q_{\rm R3} + n_{\rm H I}\,q_{\rm T3} + n_{\rm He I}\,q_{\rm T5} \right) \mathrm{,}
\end{equation}

\begin{equation}\label{eq:OII}
    v(r) \frac{df_{\rm O II}}{dr} = f_{\rm O I} \left(\Phi_{\rm O I} e^{-\tau_{\rm O I}} + n_e\,q_{\rm E3} + n_{{\rm H II}}\,q_{\rm T7}\right) - f_{\rm O II} \left(n_e\,q_{\rm R4} + + n_{\rm H I}\,q_{\rm T6} \right) \mathrm{,}
\end{equation}where $f$ is the ionization fraction of a given species, $q$ is the rate for a given reaction in Table \ref{tab:reactions}, $n$ is the number density of a given particle, $f_{\rm CI} = 1 - f_{\rm CII} - f_{\rm CIII}$ and $f_{\rm OI} = 1 - f_{\rm O II}$. The Eqs. \ref{eq:CII} and \ref{eq:CIII} are coupled and solved simultaneously using the ODEINT method of SciPy's \citep{2020SciPy-NMeth} {\tt integrate} module; Eq. \ref{eq:OII} is solved using the IVP method of the same module. The first solutions are obtained from an initial guess of the fractions $f$ provided by the user and then repeated until they reach a convergence of 1\%.

The number densities as a function of altitude for C\,{\sc ii} and O\,{\sc i} are calculated as:

\begin{equation}
    n_{\rm{C\,II}}(r) = f_{\rm{C\,II}}(r) \frac{f_{\rm C}\,\rho(r)}{(f_{\rm H} + 4 f_{\rm He} + 12 f_{\rm C})m_{\rm H}} \mathrm{\ and}
\end{equation}

\begin{equation}
    n_{\rm{O\,I}}(r) = f_{\rm{O\,I}}(r) \frac{f_{\rm O}\,\rho(r)}{(f_{\rm H} + 4 f_{\rm He} + 16 f_{\rm O})m_{\rm H}} \mathrm{,}
\end{equation}where $f_{\rm X}$ is the total fraction of X nuclei in the outflow.

We then proceed to calculate the wavelength dependent transmission spectrum by assuming that the only source of opacity in the atmosphere are the ions C\,{\sc ii}, C\,{\sc iii} and the O\,{\sc i} atoms. We use the same simplified ray tracing and line broadening descriptions from \citet{DSantos22}. The central wavelengths, oscillator strengths and Einstein coefficients of the spectral lines were obtained from the National Institute of Standards and Technology (NIST) database \citep{NIST_ASD}\footnote{\footnotesize{\url{https://www.nist.gov/pml/atomic-spectra-database}.}}.

The C\,{\sc ii} lines are composed of one transition arising from the ground state and a blended doublet arising from the first excited state (with an energy of 0.00786~eV); the three O\,{\sc i} lines present in the COS spectra arise from the ground, first and second excited states. Since the equations above do not take into account excitation, we calculate the population of excited states using the CHIANTI software \citep[version 10.0.2;][]{Dere97, DZ21} implemented in the ChiantiPy Python package\footnote{\footnotesize{\url{https://github.com/chianti-atomic/ChiantiPy/}.}} \citep[version 0.14.1;][]{Landi12}. We assume that the upper atmosphere is isothermal and calculate the excited state populations as a function of electron number density. We then multiply the fraction of each state estimated by CHIANTI by the total number densities of C\,{\sc ii} and O\,{\sc i}, which are used to calculate the transmission spectra.

\begin{deluxetable*}{l l l l}[t]
\tablecaption{New reactions used to calculate the distribution of O and C. \label{tab:reactions}}
\tablewidth{0pt}
\tablehead{
\colhead{Reaction} & \colhead{} & \colhead{Rate (cm$^3$\,s$^{-1}$)} & \colhead{Reference}
}
\startdata
\multicolumn{4}{l}{Photoionization}\\
P1 & He + $h \nu$ $\rightarrow$ He$^+$ + $e$ & See text & \citet{Yan98} \\
P2 & C + $h \nu$ $\rightarrow$ C$^+$ + $e$ & See text & \citet{Verner96} \\
P3 & C$^+$ + $h \nu$ $\rightarrow$ C$^{2+}$ + $e$ & See text & \citet{Verner96} \\
P4 & O + $h \nu$ $\rightarrow$ O$^+$ + $e$ & See text & \citet{Verner96} \\
\hline
\multicolumn{4}{l}{Recombination}\\
R1 & He$^+$ + $e$ $\rightarrow$ He + $h \nu$ & $4.6 \times 10^{-12} (300 / T_e)^{0.64}$ & \citet{Storey95} \\
R2 & C$^+$ + $e$ $\rightarrow$ C + $h \nu$ & $4.67 \times 10^{-12} (300 / T_e)^{0.60}$ & \citet{Woodall07} \\
R3 & C$^{2+}$ + $e$ $\rightarrow$ C$^+$ + $h \nu$ & $2.32 \times 10^{-12} (1000 / T_e)^{0.645}$ & \citet{Aldrovandi73} \\
R4 & O$^+$ + $e$ $\rightarrow$ O + $h \nu$ & $3.25 \times 10^{-12} (300 / T_e)^{0.66}$ & \citet{Woodall07} \\
\hline
\multicolumn{4}{l}{Electron impact ionization}\\
E1 & C + $e$ $\rightarrow$ C$^+$ + $e$ + $e$ & $6.85 \times 10^{-8} \left(\frac{1}{0.193+U}\right)^{0.25} \exp{(-U)},\ U=11.3/E_e ({\rm eV})$ & \citet{Voronov97} \\
E2 & C$^+$ + $e$ $\rightarrow$ C$^{2+}$ + $e$ + $e$ & $1.86 \times 10^{-8} \left(\frac{1}{0.286+U}\right)^{0.24} \exp{(-U)},\ U=24.4/E_e ({\rm eV})$ & \citet{Voronov97} \\
E3 & O + $e$ $\rightarrow$ O$^+$ + $e$ + $e$ & $3.59 \times 10^{-8} \left(\frac{1}{0.073+U}\right)^{0.34} \exp{(-U)},\ U=13.6/E_e ({\rm eV})$ & \citet{Voronov97} \\
\hline
\multicolumn{4}{l}{Charge transfer with H and He nuclei}\\
T1 & C$^+$ + H $\rightarrow$ C + H$^+$ & $6.30 \times 10^{-17} (300 / T)^{-1.96} \exp{(-170\,000/T)}$ & \citet{Stancil98} \\
T2 & C + H$^+$ $\rightarrow$ C$^+$ + H & $1.31 \times 10^{-15} (300 / T)^{-0.213}$ & \citet{Stancil98} \\
T3 & C$^{2+}$ + H $\rightarrow$ C$^+$ + H$^+$ & $1.67 \times 10^{-4} (10\,000 / T)^{-2.79} [1 + 304.72\,e^{(-4.07\,T / 10\,000)}]$ & \citet{Kingdon96} \\
T4 & C + He$^+$ $\rightarrow$ C$^+$ + He & $2.50 \times 10^{-15} (300 / T)^{-1.597}$ & \citet{Glover07} \\
T5 & C$^{2+}$ + He $\rightarrow$ C$^+$ + He$^+$ & $\sim 1.23 \times 10^{-9}$ for $T < 15\,000$ K & \citet{Brown72} \\
T6 & O$^+$ + H $\rightarrow$ O + H$^+$ & $5.66 \times 10^{-10} (300 / T)^{-0.36} \exp{(8.6/T)}$ & \citet{Woodall07} \\
T7 & O + H$^+$ $\rightarrow$ O$^+$ + H & $7.31 \times 10^{-10} (300 / T)^{-0.23} \exp{(-226/T)}$ & \citet{Woodall07} \\
\enddata
\tablecomments{$T_e$ is the temperature of the electrons, which we assume to be the same as the temperature of the outflow $T$. $E_e$ is the energy of the colliding electrons at a given temperature $T_e$.}
\end{deluxetable*}

\bibliography{references}{}

\begin{thebibliography}{}
\expandafter\ifx\csname natexlab\endcsname\relax\def\natexlab#1{#1}\fi
\providecommand{\url}[1]{\href{#1}{#1}}
\providecommand{\dodoi}[1]{doi:~\href{http://doi.org/#1}{\nolinkurl{#1}}}
\providecommand{\doeprint}[1]{\href{http://ascl.net/#1}{\nolinkurl{http://ascl.net/#1}}}
\providecommand{\doarXiv}[1]{\href{https://arxiv.org/abs/#1}{\nolinkurl{https://arxiv.org/abs/#1}}}

\bibitem[{{Addison} {et~al.}(2019){Addison}, {Wright}, {Wittenmyer}, {Horner},
  {Mengel}, {Johns}, {Marti}, {Nicholson}, {Soutter}, {Bowler}, {Crossfield},
  {Kane}, {Kielkopf}, {Plavchan}, {Tinney}, {Zhang}, {Clark}, {Clerte},
  {Eastman}, {Swift}, {Bottom}, {Muirhead}, {McCrady}, {Herzig}, {Hogstrom},
  {Wilson}, {Sliski}, {Johnson}, {Wright}, {Johnson}, {Blake}, {Riddle}, {Lin},
  {Cornachione}, {Bedding}, {Stello}, {Huber}, {Marsden}, \&
  {Carter}}]{2019PASP..131k5003A}
{Addison}, B., {Wright}, D.~J., {Wittenmyer}, R.~A., {et~al.} 2019, \pasp, 131,
  115003, \dodoi{10.1088/1538-3873/ab03aa}

\bibitem[{{Aldrovandi} \& {Pequignot}(1973)}]{Aldrovandi73}
{Aldrovandi}, S.~M.~V., \& {Pequignot}, D. 1973, \aap, 25, 137

\bibitem[{{Astropy Collaboration} {et~al.}(2018){Astropy Collaboration},
  {Price-Whelan}, {Sip{\H{o}}cz}, {G{\"u}nther}, {Lim}, {Crawford}, {Conseil},
  {Shupe}, {Craig}, {Dencheva}, {Ginsburg}, {Vand erPlas}, {Bradley},
  {P{\'e}rez-Su{\'a}rez}, {de Val-Borro}, {Aldcroft}, {Cruz}, {Robitaille},
  {Tollerud}, {Ardelean}, {Babej}, {Bach}, {Bachetti}, {Bakanov}, {Bamford},
  {Barentsen}, {Barmby}, {Baumbach}, {Berry}, {Biscani}, {Boquien}, {Bostroem},
  {Bouma}, {Brammer}, {Bray}, {Breytenbach}, {Buddelmeijer}, {Burke},
  {Calderone}, {Cano Rodr{\'\i}guez}, {Cara}, {Cardoso}, {Cheedella}, {Copin},
  {Corrales}, {Crichton}, {D'Avella}, {Deil}, {Depagne}, {Dietrich}, {Donath},
  {Droettboom}, {Earl}, {Erben}, {Fabbro}, {Ferreira}, {Finethy}, {Fox},
  {Garrison}, {Gibbons}, {Goldstein}, {Gommers}, {Greco}, {Greenfield},
  {Groener}, {Grollier}, {Hagen}, {Hirst}, {Homeier}, {Horton}, {Hosseinzadeh},
  {Hu}, {Hunkeler}, {Ivezi{\'c}}, {Jain}, {Jenness}, {Kanarek}, {Kendrew},
  {Kern}, {Kerzendorf}, {Khvalko}, {King}, {Kirkby}, {Kulkarni}, {Kumar},
  {Lee}, {Lenz}, {Littlefair}, {Ma}, {Macleod}, {Mastropietro}, {McCully},
  {Montagnac}, {Morris}, {Mueller}, {Mumford}, {Muna}, {Murphy}, {Nelson},
  {Nguyen}, {Ninan}, {N{\"o}the}, {Ogaz}, {Oh}, {Parejko}, {Parley}, {Pascual},
  {Patil}, {Patil}, {Plunkett}, {Prochaska}, {Rastogi}, {Reddy Janga},
  {Sabater}, {Sakurikar}, {Seifert}, {Sherbert}, {Sherwood-Taylor}, {Shih},
  {Sick}, {Silbiger}, {Singanamalla}, {Singer}, {Sladen}, {Sooley},
  {Sornarajah}, {Streicher}, {Teuben}, {Thomas}, {Tremblay}, {Turner},
  {Terr{\'o}n}, {van Kerkwijk}, {de la Vega}, {Watkins}, {Weaver}, {Whitmore},
  {Woillez}, {Zabalza}, \& {Astropy Contributors}}]{astropy:2018}
{Astropy Collaboration}, {Price-Whelan}, A.~M., {Sip{\H{o}}cz}, B.~M., {et~al.}
  2018, \aj, 156, 123, \dodoi{10.3847/1538-3881/aabc4f}

\bibitem[{{Barth} {et~al.}(2021){Barth}, {Helling}, {St{\"u}eken}, {Bourrier},
  {Mayne}, {Rimmer}, {Jardine}, {Vidotto}, {Wheatley}, \& {Fares}}]{Barth2021}
{Barth}, P., {Helling}, C., {St{\"u}eken}, E.~E., {et~al.} 2021, \mnras, 502,
  6201, \dodoi{10.1093/mnras/staa3989}

\bibitem[{{Ben-Jaffel} \& {Ballester}(2013)}]{2013A&A...553A..52B}
{Ben-Jaffel}, L., \& {Ballester}, G.~E. 2013, \aap, 553, A52,
  \dodoi{10.1051/0004-6361/201221014}

\bibitem[{{Bonomo} {et~al.}(2017){Bonomo}, {Desidera}, {Benatti}, {Borsa},
  {Crespi}, {Damasso}, {Lanza}, {Sozzetti}, {Lodato}, {Marzari}, {Boccato},
  {Claudi}, {Cosentino}, {Covino}, {Gratton}, {Maggio}, {Micela}, {Molinari},
  {Pagano}, {Piotto}, {Poretti}, {Smareglia}, {Affer}, {Biazzo}, {Bignamini},
  {Esposito}, {Giacobbe}, {H{\'e}brard}, {Malavolta}, {Maldonado}, {Mancini},
  {Martinez Fiorenzano}, {Masiero}, {Nascimbeni}, {Pedani}, {Rainer}, \&
  {Scandariato}}]{Bonomo2017}
{Bonomo}, A.~S., {Desidera}, S., {Benatti}, S., {et~al.} 2017, \aap, 602, A107,
  \dodoi{10.1051/0004-6361/201629882}

\bibitem[{{Bouchy} {et~al.}(2005){Bouchy}, {Udry}, {Mayor}, {Moutou}, {Pont},
  {Iribarne}, {da Silva}, {Ilovaisky}, {Queloz}, {Santos}, {S{\'e}gransan}, \&
  {Zucker}}]{Bouchy2005}
{Bouchy}, F., {Udry}, S., {Mayor}, M., {et~al.} 2005, \aap, 444, L15,
  \dodoi{10.1051/0004-6361:200500201}

\bibitem[{{Bourrier} \& {Lecavelier des Etangs}(2013)}]{2013A&A...557A.124B}
{Bourrier}, V., \& {Lecavelier des Etangs}, A. 2013, \aap, 557, A124,
  \dodoi{10.1051/0004-6361/201321551}

\bibitem[{{Bourrier} {et~al.}(2013){Bourrier}, {Lecavelier des Etangs},
  {Dupuy}, {Ehrenreich}, {Vidal-Madjar}, {H{\'e}brard}, {Ballester},
  {D{\'e}sert}, {Ferlet}, {Sing}, \& {Wheatley}}]{Bourrier2013}
{Bourrier}, V., {Lecavelier des Etangs}, A., {Dupuy}, H., {et~al.} 2013, \aap,
  551, A63, \dodoi{10.1051/0004-6361/201220533}

\bibitem[{{Bourrier} {et~al.}(2018){Bourrier}, {Ehrenreich}, {Lecavelier des
  Etangs}, {Louden}, {Wheatley}, {Wyttenbach}, {Vidal-Madjar}, {Lavie}, {Pepe},
  \& {Udry}}]{Bourrier2018}
{Bourrier}, V., {Ehrenreich}, D., {Lecavelier des Etangs}, A., {et~al.} 2018,
  \aap, 615, A117, \dodoi{10.1051/0004-6361/201832700}

\bibitem[{{Bourrier} {et~al.}(2020{\natexlab{a}}){Bourrier}, {Wheatley},
  {Lecavelier Des Etangs}, {King}, {Louden}, {Ehrenreich}, {Fares}, {Helling},
  {Llama}, {Jardine}, \& {Vidotto}}]{Bourrier2020}
{Bourrier}, V., {Wheatley}, P.~J., {Lecavelier Des Etangs}, A., {et~al.}
  2020{\natexlab{a}}, \mnras, 493, 559, \dodoi{10.1093/mnras/staa256}

\bibitem[{{Bourrier} {et~al.}(2020{\natexlab{b}}){Bourrier}, {Wheatley},
  {Lecavelier des Etangs}, {King}, {Louden}, {Ehrenreich}, {Fares}, {Helling},
  {Llama}, {Jardine}, \& {Vidotto}}]{Bourrier20}
{Bourrier}, V., {Wheatley}, P.~J., {Lecavelier des Etangs}, A., {et~al.}
  2020{\natexlab{b}}, \mnras, 493, 559, \dodoi{10.1093/mnras/staa256}

\bibitem[{{Bourrier} {et~al.}(2021){Bourrier}, {dos Santos}, {Sanz-Forcada},
  {Garc{\'\i}a Mu{\~n}oz}, {Henry}, {Lavvas}, {Lecavelier},
  {L{\'o}pez-Morales}, {Mikal-Evans}, {Sing}, {Wakeford}, \&
  {Ehrenreich}}]{Bourrier2021}
{Bourrier}, V., {dos Santos}, L.~A., {Sanz-Forcada}, J., {et~al.} 2021, \aap,
  650, A73, \dodoi{10.1051/0004-6361/202140487}

\bibitem[{{Brown}(1972)}]{Brown72}
{Brown}, R.~L. 1972, \apj, 174, 511, \dodoi{10.1086/151513}

\bibitem[{{Cauley} {et~al.}(2018){Cauley}, {Kuckein}, {Redfield}, {Shkolnik},
  {Denker}, {Llama}, \& {Verma}}]{Cauley18}
{Cauley}, P.~W., {Kuckein}, C., {Redfield}, S., {et~al.} 2018, \aj, 156, 189,
  \dodoi{10.3847/1538-3881/aaddf9}

\bibitem[{{Cruz Aguirre} {et~al.}(2023){Cruz Aguirre}, {Youngblood}, {France},
  \& {Bourrier}}]{Aguirre2023}
{Cruz Aguirre}, F., {Youngblood}, A., {France}, K., \& {Bourrier}, V. 2023,
  \apj, 946, 98, \dodoi{10.3847/1538-4357/acad7d}

\bibitem[{{Cubillos} {et~al.}(2023){Cubillos}, {Fossati}, {Koskinen}, {Huang},
  {Sreejith}, {France}, {Wilson Cauley}, \& {Haswell}}]{2023A&A...671A.170C}
{Cubillos}, P.~E., {Fossati}, L., {Koskinen}, T., {et~al.} 2023, \aap, 671,
  A170, \dodoi{10.1051/0004-6361/202245064}

\bibitem[{{Del Zanna} {et~al.}(2021){Del Zanna}, {Dere}, {Young}, \&
  {Landi}}]{DZ21}
{Del Zanna}, G., {Dere}, K.~P., {Young}, P.~R., \& {Landi}, E. 2021, \apj, 909,
  38, \dodoi{10.3847/1538-4357/abd8ce}

\bibitem[{{Dere} {et~al.}(1997){Dere}, {Landi}, {Mason}, {Monsignori Fossi}, \&
  {Young}}]{Dere97}
{Dere}, K.~P., {Landi}, E., {Mason}, H.~E., {Monsignori Fossi}, B.~C., \&
  {Young}, P.~R. 1997, \aaps, 125, 149, \dodoi{10.1051/aas:1997368}

\bibitem[{{Dos Santos} {et~al.}(2019){Dos Santos}, {Ehrenreich}, {Bourrier},
  {Lecavelier des Etangs}, {L{\'o}pez-Morales}, {Sing}, {Ballester},
  {Ben-Jaffel}, {Buchhave}, {Garc{\'\i}a Mu{\~n}oz}, {Henry}, {Kataria},
  {Lavie}, {Lavvas}, {Lewis}, {Mikal-Evans}, {Sanz-Forcada}, \&
  {Wakeford}}]{DSantos2019}
{Dos Santos}, L.~A., {Ehrenreich}, D., {Bourrier}, V., {et~al.} 2019, \aap,
  629, A47, \dodoi{10.1051/0004-6361/201935663}

\bibitem[{{Dos Santos} {et~al.}(2022){Dos Santos}, {Vidotto}, {Vissapragada},
  {Alam}, {Allart}, {Bourrier}, {Kirk}, {Seidel}, \& {Ehrenreich}}]{DSantos22}
{Dos Santos}, L.~A., {Vidotto}, A.~A., {Vissapragada}, S., {et~al.} 2022, \aap,
  659, A62, \dodoi{10.1051/0004-6361/202142038}

\bibitem[{{Ehrenreich} {et~al.}(2015){Ehrenreich}, {Bourrier}, {Wheatley},
  {Lecavelier des Etangs}, {H{\'e}brard}, {Udry}, {Bonfils}, {Delfosse},
  {D{\'e}sert}, {Sing}, \& {Vidal-Madjar}}]{Ehrenreich15}
{Ehrenreich}, D., {Bourrier}, V., {Wheatley}, P.~J., {et~al.} 2015, \nat, 522,
  459, \dodoi{10.1038/nature14501}

\bibitem[{{Erkaev} {et~al.}(2007){Erkaev}, {Kulikov}, {Lammer}, {Selsis},
  {Langmayr}, {Jaritz}, \& {Biernat}}]{Erkaev07}
{Erkaev}, N.~V., {Kulikov}, Y.~N., {Lammer}, H., {et~al.} 2007, \aap, 472, 329,
  \dodoi{10.1051/0004-6361:20066929}

\bibitem[{{Fortney} {et~al.}(2021){Fortney}, {Dawson}, \&
  {Komacek}}]{2021JGRE..12606629F}
{Fortney}, J.~J., {Dawson}, R.~I., \& {Komacek}, T.~D. 2021, Journal of
  Geophysical Research (Planets), 126, e06629, \dodoi{10.1029/2020JE006629}

\bibitem[{{Fossati} {et~al.}(2010){Fossati}, {Haswell}, {Froning}, {Hebb},
  {Holmes}, {Kolb}, {Helling}, {Carter}, {Wheatley}, {Collier Cameron},
  {Loeillet}, {Pollacco}, {Street}, {Stempels}, {Simpson}, {Udry}, {Joshi},
  {West}, {Skillen}, \& {Wilson}}]{Fossati2010}
{Fossati}, L., {Haswell}, C.~A., {Froning}, C.~S., {et~al.} 2010, \apjl, 714,
  L222, \dodoi{10.1088/2041-8205/714/2/L222}

\bibitem[{{Gaia Collaboration} {et~al.}(2018){Gaia Collaboration}, {Brown},
  {Vallenari}, {Prusti}, {de Bruijne}, {Babusiaux}, {Bailer-Jones}, {Biermann},
  {Evans}, {Eyer}, {Jansen}, {Jordi}, {Klioner}, {Lammers}, {Lindegren},
  {Luri}, {Mignard}, {Panem}, {Pourbaix}, {Randich}, {Sartoretti}, {Siddiqui},
  {Soubiran}, {van Leeuwen}, {Walton}, {Arenou}, {Bastian}, {Cropper},
  {Drimmel}, {Katz}, {Lattanzi}, {Bakker}, {Cacciari}, {Casta{\~n}eda},
  {Chaoul}, {Cheek}, {De Angeli}, {Fabricius}, {Guerra}, {Holl}, {Masana},
  {Messineo}, {Mowlavi}, {Nienartowicz}, {Panuzzo}, {Portell}, {Riello},
  {Seabroke}, {Tanga}, {Th{\'e}venin}, {Gracia-Abril}, {Comoretto},
  {Garcia-Reinaldos}, {Teyssier}, {Altmann}, {Andrae}, {Audard},
  {Bellas-Velidis}, {Benson}, {Berthier}, {Blomme}, {Burgess}, {Busso},
  {Carry}, {Cellino}, {Clementini}, {Clotet}, {Creevey}, {Davidson}, {De
  Ridder}, {Delchambre}, {Dell'Oro}, {Ducourant},
  {Fern{\'a}ndez-Hern{\'a}ndez}, {Fouesneau}, {Fr{\'e}mat}, {Galluccio},
  {Garc{\'\i}a-Torres}, {Gonz{\'a}lez-N{\'u}{\~n}ez}, {Gonz{\'a}lez-Vidal},
  {Gosset}, {Guy}, {Halbwachs}, {Hambly}, {Harrison}, {Hern{\'a}ndez},
  {Hestroffer}, {Hodgkin}, {Hutton}, {Jasniewicz}, {Jean-Antoine-Piccolo},
  {Jordan}, {Korn}, {Krone-Martins}, {Lanzafame}, {Lebzelter}, {L{\"o}ffler},
  {Manteiga}, {Marrese}, {Mart{\'\i}n-Fleitas}, {Moitinho}, {Mora}, {Muinonen},
  {Osinde}, {Pancino}, {Pauwels}, {Petit}, {Recio-Blanco}, {Richards},
  {Rimoldini}, {Robin}, {Sarro}, {Siopis}, {Smith}, {Sozzetti}, {S{\"u}veges},
  {Torra}, {van Reeven}, {Abbas}, {Abreu Aramburu}, {Accart}, {Aerts},
  {Altavilla}, {{\'A}lvarez}, {Alvarez}, {Alves}, {Anderson}, {Andrei},
  {Anglada Varela}, {Antiche}, {Antoja}, {Arcay}, {Astraatmadja}, {Bach},
  {Baker}, {Balaguer-N{\'u}{\~n}ez}, {Balm}, {Barache}, {Barata}, {Barbato},
  {Barblan}, {Barklem}, {Barrado}, {Barros}, {Barstow}, {Bartholom{\'e}
  Mu{\~n}oz}, {Bassilana}, {Becciani}, {Bellazzini}, {Berihuete}, {Bertone},
  {Bianchi}, {Bienaym{\'e}}, {Blanco-Cuaresma}, {Boch}, {Boeche}, {Bombrun},
  {Borrachero}, {Bossini}, {Bouquillon}, {Bourda}, {Bragaglia}, {Bramante},
  {Breddels}, {Bressan}, {Brouillet}, {Br{\"u}semeister}, {Brugaletta},
  {Bucciarelli}, {Burlacu}, {Busonero}, {Butkevich}, {Buzzi}, {Caffau},
  {Cancelliere}, {Cannizzaro}, {Cantat-Gaudin}, {Carballo}, {Carlucci},
  {Carrasco}, {Casamiquela}, {Castellani}, {Castro-Ginard}, {Charlot},
  {Chemin}, {Chiavassa}, {Cocozza}, {Costigan}, {Cowell}, {Crifo}, {Crosta},
  {Crowley}, {Cuypers}, {Dafonte}, {Damerdji}, {Dapergolas}, {David}, {David},
  {de Laverny}, {De Luise}, {De March}, {de Martino}, {de Souza}, {de Torres},
  {Debosscher}, {del Pozo}, {Delbo}, {Delgado}, {Delgado}, {Di Matteo},
  {Diakite}, {Diener}, {Distefano}, {Dolding}, {Drazinos}, {Dur{\'a}n},
  {Edvardsson}, {Enke}, {Eriksson}, {Esquej}, {Eynard Bontemps}, {Fabre},
  {Fabrizio}, {Faigler}, {Falc{\~a}o}, {Farr{\`a}s Casas}, {Federici},
  {Fedorets}, {Fernique}, {Figueras}, {Filippi}, {Findeisen}, {Fonti},
  {Fraile}, {Fraser}, {Fr{\'e}zouls}, {Gai}, {Galleti}, {Garabato},
  {Garc{\'\i}a-Sedano}, {Garofalo}, {Garralda}, {Gavel}, {Gavras}, {Gerssen},
  {Geyer}, {Giacobbe}, {Gilmore}, {Girona}, {Giuffrida}, {Glass}, {Gomes},
  {Granvik}, {Gueguen}, {Guerrier}, {Guiraud}, {Guti{\'e}rrez-S{\'a}nchez},
  {Haigron}, {Hatzidimitriou}, {Hauser}, {Haywood}, {Heiter}, {Helmi}, {Heu},
  {Hilger}, {Hobbs}, {Hofmann}, {Holland}, {Huckle}, {Hypki}, {Icardi},
  {Jan{\ss}en}, {Jevardat de Fombelle}, {Jonker}, {Juh{\'a}sz}, {Julbe},
  {Karampelas}, {Kewley}, {Klar}, {Kochoska}, {Kohley}, {Kolenberg},
  {Kontizas}, {Kontizas}, {Koposov}, {Kordopatis}, {Kostrzewa-Rutkowska},
  {Koubsky}, {Lambert}, {Lanza}, {Lasne}, {Lavigne}, {Le Fustec}, {Le
  Poncin-Lafitte}, {Lebreton}, {Leccia}, {Leclerc}, {Lecoeur-Taibi},
  {Lenhardt}, {Leroux}, {Liao}, {Licata}, {Lindstr{\o}m}, {Lister}, {Livanou},
  {Lobel}, {L{\'o}pez}, {Managau}, {Mann}, {Mantelet}, {Marchal}, {Marchant},
  {Marconi}, {Marinoni}, {Marschalk{\'o}}, {Marshall}, {Martino}, {Marton},
  {Mary}, {Massari}, {Matijevi{\v{c}}}, {Mazeh}, {McMillan}, {Messina},
  {Michalik}, {Millar}, {Molina}, {Molinaro}, {Moln{\'a}r}, {Montegriffo},
  {Mor}, {Morbidelli}, {Morel}, {Morris}, {Mulone}, {Muraveva}, {Musella},
  {Nelemans}, {Nicastro}, {Noval}, {O'Mullane}, {Ord{\'e}novic},
  {Ord{\'o}{\~n}ez-Blanco}, {Osborne}, {Pagani}, {Pagano}, {Pailler},
  {Palacin}, {Palaversa}, {Panahi}, {Pawlak}, {Piersimoni}, {Pineau}, {Plachy},
  {Plum}, {Poggio}, {Poujoulet}, {Pr{\v{s}}a}, {Pulone}, {Racero}, {Ragaini},
  {Rambaux}, {Ramos-Lerate}, {Regibo}, {Reyl{\'e}}, {Riclet}, {Ripepi}, {Riva},
  {Rivard}, {Rixon}, {Roegiers}, {Roelens}, {Romero-G{\'o}mez}, {Rowell},
  {Royer}, {Ruiz-Dern}, {Sadowski}, {Sagrist{\`a} Sell{\'e}s}, {Sahlmann},
  {Salgado}, {Salguero}, {Sanna}, {Santana-Ros}, {Sarasso}, {Savietto},
  {Schultheis}, {Sciacca}, {Segol}, {Segovia}, {S{\'e}gransan}, {Shih},
  {Siltala}, {Silva}, {Smart}, {Smith}, {Solano}, {Solitro}, {Sordo}, {Soria
  Nieto}, {Souchay}, {Spagna}, {Spoto}, {Stampa}, {Steele},
  {Steidelm{\"u}ller}, {Stephenson}, {Stoev}, {Suess}, {Surdej}, {Szabados},
  {Szegedi-Elek}, {Tapiador}, {Taris}, {Tauran}, {Taylor}, {Teixeira},
  {Terrett}, {Teyssandier}, {Thuillot}, {Titarenko}, {Torra Clotet}, {Turon},
  {Ulla}, {Utrilla}, {Uzzi}, {Vaillant}, {Valentini}, {Valette}, {van Elteren},
  {Van Hemelryck}, {van Leeuwen}, {Vaschetto}, {Vecchiato}, {Veljanoski},
  {Viala}, {Vicente}, {Vogt}, {von Essen}, {Voss}, {Votruba}, {Voutsinas},
  {Walmsley}, {Weiler}, {Wertz}, {Wevers}, {Wyrzykowski}, {Yoldas},
  {{\v{Z}}erjal}, {Ziaeepour}, {Zorec}, {Zschocke}, {Zucker}, {Zurbach}, \&
  {Zwitter}}]{Gaia2018}
{Gaia Collaboration}, {Brown}, A.~G.~A., {Vallenari}, A., {et~al.} 2018, \aap,
  616, A1, \dodoi{10.1051/0004-6361/201833051}

\bibitem[{{Garc{\'\i}a Mu{\~n}oz}(2007)}]{GMunoz2007}
{Garc{\'\i}a Mu{\~n}oz}, A. 2007, \planss, 55, 1426,
  \dodoi{10.1016/j.pss.2007.03.007}

\bibitem[{{Garc{\'\i}a Mu{\~n}oz} {et~al.}(2021){Garc{\'\i}a Mu{\~n}oz},
  {Fossati}, {Youngblood}, {Nettelmann}, {Gandolfi}, {Cabrera}, \&
  {Rauer}}]{GarciaMunoz21}
{Garc{\'\i}a Mu{\~n}oz}, A., {Fossati}, L., {Youngblood}, A., {et~al.} 2021,
  \apjl, 907, L36, \dodoi{10.3847/2041-8213/abd9b8}

\bibitem[{{Glover} \& {Jappsen}(2007)}]{Glover07}
{Glover}, S.~C.~O., \& {Jappsen}, A.~K. 2007, \apj, 666, 1,
  \dodoi{10.1086/519445}

\bibitem[{{Gray} {et~al.}(2003){Gray}, {Corbally}, {Garrison}, {McFadden}, \&
  {Robinson}}]{Gray2003}
{Gray}, R.~O., {Corbally}, C.~J., {Garrison}, R.~F., {McFadden}, M.~T., \&
  {Robinson}, P.~E. 2003, \aj, 126, 2048, \dodoi{10.1086/378365}

\bibitem[{{Green} {et~al.}(2012){Green}, {Froning}, {Osterman}, {Ebbets},
  {Heap}, {Leitherer}, {Linsky}, {Savage}, {Sembach}, {Shull}, {Siegmund},
  {Snow}, {Spencer}, {Stern}, {Stocke}, {Welsh}, {B{\'e}land}, {Burgh},
  {Danforth}, {France}, {Keeney}, {McPhate}, {Penton}, {Andrews},
  {Brownsberger}, {Morse}, \& {Wilkinson}}]{Green2012}
{Green}, J.~C., {Froning}, C.~S., {Osterman}, S., {et~al.} 2012, \apj, 744, 60,
  \dodoi{10.1088/0004-637X/744/1/6010.1086/141956}

\bibitem[{Harris {et~al.}(2020)Harris, Millman, van~der Walt, Gommers,
  Virtanen, Cournapeau, Wieser, Taylor, Berg, Smith, Kern, Picus, Hoyer, van
  Kerkwijk, Brett, Haldane, del R{\'{i}}o, Wiebe, Peterson,
  G{\'{e}}rard-Marchant, Sheppard, Reddy, Weckesser, Abbasi, Gohlke, \&
  Oliphant}]{harris2020array}
Harris, C.~R., Millman, K.~J., van~der Walt, S.~J., {et~al.} 2020, Nature, 585,
  357, \dodoi{10.1038/s41586-020-2649-2}

\bibitem[{{Henry}(1999)}]{Henry1999}
{Henry}, G.~W. 1999, \pasp, 111, 845, \dodoi{10.1086/316388}

\bibitem[{{Hundhausen}(1997)}]{Hundhausen97}
{Hundhausen}, A.~J. 1997, in Cosmic Winds and the Heliosphere, 259

\bibitem[{Hunter(2007)}]{Hunter:2007}
Hunter, J.~D. 2007, Computing in Science \& Engineering, 9, 90,
  \dodoi{10.1109/MCSE.2007.55}

\bibitem[{{Ito} \& {Ikoma}(2021)}]{Ito2021}
{Ito}, Y., \& {Ikoma}, M. 2021, \mnras, 502, 750,
  \dodoi{10.1093/mnras/staa3962}

\bibitem[{{Johnson} {et~al.}(2021){Johnson}, {Plesha}, {Jedrzejewski},
  {Frazer}, \& {Dashtamirova}}]{Johnson2021}
{Johnson}, C.~I., {Plesha}, R., {Jedrzejewski}, R., {Frazer}, E., \&
  {Dashtamirova}, D. 2021, {Updated Flux Error Calculations for CalCOS},
  Instrument Science Report COS 2021-03

\bibitem[{{Kingdon} \& {Ferland}(1996)}]{Kingdon96}
{Kingdon}, J.~B., \& {Ferland}, G.~J. 1996, \apjs, 106, 205,
  \dodoi{10.1086/192335}

\bibitem[{{Kirk} {et~al.}(2022){Kirk}, {Dos Santos}, {L{\'o}pez-Morales},
  {Alam}, {Oklop{\v{c}}i{\'c}}, {MacLeod}, {Zeng}, \& {Zhou}}]{Kirk22}
{Kirk}, J., {Dos Santos}, L.~A., {L{\'o}pez-Morales}, M., {et~al.} 2022, \aj,
  164, 24, \dodoi{10.3847/1538-3881/ac722f}

\bibitem[{Kluyver {et~al.}(2016)Kluyver, Ragan-Kelley, P{\'e}rez, Granger,
  Bussonnier, Frederic, Kelley, Hamrick, Grout, Corlay, Ivanov, Avila, Abdalla,
  Willing, \& development team}]{jupyter}
Kluyver, T., Ragan-Kelley, B., P{\'e}rez, F., {et~al.} 2016, in Positioning and
  Power in Academic Publishing: Players, Agents and Agendas, ed. F.~Loizides \&
  B.~Scmidt (Netherlands: IOS Press), 87--90.
\newblock \url{https://eprints.soton.ac.uk/403913/}

\bibitem[{{Koskinen} {et~al.}(2013){Koskinen}, {Harris}, {Yelle}, \&
  {Lavvas}}]{Koskinen13}
{Koskinen}, T.~T., {Harris}, M.~J., {Yelle}, R.~V., \& {Lavvas}, P. 2013,
  \icarus, 226, 1678, \dodoi{10.1016/j.icarus.2012.09.027}

\bibitem[{Kramida {et~al.}(2022)Kramida, {Yu.~Ralchenko}, Reader, \& {and NIST
  ASD Team}}]{NIST_ASD}
Kramida, A., {Yu.~Ralchenko}, Reader, J., \& {and NIST ASD Team}. 2022, {NIST
  Atomic Spectra Database (ver. 5.10), [Online]. Available:
  {\tt{https://physics.nist.gov/asd}} [2022, December 1]. National Institute of
  Standards and Technology, Gaithersburg, MD.}

\bibitem[{{Kreidberg}(2015)}]{Kreidberg2015}
{Kreidberg}, L. 2015, \pasp, 127, 1161, \dodoi{10.1086/683602}

\bibitem[{{Lamp{\'o}n} {et~al.}(2021){Lamp{\'o}n}, {L{\'o}pez-Puertas},
  {Sanz-Forcada}, {S{\'a}nchez-L{\'o}pez}, {Molaverdikhani}, {Czesla},
  {Quirrenbach}, {Pall{\'e}}, {Caballero}, {Henning}, {Salz}, {Nortmann},
  {Aceituno}, {Amado}, {Bauer}, {Montes}, {Nagel}, {Reiners}, \&
  {Ribas}}]{Lampon21}
{Lamp{\'o}n}, M., {L{\'o}pez-Puertas}, M., {Sanz-Forcada}, J., {et~al.} 2021,
  \aap, 647, A129, \dodoi{10.1051/0004-6361/202039417}

\bibitem[{{Landi} {et~al.}(2012){Landi}, {Del Zanna}, {Young}, {Dere}, \&
  {Mason}}]{Landi12}
{Landi}, E., {Del Zanna}, G., {Young}, P.~R., {Dere}, K.~P., \& {Mason}, H.~E.
  2012, \apj, 744, 99, \dodoi{10.1088/0004-637X/744/2/99}

\bibitem[{{Lecavelier Des Etangs} {et~al.}(2008){Lecavelier Des Etangs},
  {Pont}, {Vidal-Madjar}, \& {Sing}}]{Lecavelier2008a}
{Lecavelier Des Etangs}, A., {Pont}, F., {Vidal-Madjar}, A., \& {Sing}, D.
  2008, \aap, 481, L83, \dodoi{10.1051/0004-6361:200809388}

\bibitem[{{Lecavelier Des Etangs} {et~al.}(2010){Lecavelier Des Etangs},
  {Ehrenreich}, {Vidal-Madjar}, {Ballester}, {D{\'e}sert}, {Ferlet},
  {H{\'e}brard}, {Sing}, {Tchakoumegni}, \& {Udry}}]{Lecavelier2010}
{Lecavelier Des Etangs}, A., {Ehrenreich}, D., {Vidal-Madjar}, A., {et~al.}
  2010, \aap, 514, A72, \dodoi{10.1051/0004-6361/200913347}

\bibitem[{{Lecavelier Des Etangs} {et~al.}(2012){Lecavelier Des Etangs},
  {Bourrier}, {Wheatley}, {Dupuy}, {Ehrenreich}, {Vidal-Madjar}, {H{\'e}brard},
  {Ballester}, {D{\'e}sert}, {Ferlet}, \& {Sing}}]{Lecavelier2012}
{Lecavelier Des Etangs}, A., {Bourrier}, V., {Wheatley}, P.~J., {et~al.} 2012,
  \aap, 543, L4, \dodoi{10.1051/0004-6361/201219363}

\bibitem[{{Linsky} {et~al.}(2012){Linsky}, {Bushinsky}, {Ayres}, {Fontenla}, \&
  {France}}]{Linsky2012a}
{Linsky}, J.~L., {Bushinsky}, R., {Ayres}, T., {Fontenla}, J., \& {France}, K.
  2012, \apj, 745, 25, \dodoi{10.1088/0004-637X/745/1/25}

\bibitem[{{Loyd} \& {France}(2014)}]{Loyd2014}
{Loyd}, R.~O.~P., \& {France}, K. 2014, \apjs, 211, 9,
  \dodoi{10.1088/0067-0049/211/1/9}

\bibitem[{{Madhusudhan} {et~al.}(2014){Madhusudhan}, {Crouzet}, {McCullough},
  {Deming}, \& {Hedges}}]{2014ApJ...791L...9M}
{Madhusudhan}, N., {Crouzet}, N., {McCullough}, P.~R., {Deming}, D., \&
  {Hedges}, C. 2014, \apjl, 791, L9, \dodoi{10.1088/2041-8205/791/1/L9}

\bibitem[{{McCullough} {et~al.}(2014){McCullough}, {Crouzet}, {Deming}, \&
  {Madhusudhan}}]{2014ApJ...791...55M}
{McCullough}, P.~R., {Crouzet}, N., {Deming}, D., \& {Madhusudhan}, N. 2014,
  \apj, 791, 55, \dodoi{10.1088/0004-637X/791/1/55}

\bibitem[{{Nakayama} {et~al.}(2022){Nakayama}, {Ikoma}, \&
  {Terada}}]{Nakayama2022}
{Nakayama}, A., {Ikoma}, M., \& {Terada}, N. 2022, \apj, 937, 72,
  \dodoi{10.3847/1538-4357/ac86ca}

\bibitem[{{Odert} {et~al.}(2020){Odert}, {Erkaev}, {Kislyakova}, {Lammer},
  {Mezentsev}, {Ivanov}, {Fossati}, {Leitzinger}, {Kubyshkina}, \&
  {Holmstr{\"o}m}}]{2020A&A...638A..49O}
{Odert}, P., {Erkaev}, N.~V., {Kislyakova}, K.~G., {et~al.} 2020, \aap, 638,
  A49, \dodoi{10.1051/0004-6361/201834814}

\bibitem[{{Oklop{\v{c}}i{\'c}} \& {Hirata}(2018)}]{Oklopcic18}
{Oklop{\v{c}}i{\'c}}, A., \& {Hirata}, C.~M. 2018, \apjl, 855, L11,
  \dodoi{10.3847/2041-8213/aaada9}

\bibitem[{{Owen}(2019)}]{Owen2019}
{Owen}, J.~E. 2019, Annual Review of Earth and Planetary Sciences, 47, 67,
  \dodoi{10.1146/annurev-earth-053018-060246}

\bibitem[{{Owen} {et~al.}(2023){Owen}, {Murray-Clay}, {Schreyer},
  {Schlichting}, {Ardila}, {Gupta}, {Loyd}, {Shkolnik}, {Sing}, \&
  {Swain}}]{Owen2023}
{Owen}, J.~E., {Murray-Clay}, R.~A., {Schreyer}, E., {et~al.} 2023, \mnras,
  518, 4357, \dodoi{10.1093/mnras/stac3414}

\bibitem[{{Parker}(1958)}]{Parker58}
{Parker}, E.~N. 1958, \apj, 128, 664, \dodoi{10.1086/146579}

\bibitem[{{Pillitteri} {et~al.}(2022){Pillitteri}, {Micela}, {Maggio},
  {Sciortino}, \& {Lopez-Santiago}}]{Pillitteri22}
{Pillitteri}, I., {Micela}, G., {Maggio}, A., {Sciortino}, S., \&
  {Lopez-Santiago}, J. 2022, \aap, 660, A75,
  \dodoi{10.1051/0004-6361/202142232}

\bibitem[{{Rumenskikh} {et~al.}(2022){Rumenskikh}, {Shaikhislamov},
  {Khodachenko}, {Lammer}, {Miroshnichenko}, {Berezutsky}, \&
  {Fossati}}]{2022ApJ...927..238R}
{Rumenskikh}, M.~S., {Shaikhislamov}, I.~F., {Khodachenko}, M.~L., {et~al.}
  2022, \apj, 927, 238, \dodoi{10.3847/1538-4357/ac441d}

\bibitem[{{Salz} {et~al.}(2016{\natexlab{a}}){Salz}, {Czesla}, {Schneider}, \&
  {Schmitt}}]{Salz16b}
{Salz}, M., {Czesla}, S., {Schneider}, P.~C., \& {Schmitt}, J.~H.~M.~M.
  2016{\natexlab{a}}, \aap, 586, A75, \dodoi{10.1051/0004-6361/201526109}

\bibitem[{{Salz} {et~al.}(2016{\natexlab{b}}){Salz}, {Schneider}, {Czesla}, \&
  {Schmitt}}]{Salz16a}
{Salz}, M., {Schneider}, P.~C., {Czesla}, S., \& {Schmitt}, J.~H.~M.~M.
  2016{\natexlab{b}}, \aap, 585, L2, \dodoi{10.1051/0004-6361/201527042}

\bibitem[{{Salz} {et~al.}(2018){Salz}, {Czesla}, {Schneider}, {Nagel},
  {Schmitt}, {Nortmann}, {Alonso-Floriano}, {L{\'o}pez-Puertas}, {Lamp{\'o}n},
  {Bauer}, {Snellen}, {Pall{\'e}}, {Caballero}, {Yan}, {Chen}, {Sanz-Forcada},
  {Amado}, {Quirrenbach}, {Ribas}, {Reiners}, {B{\'e}jar}, {Casasayas-Barris},
  {Cort{\'e}s-Contreras}, {Dreizler}, {Guenther}, {Henning}, {Jeffers},
  {Kaminski}, {K{\"u}rster}, {Lafarga}, {Lara}, {Molaverdikhani}, {Montes},
  {Morales}, {S{\'a}nchez-L{\'o}pez}, {Seifert}, {Zapatero Osorio}, \&
  {Zechmeister}}]{Salz18}
{Salz}, M., {Czesla}, S., {Schneider}, P.~C., {et~al.} 2018, \aap, 620, A97,
  \dodoi{10.1051/0004-6361/201833694}

\bibitem[{{Sanz-Forcada} {et~al.}(2011){Sanz-Forcada}, {Micela}, {Ribas},
  {Pollock}, {Eiroa}, {Velasco}, {Solano}, \&
  {Garc{\'\i}a-{\'A}lvarez}}]{Sanz2011}
{Sanz-Forcada}, J., {Micela}, G., {Ribas}, I., {et~al.} 2011, \aap, 532, A6,
  \dodoi{10.1051/0004-6361/201116594}

\bibitem[{{Sanz-Forcada} {et~al.}(2010){Sanz-Forcada}, {Ribas}, {Micela},
  {Pollock}, {Garc{\'\i}a-{\'A}lvarez}, {Solano}, \& {Eiroa}}]{Sanz2010}
{Sanz-Forcada}, J., {Ribas}, I., {Micela}, G., {et~al.} 2010, \aap, 511, L8,
  \dodoi{10.1051/0004-6361/200913670}

\bibitem[{{Sing} {et~al.}(2011){Sing}, {Pont}, {Aigrain}, {Charbonneau},
  {D{\'e}sert}, {Gibson}, {Gilliland}, {Hayek}, {Henry}, {Knutson}, {Lecavelier
  Des Etangs}, {Mazeh}, \& {Shporer}}]{2011MNRAS.416.1443S}
{Sing}, D.~K., {Pont}, F., {Aigrain}, S., {et~al.} 2011, \mnras, 416, 1443,
  \dodoi{10.1111/j.1365-2966.2011.19142.x}

\bibitem[{{Sing} {et~al.}(2016){Sing}, {Fortney}, {Nikolov}, {Wakeford},
  {Kataria}, {Evans}, {Aigrain}, {Ballester}, {Burrows}, {Deming},
  {D{\'e}sert}, {Gibson}, {Henry}, {Huitson}, {Knutson}, {Lecavelier Des
  Etangs}, {Pont}, {Showman}, {Vidal-Madjar}, {Williamson}, \&
  {Wilson}}]{2016Natur.529...59S}
{Sing}, D.~K., {Fortney}, J.~J., {Nikolov}, N., {et~al.} 2016, \nat, 529, 59,
  \dodoi{10.1038/nature16068}

\bibitem[{{Sing} {et~al.}(2019){Sing}, {Lavvas}, {Ballester}, {Lecavelier des
  Etangs}, {Marley}, {Nikolov}, {Ben-Jaffel}, {Bourrier}, {Buchhave}, {Deming},
  {Ehrenreich}, {Mikal-Evans}, {Kataria}, {Lewis}, {L{\'o}pez-Morales},
  {Garc{\'\i}a Mu{\~n}oz}, {Henry}, {Sanz-Forcada}, {Spake}, {Wakeford}, \&
  {PanCET Collaboration}}]{Sing2019}
{Sing}, D.~K., {Lavvas}, P., {Ballester}, G.~E., {et~al.} 2019, \aj, 158, 91,
  \dodoi{10.3847/1538-3881/ab2986}

\bibitem[{{Stancil} {et~al.}(1998){Stancil}, {Havener}, {Krsti{\'c}},
  {Schultz}, {Kimura}, {Gu}, {Hirsch}, {Buenker}, \& {Zygelman}}]{Stancil98}
{Stancil}, P.~C., {Havener}, C.~C., {Krsti{\'c}}, P.~S., {et~al.} 1998, \apj,
  502, 1006, \dodoi{10.1086/305937}

\bibitem[{{Storey} \& {Hummer}(1995)}]{Storey95}
{Storey}, P.~J., \& {Hummer}, D.~G. 1995, \mnras, 272, 41,
  \dodoi{10.1093/mnras/272.1.41}

\bibitem[{{Verner} {et~al.}(1996){Verner}, {Ferland}, {Korista}, \&
  {Yakovlev}}]{Verner96}
{Verner}, D.~A., {Ferland}, G.~J., {Korista}, K.~T., \& {Yakovlev}, D.~G. 1996,
  \apj, 465, 487, \dodoi{10.1086/177435}

\bibitem[{{Vidal-Madjar} {et~al.}(2003){Vidal-Madjar}, {Lecavelier des Etangs},
  {D{\'e}sert}, {Ballester}, {Ferlet}, {H{\'e}brard}, \& {Mayor}}]{Vidal03}
{Vidal-Madjar}, A., {Lecavelier des Etangs}, A., {D{\'e}sert}, J.~M., {et~al.}
  2003, \nat, 422, 143, \dodoi{10.1038/nature01448}

\bibitem[{{Vidal-Madjar} {et~al.}(2004){Vidal-Madjar}, {D{\'e}sert},
  {Lecavelier des Etangs}, {H{\'e}brard}, {Ballester}, {Ehrenreich}, {Ferlet},
  {McConnell}, {Mayor}, \& {Parkinson}}]{Vidal2004}
{Vidal-Madjar}, A., {D{\'e}sert}, J.~M., {Lecavelier des Etangs}, A., {et~al.}
  2004, \apjl, 604, L69, \dodoi{10.1086/383347}

\bibitem[{Virtanen {et~al.}(2020)Virtanen, Gommers, Oliphant, Haberland, Reddy,
  Cournapeau, Burovski, Peterson, Weckesser, Bright, {van der Walt}, Brett,
  Wilson, Millman, Mayorov, Nelson, Jones, Kern, Larson, Carey, Polat, Feng,
  Moore, {VanderPlas}, Laxalde, Perktold, Cimrman, Henriksen, Quintero, Harris,
  Archibald, Ribeiro, Pedregosa, {van Mulbregt}, \& {SciPy 1.0
  Contributors}}]{2020SciPy-NMeth}
Virtanen, P., Gommers, R., Oliphant, T.~E., {et~al.} 2020, Nature Methods, 17,
  261, \dodoi{10.1038/s41592-019-0686-2}

\bibitem[{{Vissapragada} {et~al.}(2022{\natexlab{a}}){Vissapragada}, {Knutson},
  {dos Santos}, {Wang}, \& {Dai}}]{Vissa22a}
{Vissapragada}, S., {Knutson}, H.~A., {dos Santos}, L.~A., {Wang}, L., \&
  {Dai}, F. 2022{\natexlab{a}}, \apj, 927, 96, \dodoi{10.3847/1538-4357/ac4e8a}

\bibitem[{{Vissapragada} {et~al.}(2022{\natexlab{b}}){Vissapragada}, {Knutson},
  {Greklek-McKeon}, {Oklop{\v{c}}i{\'c}}, {Dai}, {dos Santos}, {Jovanovic},
  {Mawet}, {Millar-Blanchaer}, {Paragas}, {Spake}, {Tinyanont}, \&
  {Vasisht}}]{Vissa22b}
{Vissapragada}, S., {Knutson}, H.~A., {Greklek-McKeon}, M., {et~al.}
  2022{\natexlab{b}}, \aj, 164, 234, \dodoi{10.3847/1538-3881/ac92f2}

\bibitem[{{Voronov}(1997)}]{Voronov97}
{Voronov}, G.~S. 1997, Atomic Data and Nuclear Data Tables, 65, 1,
  \dodoi{10.1006/adnd.1997.0732}

\bibitem[{{Watson} {et~al.}(1981){Watson}, {Donahue}, \& {Walker}}]{Watson1981}
{Watson}, A.~J., {Donahue}, T.~M., \& {Walker}, J.~C.~G. 1981, \icarus, 48,
  150, \dodoi{10.1016/0019-1035(81)90101-9}

\bibitem[{{Woodall} {et~al.}(2007){Woodall}, {Ag{\'u}ndez}, {Markwick-Kemper},
  \& {Millar}}]{Woodall07}
{Woodall}, J., {Ag{\'u}ndez}, M., {Markwick-Kemper}, A.~J., \& {Millar}, T.~J.
  2007, \aap, 466, 1197, \dodoi{10.1051/0004-6361:20064981}

\bibitem[{{Yan} {et~al.}(1998){Yan}, {Sadeghpour}, \& {Dalgarno}}]{Yan98}
{Yan}, M., {Sadeghpour}, H.~R., \& {Dalgarno}, A. 1998, \apj, 496, 1044,
  \dodoi{10.1086/305420}

\bibitem[{{Yee} {et~al.}(2021){Yee}, {Winn}, \& {Hartman}}]{Yee2021}
{Yee}, S.~W., {Winn}, J.~N., \& {Hartman}, J.~D. 2021, \aj, 162, 240,
  \dodoi{10.3847/1538-3881/ac2958}

\bibitem[{{Zhang} {et~al.}(2022){Zhang}, {Cauley}, {Knutson}, {France},
  {Kreidberg}, {Oklop{\v{c}}i{\'c}}, {Redfield}, \& {Shkolnik}}]{Zhang22}
{Zhang}, M., {Cauley}, P.~W., {Knutson}, H.~A., {et~al.} 2022, \aj, 164, 237,
  \dodoi{10.3847/1538-3881/ac9675}

\bibitem[{{Zhang} {et~al.}(2020){Zhang}, {Chachan}, {Kempton}, {Knutson}, \&
  {Chang}}]{2020ApJ...899...27Z}
{Zhang}, M., {Chachan}, Y., {Kempton}, E. M.~R., {Knutson}, H.~A., \& {Chang},
  W.~H. 2020, \apj, 899, 27, \dodoi{10.3847/1538-4357/aba1e6}

\bibitem[{{Zhu} \& {Dong}(2021)}]{2021ARA&A..59..291Z}
{Zhu}, W., \& {Dong}, S. 2021, \araa, 59,
  \dodoi{10.1146/annurev-astro-112420-020055}

\end{thebibliography}
\bibliographystyle{aasjournal}

\end{document}